\documentclass[useAMS,usenatbib]{mn2e}
\usepackage[english]{babel} 
\usepackage{amsmath}
\usepackage{amssymb}
\usepackage{calc}
\usepackage[dvipdfm]{graphicx}
\usepackage{bmpsize}
\usepackage{longtable}
\usepackage{rotating}
\usepackage{ctable}
\usepackage{lscape}
\usepackage{url}
\usepackage{blindtext}
\usepackage{listings}
\usepackage{multicol}
\usepackage{caption}
\usepackage{appendix}
\newcommand{\lsim}{\mathrel{\hbox{\rlap{\lower.55ex 
\hbox{$\sim$}} \kern-.3em \raise.4ex \hbox{$<$}}}}
\newcommand{\gsim}{\mathrel{\hbox{\rlap{\lower.55ex 
\hbox{$\sim$}} \kern-.3em \raise.4ex \hbox{$>$}}}}
\newcommand{\figdir}{./}

\newcommand{\myincludegraphics}[2]{\includegraphics*[#1]{#2}}

\newcommand{\catnameshort}{Spitzer Data Fusion}
\title[The HerMES sub-millimetre local and low-redshift luminosity functions]
{The HerMES sub-millimetre local and low-redshift luminosity functions \thanks{\textit{Herschel} is an ESA space observatory with science instruments provided by European-led Principal Investigator consortia and with important participation from NASA.}}
\author[L.~Marchetti et al.]
{\parbox{\textwidth}{\raggedright L.~Marchetti,$^{1,2}$
M.~Vaccari,$^{3,2,32}$
A.~Franceschini,$^{2}$
V.~Arumugam,$^{4}$
H.~Aussel,$^{5}$
M.~B{\'e}thermin,$^{5,6,18}$
J.~Bock,$^{7,8}$
A.~Boselli,$^{9}$
V.~Buat,$^{9}$
D.~Burgarella,$^{9}$
D.L.~Clements,$^{10}$
A.~Conley,$^{11}$
L.~Conversi,$^{12}$
A.~Cooray,$^{13,7}$
C.D.~Dowell,$^{7,8}$
D.~Farrah,$^{15}$
A.~Feltre,$^{34}$
J.~Glenn,$^{16,11}$
M.~Griffin,$^{17}$
E.~Hatziminaoglou,$^{18}$
S.~Heinis,$^{9}$
E.~Ibar,$^{20}$
R.J.~Ivison,$^{19,4}$
H.T.~Nguyen,$^{8,7}$
B.~O'Halloran,$^{10}$
S.J.~Oliver,$^{14}$
A.~Omont,$^{21}$
M.J.~Page,$^{22}$
A.~Papageorgiou,$^{17}$
C.P.~Pearson,$^{23,1}$
I.~P{\'e}rez-Fournon,$^{24,25}$
M.~Pohlen,$^{17}$
D.~Rigopoulou,$^{23,26}$
I.G.~Roseboom,$^{14,4}$
M.~Rowan-Robinson,$^{10}$
B.~Schulz,$^{7,27}$
Douglas~Scott,$^{28}$
N.~Seymour,$^{29}$
D.L.~Shupe,$^{7,27}$
A.J.~Smith,$^{14}$
M.~Symeonidis,$^{22}$
I.~Valtchanov,$^{12}$
M.~Viero,$^{7}$
L.~Wang,$^{30,33}$
J.~Wardlow,$^{31}$
C.K.~Xu$^{7,27}$ and
M.~Zemcov$^{7,8}$}\vspace{0.4cm}\\
\parbox{\textwidth}{\raggedright $^{1}$Department of Physical Sciences, The Open University, Milton Keynes MK7 6AA, UK\\
$^{2}$Dipartimento di Fisica e Astronomia, Universit\`{a} di Padova, vicolo Osservatorio, 3, 35122 Padova, Italy\\
$^{3}$Astrophysics Group, Physics Department, University of the Western Cape, Private Bag X17, 7535, Bellville, Cape Town, South Africa\\
$^{4}$Institute for Astronomy, University of Edinburgh, Royal Observatory, Blackford Hill, Edinburgh EH9 3HJ, UK\\
$^{5}$Laboratoire AIM-Paris-Saclay, CEA/DSM/Irfu - CNRS - Universit\'e Paris Diderot, CE-Saclay, pt courrier 131, F-91191 Gif-sur-Yvette, France\\
$^{6}$Institut d'Astrophysique Spatiale (IAS), b\^atiment 121, Universit\'e Paris-Sud 11 and CNRS (UMR 8617), 91405 Orsay, France\\
$^{7}$California Institute of Technology, 1200 E. California Blvd., Pasadena, CA 91125, USA\\
$^{8}$Jet Propulsion Laboratory, 4800 Oak Grove Drive, Pasadena, CA 91109, USA\\
$^{9}$Aix-Marseille Universit\'e, CNRS, LAM (Laboratoire d'Astrophysique de Marseille) UMR7326, 13388, France\\
$^{10}$Astrophysics Group, Imperial College London, Blackett Laboratory, Prince Consort Road, London SW7 2AZ, UK\\
$^{11}$Center for Astrophysics and Space Astronomy 389-UCB, University of Colorado, Boulder, CO 80309, USA\\
$^{12}$Herschel Science Centre, European Space Astronomy Centre, Villanueva de la Ca\~nada, 28691 Madrid, Spain\\
$^{13}$Dept. of Physics \& Astronomy, University of California, Irvine, CA 92697, USA\\
$^{14}$Astronomy Centre, Dept. of Physics \& Astronomy, University of Sussex, Brighton BN1 9QH, UK\\
$^{15}$Department of Physics, Virginia Tech, Blacksburg, VA 24061, USA\\
$^{16}$Dept. of Astrophysical and Planetary Sciences, CASA 389-UCB, University of Colorado, Boulder, CO 80309, USA\\
$^{17}$School of Physics and Astronomy, Cardiff University, Queens Buildings, The Parade, Cardiff CF24 3AA, UK\\
$^{18}$ESO, Karl-Schwarzschild-Str. 2, 85748 Garching bei M\"unchen, Germany\\
$^{19}$UK Astronomy Technology Centre, Royal Observatory, Blackford Hill, Edinburgh EH9 3HJ, UK\\
$^{20}$Instituto de F\'isica y Astronom\'ia, Universidad de Valpara\'iso, Avda. Gran Breta\~na 1111, Valpara\'iso, Chile\\
$^{21}$Institut d'Astrophysique de Paris, UMR 7095, CNRS, UPMC Univ. Paris 06, 98bis boulevard Arago, F-75014 Paris, France\\
$^{22}$Mullard Space Science Laboratory, University College London, Holmbury St. Mary, Dorking, Surrey RH5 6NT, UK\\
$^{23}$RAL Space, Rutherford Appleton Laboratory, Chilton, Didcot, Oxfordshire OX11 0QX, UK\\
$^{24}$Instituto de Astrof{\'\i}sica de Canarias (IAC), E-38200 La Laguna, Tenerife, Spain\\
$^{25}$Departamento de Astrof{\'\i}sica, Universidad de La Laguna (ULL), E-38205 La Laguna, Tenerife, Spain\\
$^{26}$Department of Physics, Denys Wilkinson Building, University of Oxford, Keble Road, Oxford OX1 3RH, UK\\
$^{27}$Infrared Processing and Analysis Center, MS 100-22, California Institute of Technology, JPL, Pasadena, CA 91125, USA\\
$^{28}$Department of Physics \& Astronomy, University of British Columbia, 6224 Agricultural Road, Vancouver, BC V6T~1Z1, Canada\\
$^{29}$International Centre for Radio Astronomy Research, Curtin University, Perth, Australia\\
$^{30}$Institue for Computational Cosmology, Durham University, South Road, Durham, DH1 3LE, UK\\
$^{31}$Dark Cosmology Centre, Niels Bohr Institute, University of Copenhagen, Juliane Maries Vej 30, 2100 Copenhagen, Denmark\\
$^{32}$INAF - Istituto di Radioastronomia, via Gobetti 101, 40129 Bologna, Italy\\
$^{33}$SRON Netherlands Institute for Space Research, Landleven 12, 9747 AD, Groningen, The Netherlands\\
$^{34}$Sorbonne Universit\'es, UPMC-CNRS, UMR7095, Institut d'Astrophysique de Paris, F-75014, Paris, France.}}

\begin{document}
%
\date{Accepted ??. Received ??; in original form 2015 December 31}
\pagerange{\pageref{firstpage}--\pageref{lastpage}} \pubyear{2015}
\maketitle
\label{firstpage}
\clearpage
  \begin{abstract}
We used wide area surveys over 39 deg$^2$ by the HerMES collaboration, performed with the \textit{Herschel} Observatory SPIRE multi-wavelength camera, to estimate the low-redshift, $0.02<z<0.5$, monochromatic luminosity functions (LFs) of galaxies at 250, 350 and 500$\,\mu$m. Within this redshift interval we detected 7087 sources in 5 independent sky areas, $\sim$40\% of which have spectroscopic redshifts, while for the remaining objects photometric redshifts were used.
The SPIRE LFs in different fields did not show any field-to-field variations beyond the small differences to be expected from cosmic variance. 
SPIRE flux densities were also combined with \textit{Spitzer} photometry and multi-wavelength archival data to perform a complete SED fitting analysis of SPIRE detected sources to calculate precise k-corrections, as well as the bolometric infrared (8-1000$\,\mu$m) luminosity functions and their low-$z$ evolution from a combination of statistical estimators.
Integration of the latter prompted us to also compute the local luminosity density (LLD) and the comoving star formation rate density (SFRD) for our sources, and to compare them with theoretical predictions of galaxy formation models.
The luminosity functions show significant and rapid luminosity evolution already at low redshifts, $0.02<z<0.2$, with L$_{IR}^* \propto (1+z)^{6.0\pm0.4}$ and $\Phi_{IR}^* \propto (1+z)^{-2.1\pm0.4}$, L$_{250}^* \propto (1+z)^{5.3\pm0.2}$ and $\Phi_{250}^* \propto (1+z)^{-0.6\pm0.4}$ estimated using the IR bolometric and the 250$\,\mu$m LFs respectively. Converting our IR LD estimate into an SFRD assuming a standard Salpeter IMF and including the unobscured contribution based on the UV dust-uncorrected emission from local galaxies, we estimate a SFRD scaling of SFRD$_0+0.08 z$,  where SFRD$_0\simeq (1.9\pm 0.03)\times 10^{-2} [\mathrm{M}_\odot\,\mathrm{Mpc}^{-3}]$ is our total SFRD estimate at $z\sim0.02$.
  \end{abstract}
\section{Introduction}
Observations carried out in roughly the past twenty years have revealed a rapid evolution of cosmic sources, both normal, actively star-forming and AGN-dominated galaxies, over the last several billion years. This was mostly achieved from continuum rest-frame UV photometric imaging in the optical (e.g. \citealt{Lilly1995}), and H$\alpha$ or [OII] line spectroscopy (e.g. \citealt{Gallego1995}), all, however, including very uncertain dust extinction corrections. \textit{Galex} has also been used for similar purposes by \cite{Martin2005.59M} and \cite{Bothwell2011}, among others.

In the far IR, the pioneering exploration by the \textit{IRAS} satellite revealed a particularly dramatic evolution of the galaxy LFs (\citealt{Saunders1990}), illustrating the importance of local studies at infrared (IR) wavelengths. This result was later confirmed up to $z\simeq1$ by \textit{ISO} (\citealt{Pozzi2004}), and \textit{Spitzer} studies using the MIPS 24$\,\mu$m (\citealt{LeFloch2005}, \citealt{Marleau2007}, \citealt{Rodighiero2010}) and 70$\,\mu$m (\citealt{Frayer2006}, \citealt{Huynh2007}, \citealt{Magnelli2009},\citealt{Patel2013}) channels. At longer, sub-millimetre wavelengths the balloon borne telescope BLAST was able to estimate the galaxy LF at low \textit{z} and map its evolution (\citealt{Eales2009}), although with limited statistics and uncertain identification of the sources.
Finally, surveys in the radio bands have also been exploited, with the necessity to include large bolometric corrections, for luminosity function estimates (\citealt{Condon1989}; \citealt{Serjeant2002}).

Interpretations of these fast evolutionary rates are actively debated in the literature, with various processes being claimed as responsible (like gas fuel consumption in galaxies, heating of the gas so as to prevent cooling and collapse, decreasing rates of galaxy interactions with time, etc.). Indeed, galaxy evolution codes have often found it difficult to reproduce these data, and slower evolution seems predicted by the models than it is observed.

However, the estimates of the low-redshift luminosity functions of galaxies, and correspondingly the total star-formation and AGN accretion rates, still contain some significant uncertainties. 
In particular, due to the moderate volumes sampled at low redshift, an essential pre-requisite for determining the LLFs is the imaging of large fields, where it is difficult however to achieve the required multi-wavelength homogeneous coverage and complete redshift information.

In the very local universe, at $z<0.02$, a sample of a few hundreds sources from the Early Release Compact Source Catalogue by the \textit{Planck} all-sky survey (Planck Collaboration VII, 2011) have been used by \cite{Negrello2013} to estimate luminosity functions at 350, 500, and 850$\,\mu$m. Although the authors were very careful to account for various potentially problematic factors, namely the photometric calibration from the large \textit{Planck} beam, removal of Galactic emission and CO line contribution to the photometry, their estimate might not be completely immune to the effects of large inhomogeneities (like the Virgo cluster) inherent in their very local spatial sampling (see Sec. \ref{discussion} for further details).

\cite{Vaccari2010} report a preliminary determination of the local sub-millimetre luminosity functions of galaxies, exploiting the much improved angular resolution and mapping speed of the SPIRE instrument \citep{Griffin2010} on the \textit{Herschel Space Observatory} \citep{Pilbratt2010}. They used data from the Lockman Hole and Extragalactic First Look Survey (XFLS) fields of the \textit{Herschel} Multi-tiered Extragalactic Survey program (HerMES, \texttt{http://hermes.sussex.ac.uk},  \citealt{Oliver2012}) over about 15 deg$^2$ observed during the \textit{Herschel} Science Demonstration Phase (SDP), and including a few hundred sources to a flux limit of about 40 mJy in all three SPIRE bands (250, 350, 500$\,\mu$m). Their published functions were integrated over a wide redshift interval at $0<z<0.2$.  Because of the limited source statistics, \cite{Vaccari2010} could not
take into account any evolutionary corrections, while significant evolution is expected to be present over this large redshift bin.

Still based on the HerMES database, but using a much larger total area, many more independent sky fields, and deeper fluxes, the present paper reports on a systematic effort to characterise the local and low-redshift luminosity functions of galaxies in the sub-millimetre bins. The \textit{Herschel} survey catalogue has been cross-correlated with existing optical photometry and spectroscopy in the fields, as well as with photometric data in the mid- and far-IR from \textit{Spitzer} (\citealt{Werner2004}). By fitting the source-by-source multi-wavelength photometry with spectral templates, the bolometric IR luminosities and bolometric luminosity functions can also be estimated.
Importantly, the much improved statistics allows us to work in narrow redshift bins, so as to disentangle luminosity function shapes from evolution, and to obtain the most robust and complete statistical characterisation over the last few Gyrs of galaxy evolution.
By combining this long-wavelength information with similar analyses in the optical-UV, we can determine the local bolometric luminosity density and comoving star-formation rate and their low-$z$ evolution.

The paper is structured as follows. 
In Section 2 we describe the multi-wavelength data set that we use, as well as the selection of the samples, source identification, and SED fitting. 
In Section 3 we detail the statistical methods used in our data analysis and the various adopted luminosity function estimators, including the Bayesian parametric recipe that we develop.
Our results are reported in Section 4, including the multi-wavelength luminosity functions, the local luminosity densities and the star-formation rates. Our results are then discussed in Section 5 and our main conclusions summarised in Section 6.

Throughout the paper we adopt a standard cosmology with $\Omega_\mathrm{M}=0.3$, $\Omega_\Lambda=0.7$ and $H_0=70~\mathrm{km\,s^{-1}\,Mpc^{-1}}$.

\section{The HerMES Wide Sample}\label{sample.sec}
The Herschel Multi-tiered Extragalactic Survey, or HerMES, is a \textit{Herschel} Guarantee Time (GT) Key Programme (\citealt{Oliver2012} \footnote{\url{http://hermes.sussex.ac.uk/}}) and the largest single project on \textit{Herschel}, for a total 900 hours of observing time. HerMES was designed to comprise a number of tiers of different depths and areas, and has performed observations with both SPIRE \citep{Griffin2010} and PACS \citep{Poglitsch2010}, surveying approximately 70 deg$^2$ over 12 fields whose sizes range from 0.01 to 20 deg$^2$.

To estimate the SPIRE LLF we use HerMES L5/L6 SPIRE observations (see Tab. 1 in \citealt{Oliver2012} for more details on the observations) covering five fields: Lockman Hole (LH); Extragalactic First Look Survey (XFLS); Bootes, ELAIS-N1 (EN1) and XMM-LSS. In the following, these fields and the SPIRE sample arising from them will collectively be referred to as the HerMES Wide Fields and Sample respectively. These fields are the widest \textit{Herschel} HerMES fields where imaging data are available with both \textit{Spitzer} IRAC and MIPS cameras, thus enabling the detailed study of the full infrared SED of a significant number of sources in the local Universe. 

\begin{table}
\centering
\begin{tabular}{cccc}
\hline
\textbf{Field} & \textbf{250$\,\mu$m detections} & \textbf{Area [deg$^2$]} & \textbf{Set}\\
\hline\hline
LH & 2336 (942/1394) & $~11.29$ & 34 \\
XFLS & 801 (427/374) & $~4.19$ & 40  \\
BOOTES & 1792 (1220/572) & $~9.93$ & 37 \\
EN1 & 693 (246/447) & $~3.91$ & 35 \\
XMM & 1606 (367/1239)& $~9.59$ & 36 \\
Total &  7087 (3195/3892) & $~38.9$ &  \\ 
\hline
\end{tabular}
\caption[]{Number of $0.02 < z \lsim0.5 $ sources used to estimate the SPIRE LLFs. The number of sources with spectroscopic/photometric redshifts is indicated in brackets after the total number of sources. The 250$\,\mu$m sample is cut at $S_{\mathrm{250}} > 30$~mJy, according to the SPIRE 250$\,\mu$m completeness (see text for details).''Set'' refers to Tab. 1 in \cite{Oliver2012} and identifies the HerMES specific observing mode in each field.}
\label{spire-llf-numbers.tab}
\end{table}
 
\subsection{SPIRE source extraction}\label{extraction.sec}

Source confusion is the most serious challenge for \textit{Herschel} and SPIRE source extraction and identification. In particular, confusion is an important driver in determining the optimal survey depth. By making, a maximum use of the full spectrum of ancillary data it is possible to limit the confusion problem at the source detection and identification steps. For this reason the choice of HerMES survey fields has been largely driven by the availability of extensive multi-wavelength ancillary data at both optical and infrared wavelengths. In particular, \cite{Roseboom2010} (and \citealt{Roseboom2012}) developed a new method for SPIRE source extraction, hereafter referred to as XID, which improves upon more conventional techniques (e.g., \citealt{Smith2012}; \citealt{Wang2014}) based on existing source extraction algorithms optimised for \textit{Herschel} purposes.

The XID technique makes use of a combination of linear inversion and model selection techniques to produce reliable cross-identification catalogues based on \textit{Spitzer} MIPS 24$\,\mu$m source positions. The tiered nature of HerMES is well matched to the variable quality of the \textit{Spitzer} data, in particular the MIPS 24$\,\mu$m observations. This is confirmed by simulation performed using pre-\textit{Herschel} empirical models (e.g., \citealt{FernandezConde2008}; \citealt{LeBorgne2009}; \citealt{Franceschini2010}) which shared the comparable sensitivities of the 250 and 24$\,\mu$m source densities. Since the HerMES Wide fields are homogeneously covered by the \catnameshort\ (described in Sec~\ref{df.sec}), which provides homogeneous MIPS 24$\,\mu$m source lists, the SPIRE flux densities used in this paper are obtained with the XID technique using the \catnameshort\ MIPS 24$\,\mu$m positional priors (or, in other words, the MIPS 24$\,\mu$m positions are used as a prior to guide the SPIRE flux extraction on the SPIRE maps).

As reported in \cite{Roseboom2010}, using a prior for the SPIRE source identification based on MIPS 24$\,\mu$m detections could, in principle, introduce an additional incompleteness related to the relative depth at 24$\,\mu$m catalogues used in the process and the distribution of intrinsic SED shapes. However, \cite{Roseboom2010} show how incompleteness would affect only the fainter SPIRE sources with the higher 250$\,\mu$m / 24$\,\mu$m flux density ratios, which are very likely to be ultra-red high-redshift objects. We can therefore be confident that for relatively nearby sources the XID catalogues are complete at the relatively bright flux limits used in this work. This relatively complex procedure of association is reported in the dedicated papers by \cite{Roseboom2010} and \cite{Roseboom2012}, to which we refer the reader for further details about this method.

\subsection{SPIRE sample selection}\label{sam.sec.}

To define the sample to be used for our LLF determinations, we use SPIRE flux density estimates obtained using the XID method (\citealt{Roseboom2010} and \citealt{Roseboom2012}), applied to SCAT maps produced by \cite{Smith2012} and using MIPS 24$\,\mu$m positional priors based on the \catnameshort\ detailed in Sec.~\ref{df.sec}. The SPIRE 250$\,\mu$m channel is the most sensitive of the three SPIRE bands and thus we select sources on the basis of a SPIRE 250$\,\mu$m reliability criterion (discussed in \citealt{Roseboom2010}) defined as $\chi^2_{250} < 5$ and $\mathrm{SNRT}_{250} > 4$, where the first criterion is the $\chi^2$ of the source solution in the neighbourhood of a source (7 pixel radius) and the second is the signal to noise ratio (SNR) at a given selection $\lambda$, including confusion, and referred to as 'Total' SNR$_{\lambda}$ or SNRT$_{\lambda}$.

\begin{figure*}
\begin{center}
   {\myincludegraphics{width=0.43\textwidth}{\figdir{}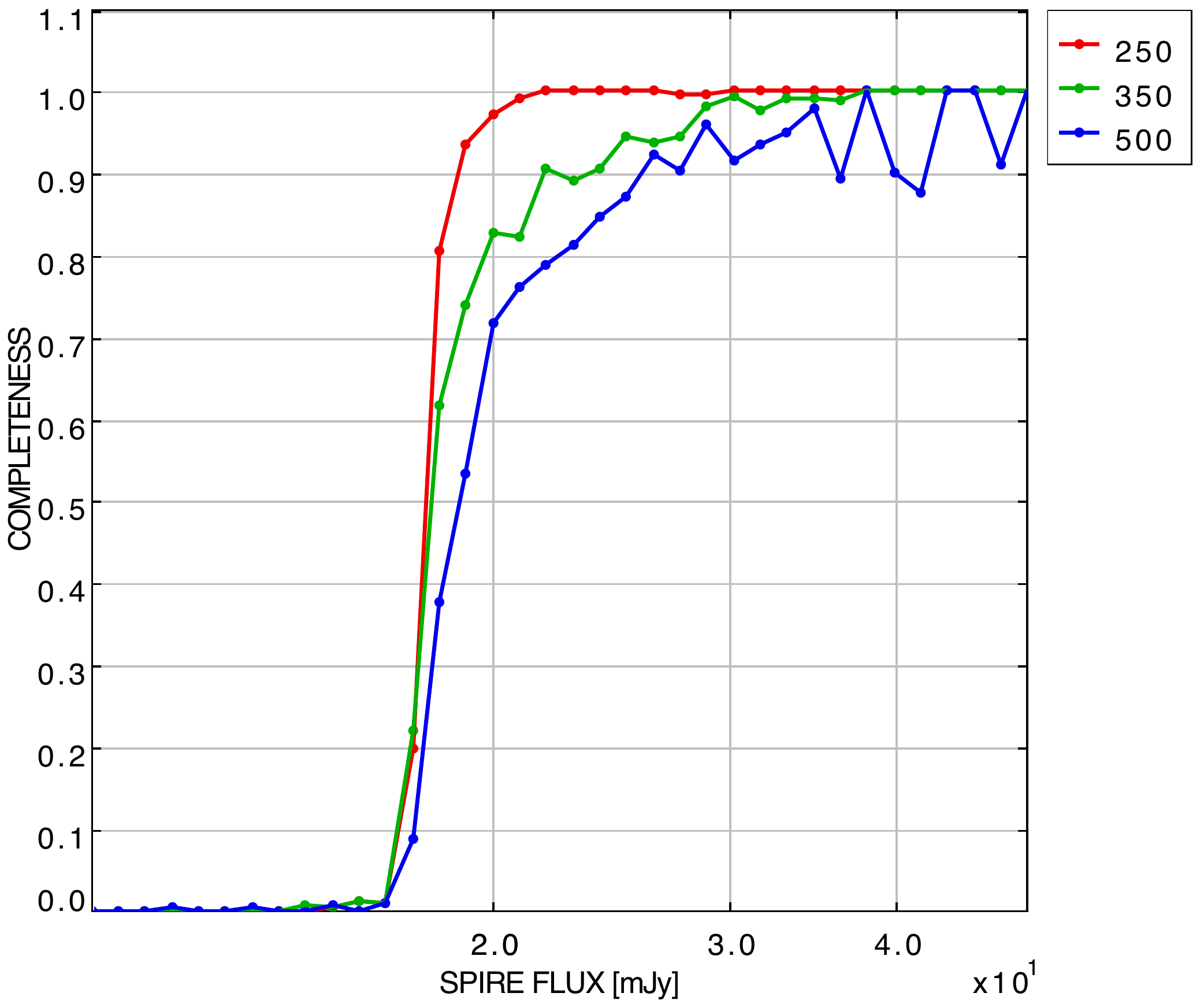}}
   {\myincludegraphics{width=0.53\textwidth}{\figdir{}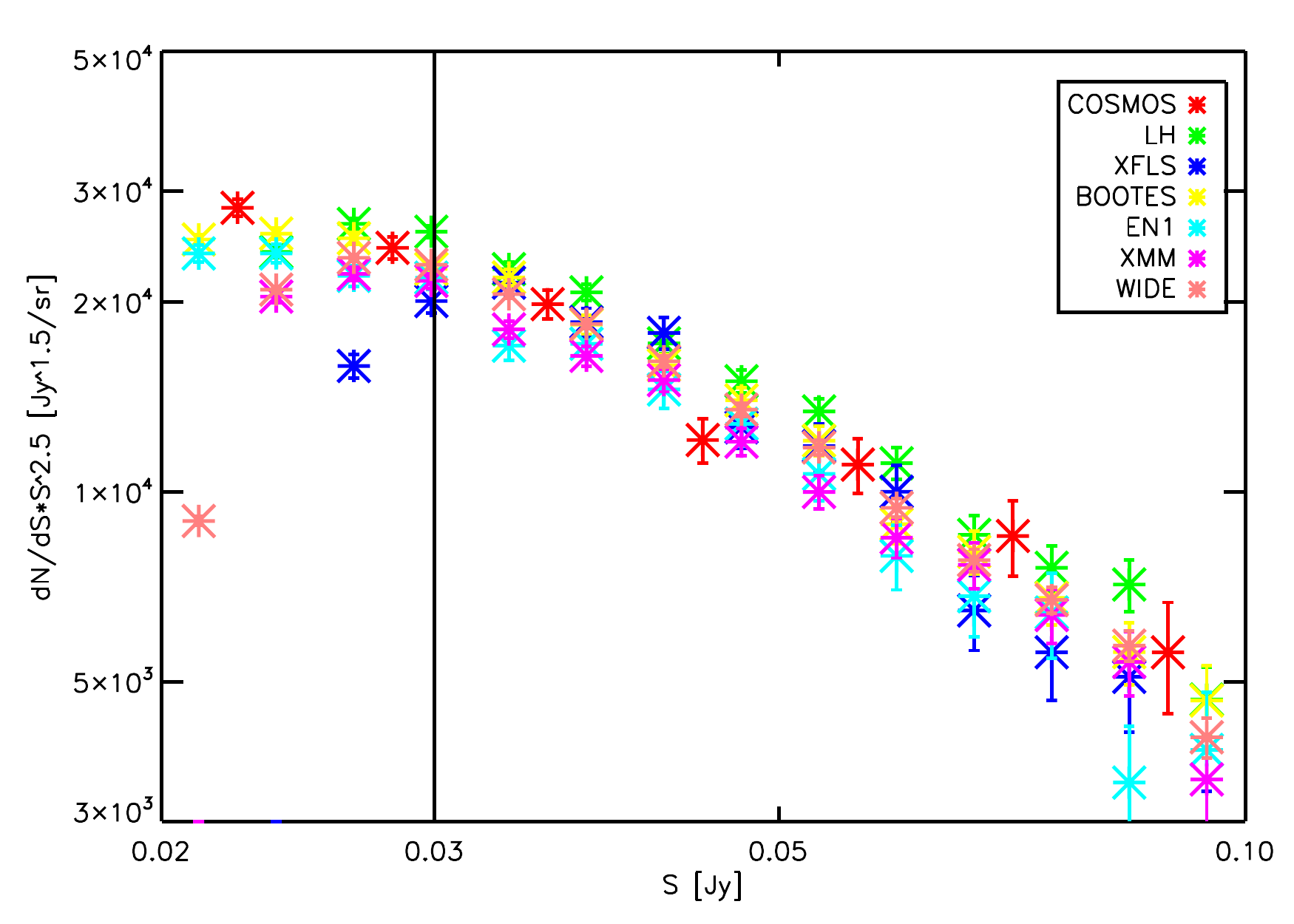}}
\end{center}
\caption[]{SPIRE 250$\,\mu$m source counts (right) and completeness (left) based on XID catalogues from \cite{Roseboom2012} for the HerMES Wide Fields sample used to estimate the SPIRE LLFs, compared with COSMOS estimates from Vaccari et al. (in prep.). The black solid line signs the flux limit of our selection.}
\label{wide.comp}
\end{figure*}

The SPIRE 250$\,\mu$m catalogues of L5/L6 HerMES observations are highly complete and reliable down to approximately 25/30/35$\,$mJy at 250/350/500$\,\mu$m, respectively, as shown in Fig. \ref{wide.comp} (left). In order to combine the data collected in these different fields we have to ensure a uniform completeness both in flux and in redshift coverage across fields, thus, due to some minor differences across the fields we decided to cut our sample at 30$\,$mJy at 250 $\mu$m. These minor differences are visible in Fig. \ref{wide.comp} (right) where we compare SPIRE 250$\,\mu$m number counts estimated for the 5 wide fields and for the COSMOS deep field (the COSMOS sample is from Vaccari et al., in prep). 
These discrepancies  are consistent with the levels of cosmic variance predicted by theoretical models for fields of this size \citep{Moster2011}, as well as with the slightly different depths of MIPS 24$\,\mu$m observations available for these fields, which were used to guide HerMES XID source extraction. In any case, differences are on the whole small and have major effects only at low flux densities, well below our selected limit. The greatest discrepancy is shown in XFLS, where the SPIRE 250$\,\mu$m completeness reflects the slightly brighter flux limit of the XFLS MIPS 24$\,\mu$m and IRAC catalogues, due to a shorter exposure time in comparison with the other fields. 

\subsection{The \textit{\catnameshort}}\label{df.sec}

As previously mentioned, the HerMES fields were chosen so as to have the best multi-wavelength data for sky areas of a given size. In particular, the fields used in this work are covered by Spitzer 7-band IRAC and MIPS imaging data which enable not only an improved identification process but also the detailed characterisation of the infrared SEDs of \textit{Herschel} sources.

In this work we exploit the \textit{Spitzer Data Fusion} (\citealt{Vaccari2010} and Vaccari et al., in prep., \url{http://www.mattiavaccari.net/df}). The \catnameshort\ combines \textit{Spitzer} mid- and far-infrared data from the Spitzer Wide-area InfraRed Extragalactic (SWIRE, \citealt{Lonsdale2003}) survey in six fields, the Spitzer Deep-Wide Field Survey (SDWFS, PI Daniel Stern, Spitzer PID 40839), the Spitzer Extragalactic First Look Survey (XFLS, PI Tom Soifer, Spitzer PID 26), with photometric data at UV, optical and NIR wavelengths, as well as optical spectroscopy over about 70 deg$^2$ in total. It thus makes full use of public survey data from the GALEX, SDSS, INT WFS, 2MASS, UKIDSS and VISTA projects, as well as further optical imaging obtained by the SWIRE, SDWFS and XFLS teams. It also provides spectroscopic information variously available from SDSS, NASA/IPAC Extragalactic Database (NED \url{http://ned.ipac.caltech.edu}), recent literature and proprietary follow-up programmes.

The \catnameshort\ thus represents an ideal starting point to perform statistical studies of infrared galaxy populations, such as detailed SED fitting analyses to estimate photometric redshifts and masses, as well as star formation rates (SFRs); an early version of the database has already been used to that effect by \cite{RowanRobinson2013}. It has been used to validate \textit{Herschel} SDP observations within the HerMES consortium team and to produce current and future public HerMES catalogues \footnote{available at \url{http://hedam.oamp.fr/HerMES/}}. Since this paper only uses the \catnameshort\ to derive SPIRE local luminosity function estimates, we refer the reader to Vaccari et al. (in prep.) for a complete description of the database and in the following we only summarise its basic properties as they relate to this work.

The \catnameshort\  is constructed by combining \textit{Spitzer} IRAC and MIPS source lists, as well as ancillary catalogues, following a positional association procedure. Source extraction of IRAC 4-band images and of MIPS 24$\,\mu$m images is carried out using Sextractor \citep{Bertin1996}, whereas MIPS 70 and 160$\,\mu$m source extraction is carried out using APEX \citep{Makovoz2005}. Catalogue selection is determined by a reliable IRAC 3.6 or IRAC 4.5$\,\mu$m detection. We then associate MIPS 24$\,\mu$m detections to IRAC detections using a 3 arcsec search radius, while MIPS 70 and 160$\,\mu$m catalogues are matched against MIPS 24$\,\mu$m positions using a search radius of 6 and 12 arcsec, respectively. UV, optical and near-infrared catalogues are then matched against IRAC positions using a 1 arcsec search radius. This multi-step approach increases the completeness and reliability of the longer-wavelength associations, while better pin-pointing MIPS sources using their IRAC positions.

\begin{figure*}
\begin{center}
   {\myincludegraphics{width=0.45\textwidth}{\figdir{}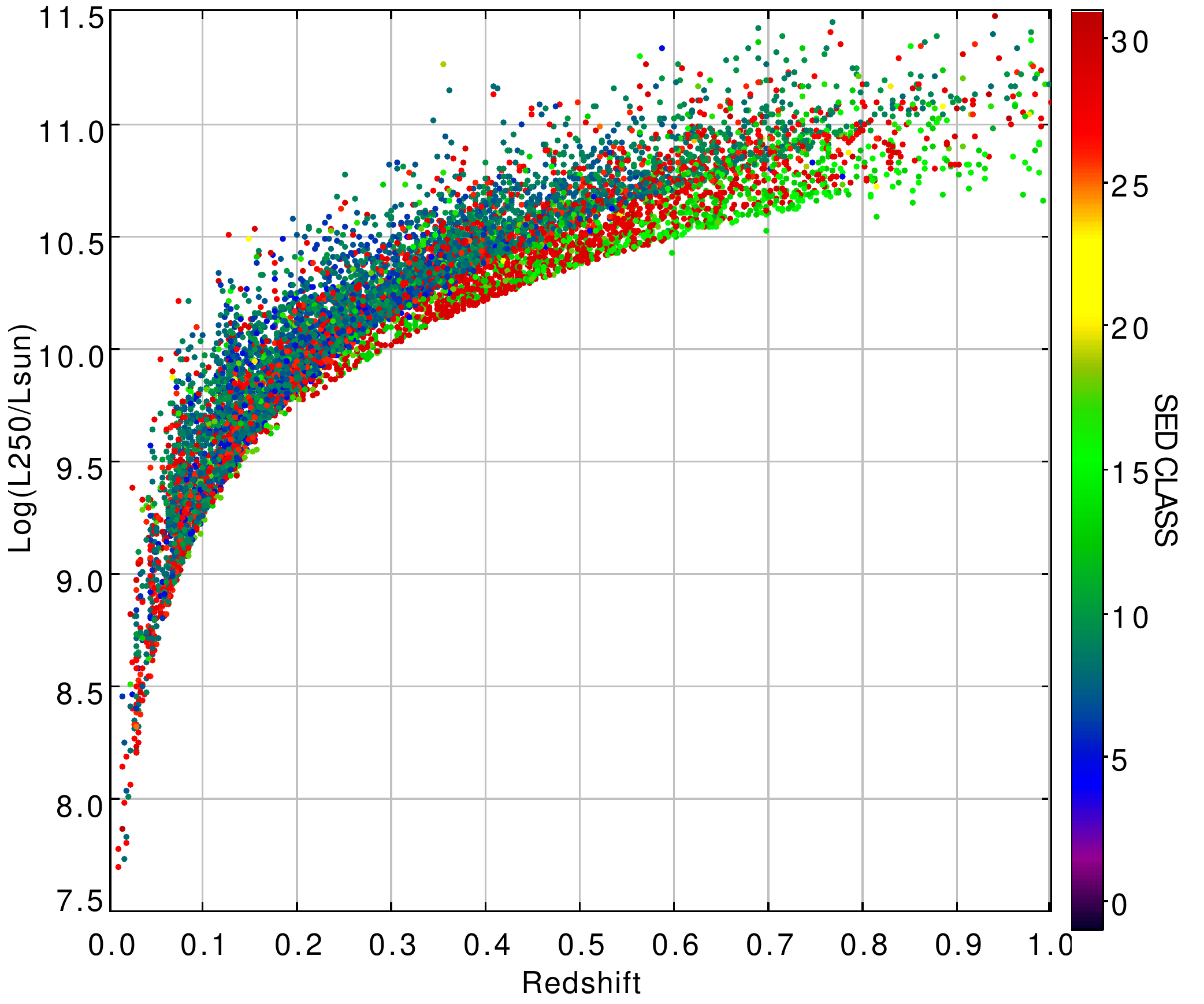}}
   {\myincludegraphics{width=0.45\textwidth}{\figdir{}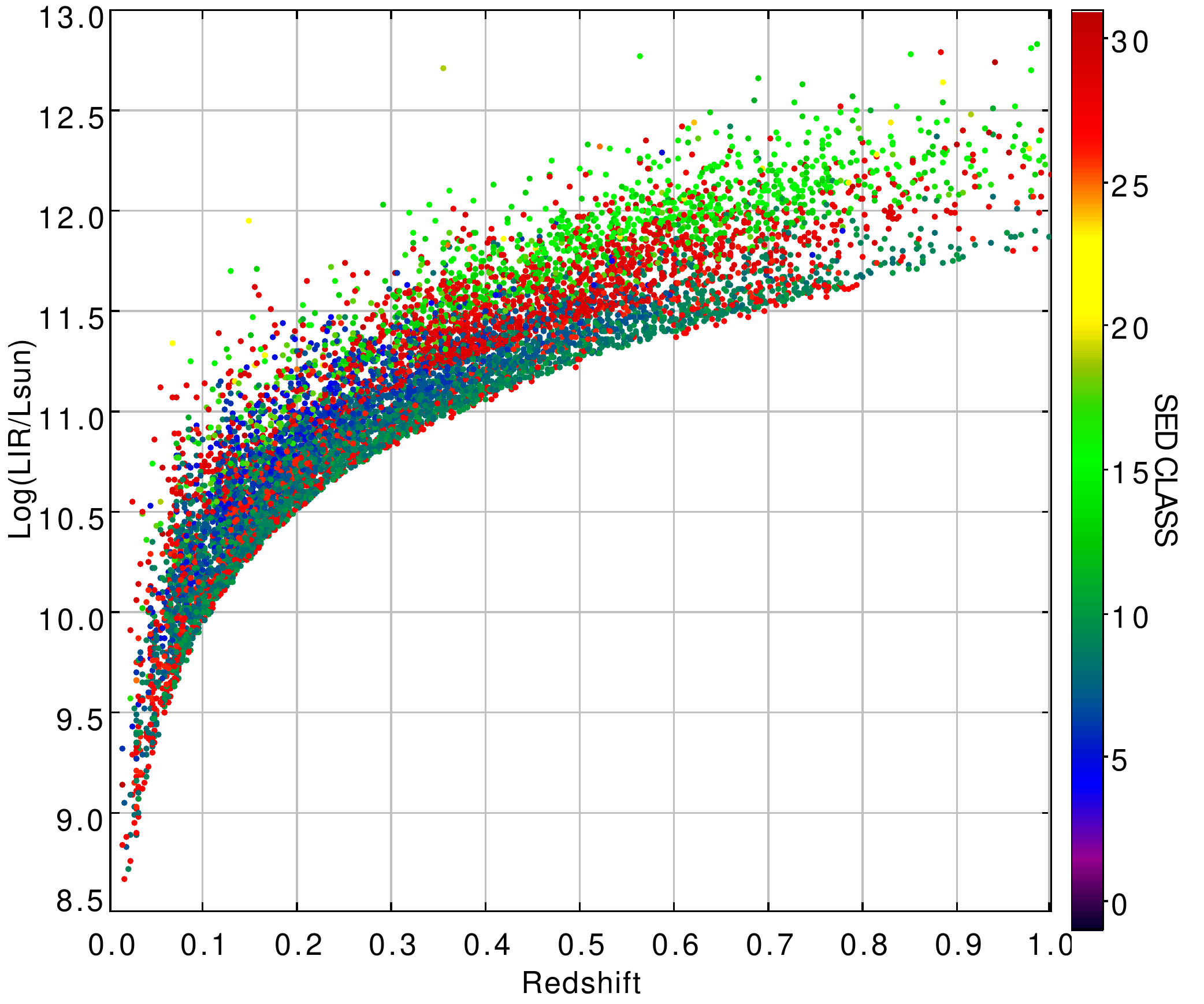}}
   {\myincludegraphics{width=0.45\textwidth}{\figdir{}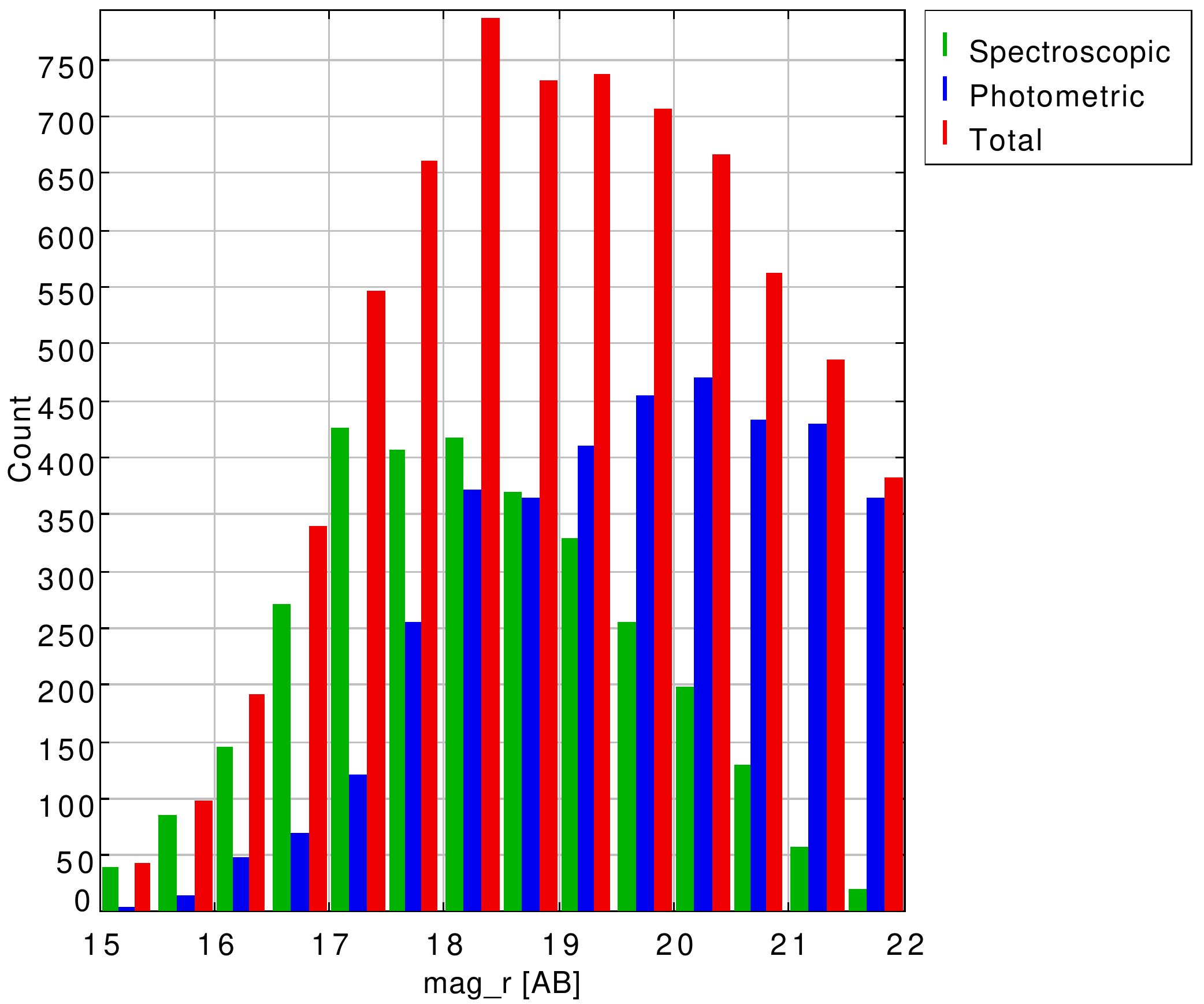}}
   {\myincludegraphics{width=0.45\textwidth}{\figdir{}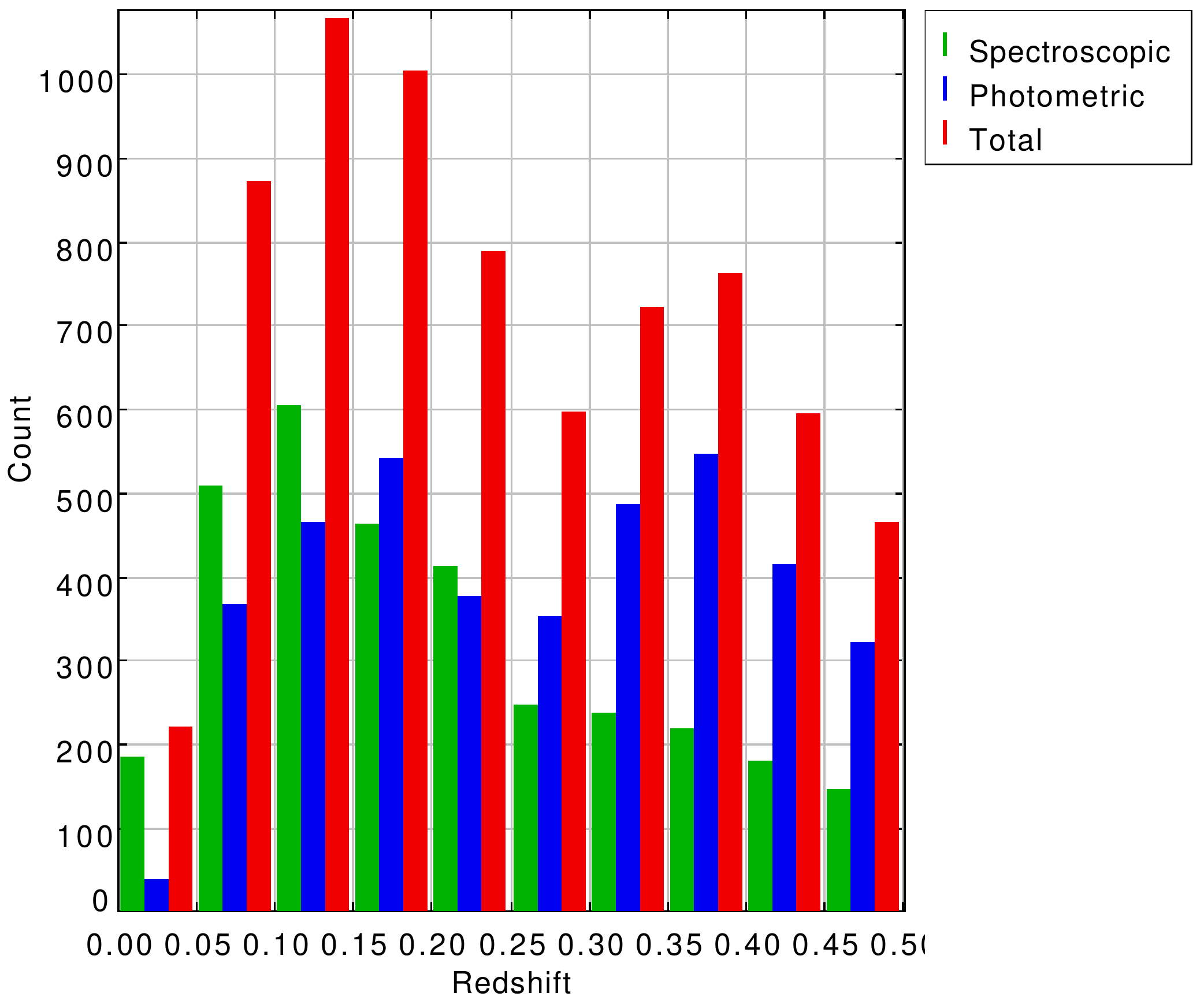}}
\end{center}
\caption[]{\textbf{Top}: SPIRE 250$\,\mu$m (expressed as $\nu L_\nu$) and IR bolometric luminosity versus redshift; \textbf{Bottom}: SDSS $r_{\rm AB}$ (left) and redshift histograms (right) for the HerMES Wide Fields sample used to estimate the SPIRE LLFs. The $L-z$ plots are colour-coded according to the SED best fit class obtained by the SED fitting procedure following the list reported in Sec. \ref{sedfit.sec}. The histograms report the relative quantities for the photometric and spectroscopic samples in blue and in green, respectively, with the total sample illustrated in red.}
\label{Lz.wide}
\end{figure*}

The HerMES wide fields used in this work are part of the \catnameshort\ and are all covered both by \textit{Spitzer} 7-band infrared imaging and by SDSS 5-band optical imaging and optical spectroscopy \citep{Csabai2007,Abazajian2009, Carliles2010, Bolton2012}. They also benefit by a vast quantity of additional homogeneous multi-wavelength observations and additional spectroscopic redshifts available from NED, as well as the recent literature, and our own \textit{Spitzer}/\textit{Herschel} proprietary follow-up programmes. We thus associate a reliable spectroscopic redshift to our sources whenever this is available and otherwise rely on SDSS photometric redshift estimates based on a KD-tree nearest neighbour search (see \citealt{Csabai2007} for more details). In so doing we follow a commonly adopted photometric reliability criterion for SDSS good photometry, only selecting detections with SDSS \textit{cmodelmag} $r_{\mathrm{AB}} < 22.2$, thus avoiding unreliable photometric redshifts. In Fig. \ref{Lz.wide} we report SDSS $r_{\mathrm{AB}}$ and redshift histograms of the HerMES Wide sample. In order to avoid effects of incompleteness in redshift, we limit our HerMES Wide sample to $z \lsim 0.5$, below the completeness and reliability limit of SDSS redshift estimates. Moreover, to avoid the possible redshift incompleteness that affects the very bright and nearby galaxies in SDSS data, we cut our sample to the lowest redshift of $z = 0.02$, as suggested by e.g. \cite{MonteroPrada2009}. As discussed in \cite{Roseboom2010} the SPIRE source extraction works very well for point-like sources, but can underestimate the fluxes of the extended sources; cutting the sample at $z > 0.02$, also avoids this problem since the vast majority of extended sources are located at lower redshifts.
The numbers of sources of the HerMES Wide sample are detailed in Tab. \ref{spire-llf-numbers.tab}. 

%
\subsection{SED fitting}
\label{sedfit.sec}
\begin{figure*}
\begin{center}
   {\myincludegraphics{width=0.45\textwidth}{\figdir{}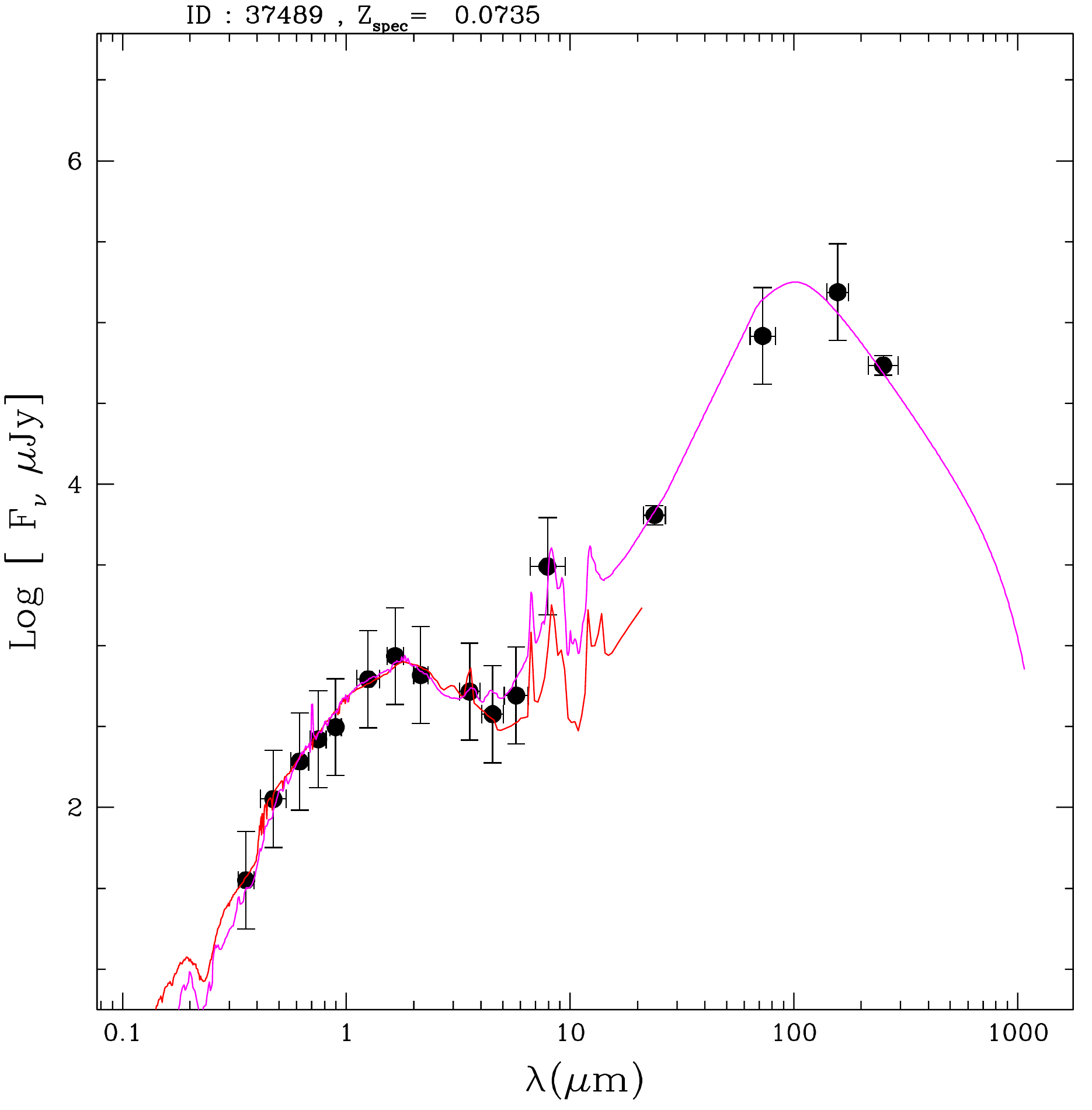}}
   {\myincludegraphics{width=0.45\textwidth}{\figdir{}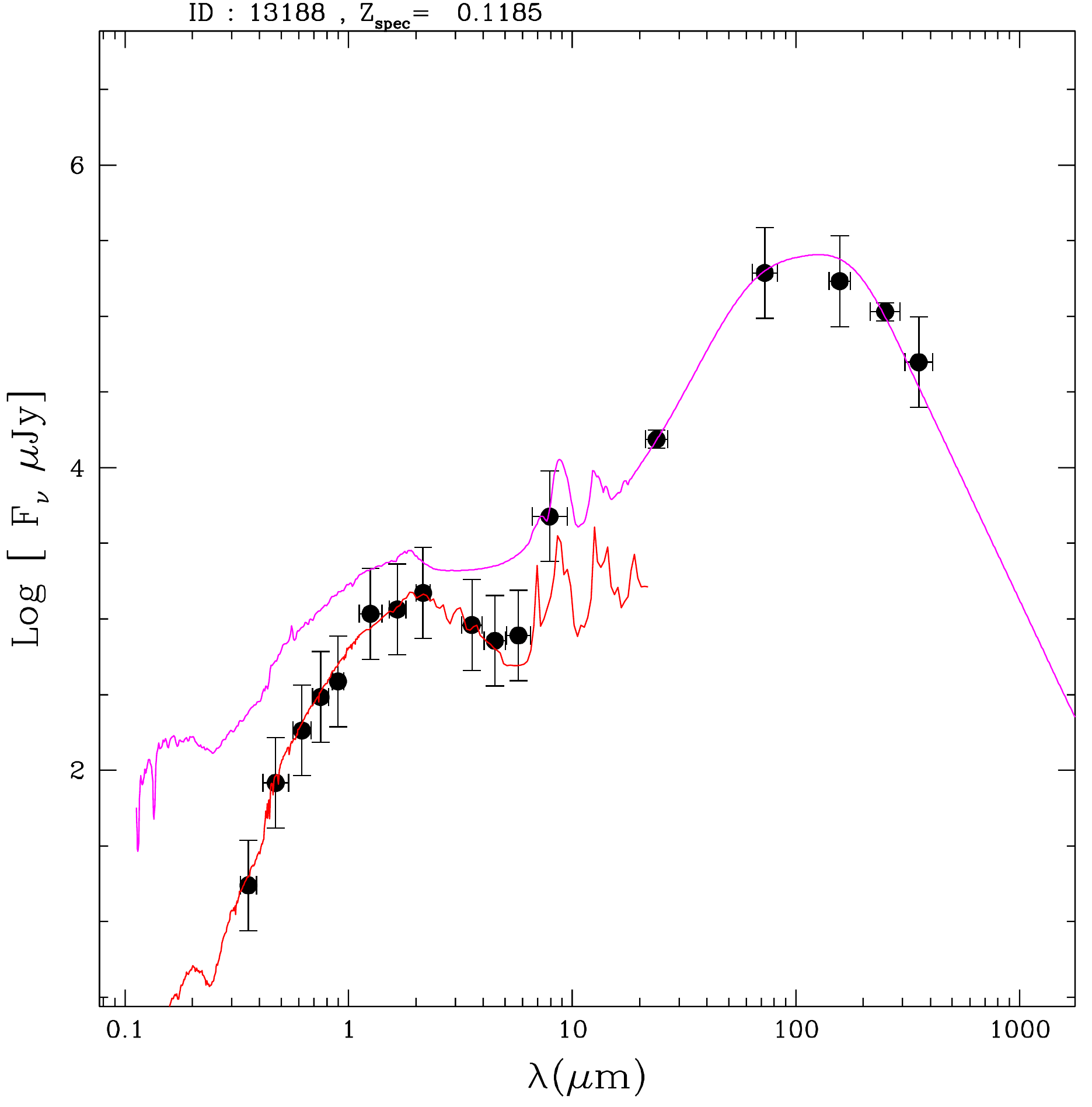}}
\end{center}
\caption[Typical \texttt{Le Phare} best fit results]{Typical \texttt{Le Phare} SED fits. The two best-fit SEDs used to fit short- and long-wavelength photometry are shown by the red and magenta solid line, respectively. The black solid circles are the photometric data used to perform the fit. The ID and the redshift of the source are reported on top of each panel.}
\label{Lephare-fit}
\end{figure*}

Thanks to the \catnameshort\  we are able to perform the multi-wavelength SED fitting analysis of our HerMES Wide Fields sample and thus estimate the IR bolometric ($8-1000\,\mu$m) and monochromatic rest-frame luminosities and relative k-corrections. We perform the SED fitting analysis using \texttt{Le Phare} (\citealt{Arnouts99} and \citealt{Ilbert2006}). To perform the fit we use SDSS $ugriz$, 2MASS $JHK_\mathrm{s}$, IRAC-3.6, IRAC-4.5, IRAC-5.8, IRAC-8.0, MIPS-24, MIPS-70, MIPS-160, SPIRE-250, SPIRE-350 and SPIRE-500 flux densities, which are available over the whole area covered by our sample.
As template SEDs we use two different set of empirical templates according to the range of wavelengths we are fitting: in the optical-MIR range (up to 7$\,\mu$m rest-frame) we use the same templates and extinction laws exploited by the COSMOS team to estimate the COSMOS photometric redshifts as in \cite{Ilbert2009}, while to fit the IR/submm range (from 7$\,\mu$m rest-frame upwards) we use the SWIRE templates of \cite{Polletta2007} and their slightly modified version described in \cite{Gruppioni2010}, for a total of 32 and 31 SEDs respectively; this includes Elliptical, Spiral, AGN, Irregular and Starburst spectral types as summarised in Tab. \ref{sed.list}. As an example we report two typical examples of our SED fitting results in Fig. \ref{Lephare-fit}. Splitting the overall wavelength coverage into two provides us with a particularly good fit to the FIR bump and a reasonably good fit at all other wavelengths for all sources, with a mean value of the reduced $\chi^2$ of around 0.5. Fig. \ref{Lz.wide} (upper panels) we report the $L-z$ distribution for both the $L_{250}$ and the $L_{\mathrm{IR}}$ rest-frame luminosities obtained through the SED fitting procedure. 

\begin{figure*}
\begin{center}
   {\myincludegraphics{width=0.45\textwidth}{\figdir{}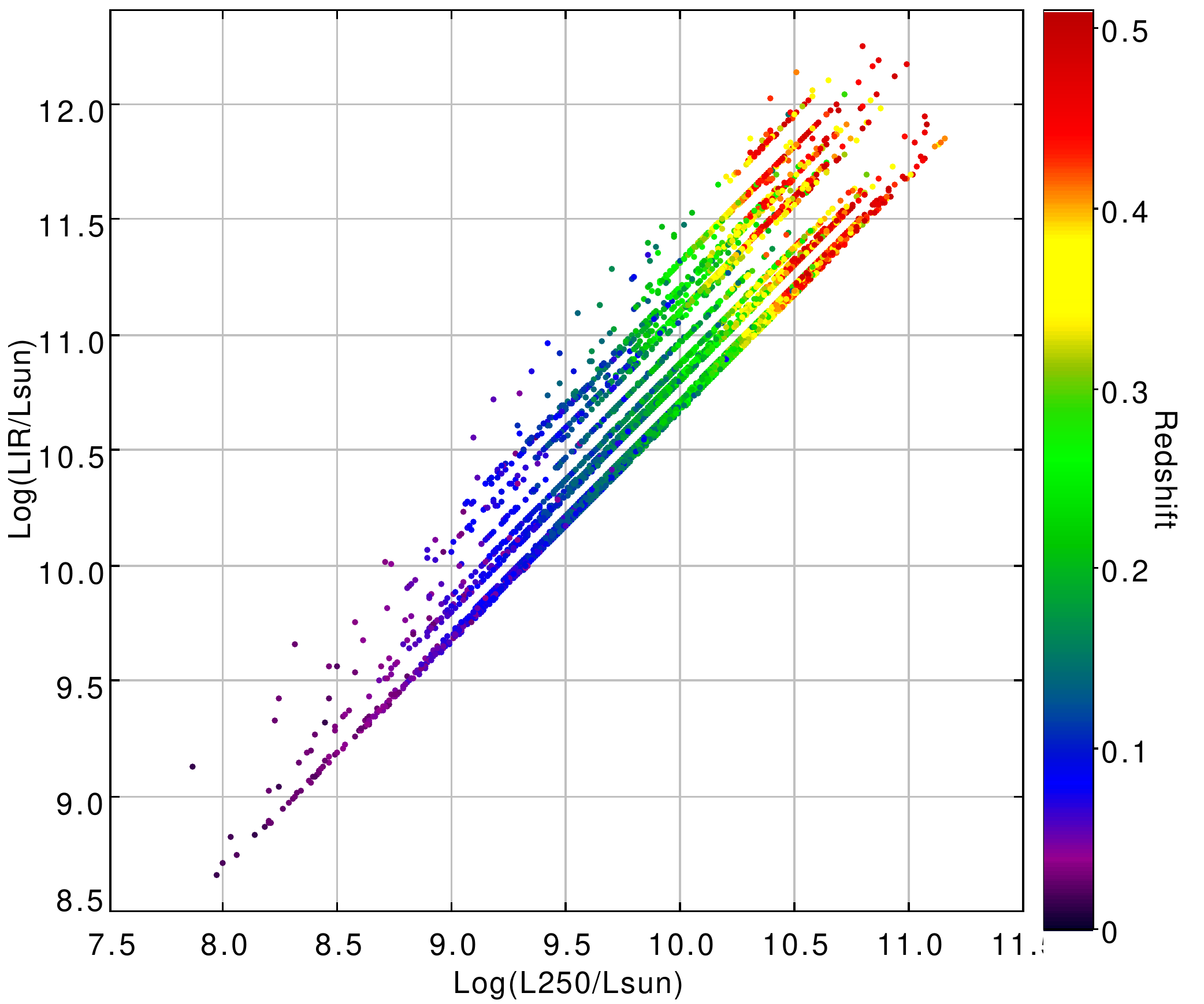}}
   {\myincludegraphics{width=0.45\textwidth}{\figdir{}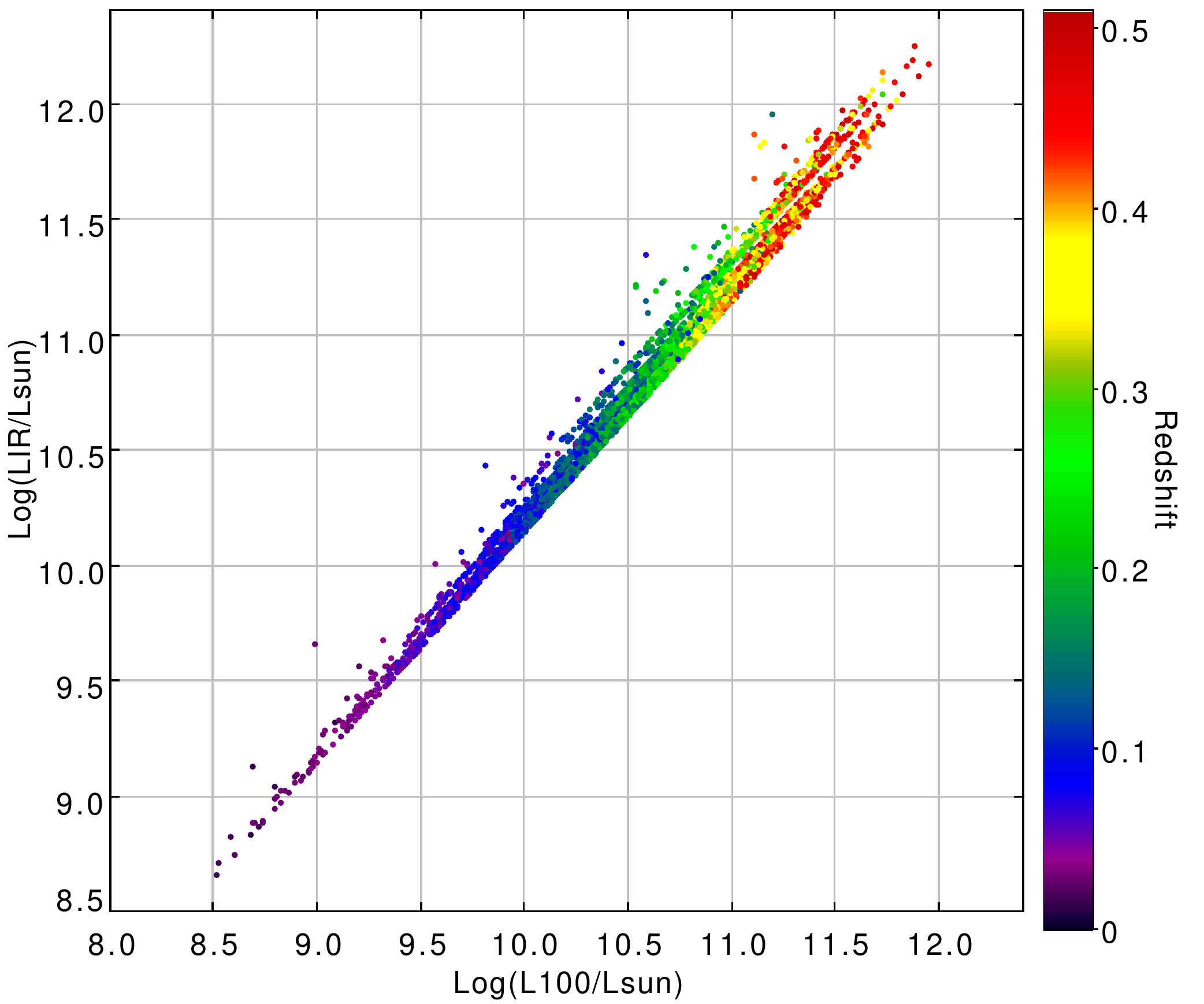}}
   {\myincludegraphics{width=0.45\textwidth}{\figdir{}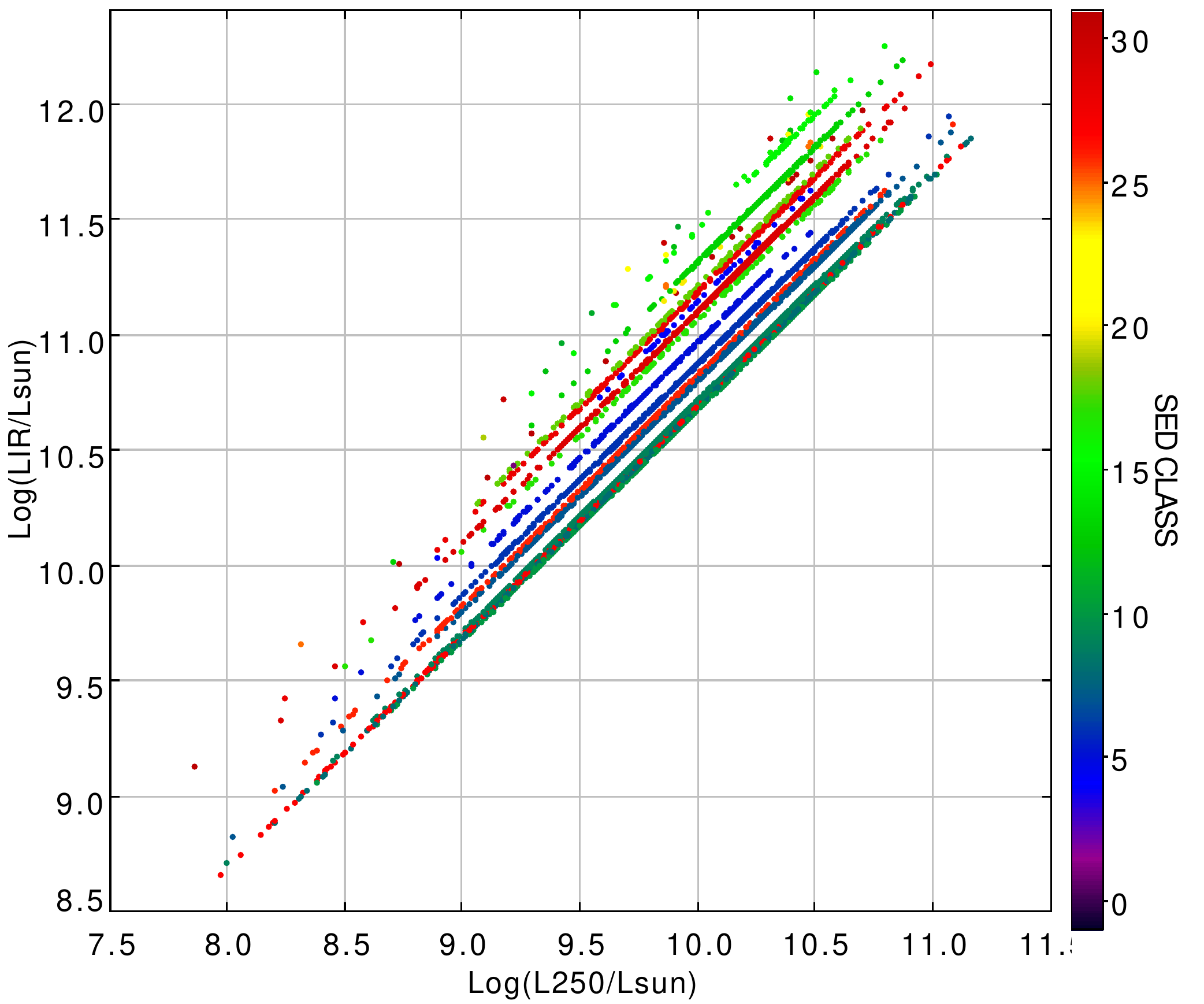}}
   {\myincludegraphics{width=0.45\textwidth}{\figdir{}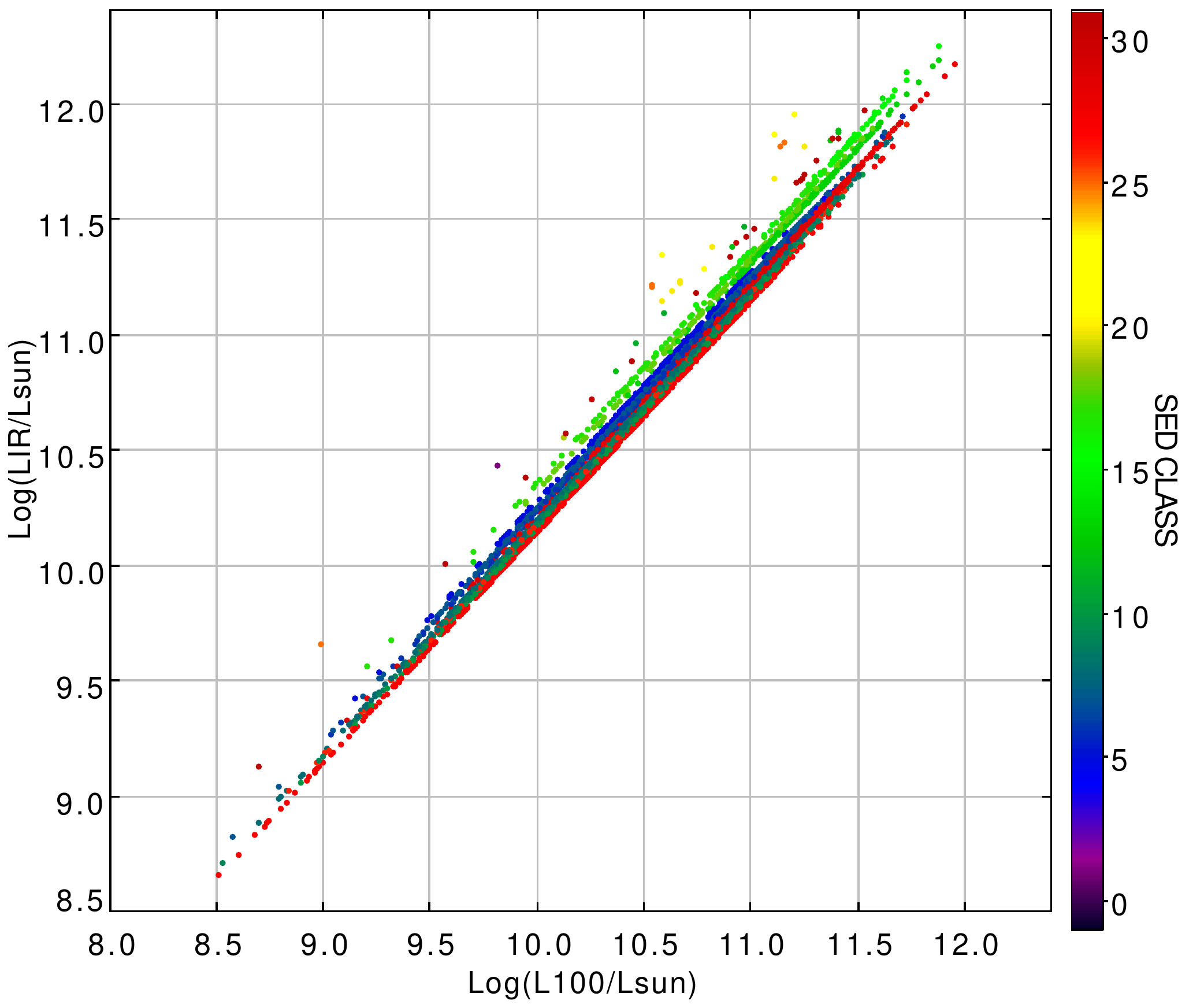}}
   {\myincludegraphics{width=0.45\textwidth}{\figdir{}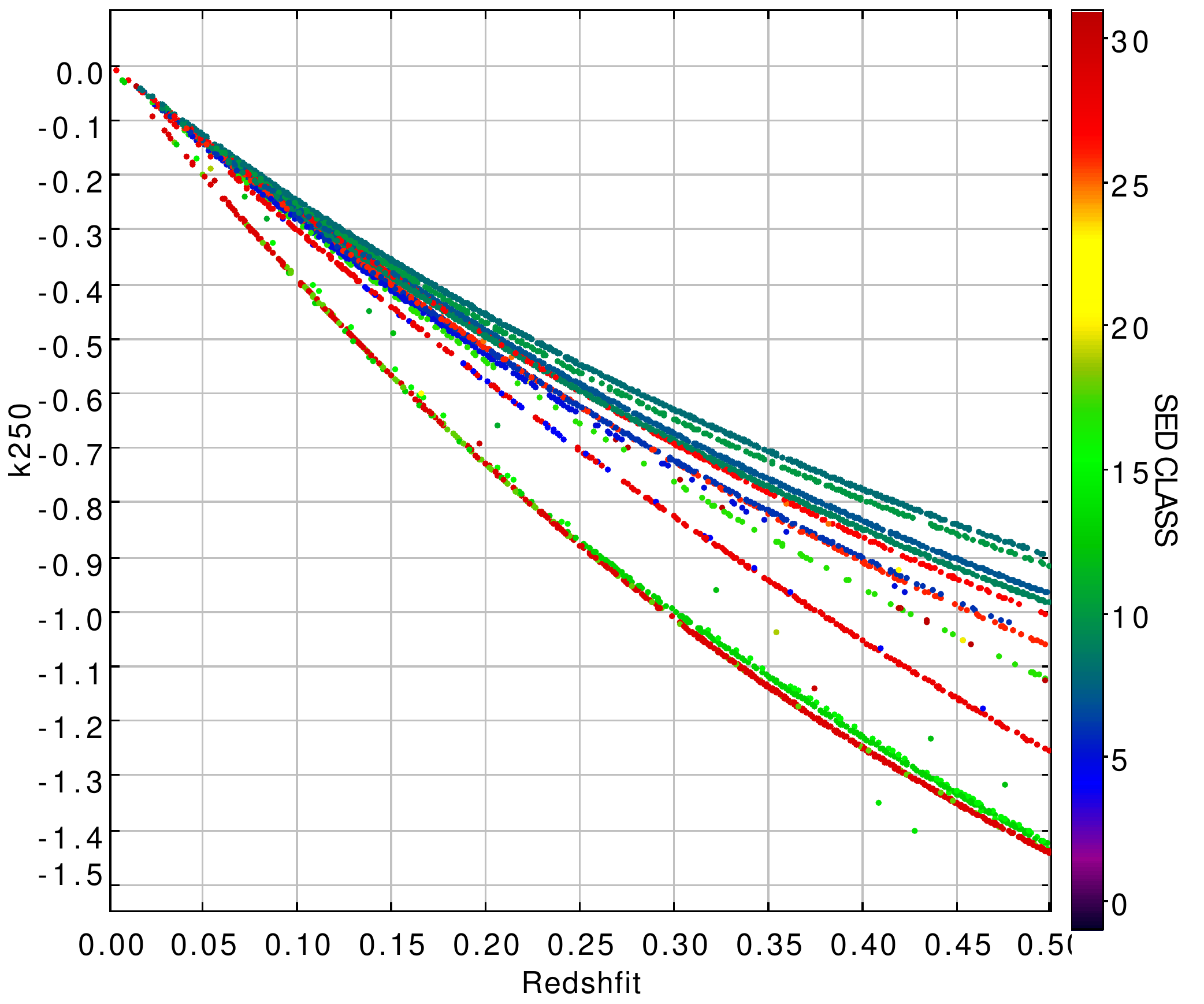}}
   {\myincludegraphics{width=0.45\textwidth}{\figdir{}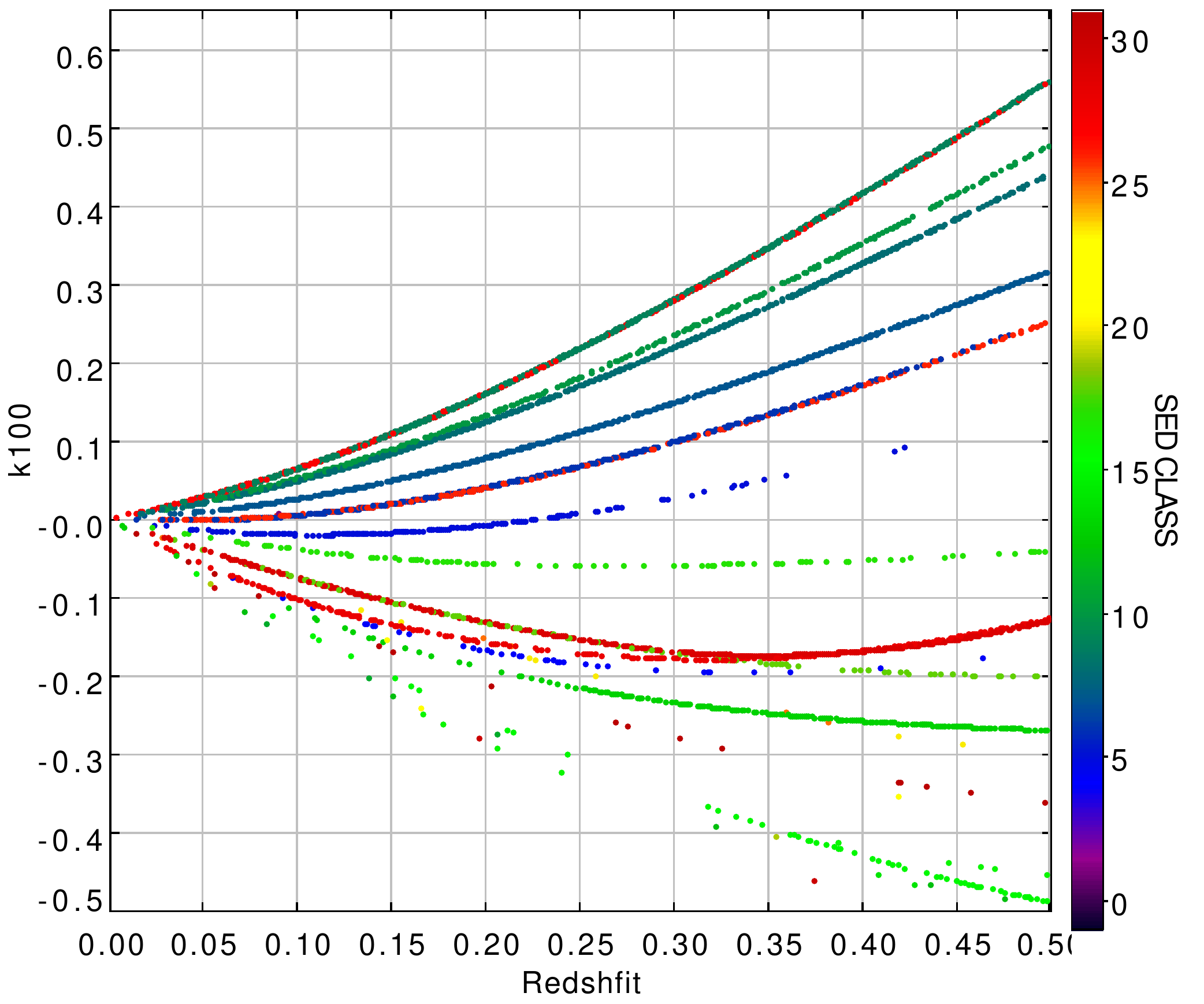}}
\end{center}
\caption[]{\textbf{Top}: relation between rest frame SPIRE 250$\,\mu$m or PACS 100$\,\mu$m luminosities and IR bolometric luminosity colour-coded as a function of redshift. \textbf{Middle}: relations between rest frame SPIRE 250$\,\mu$m/PACS 100$\,\mu$m luminosities and IR bolometric luminosity colour-coded according to the SED best fit class obtained by the SED fitting procedure following the list reported in Tab. \ref{sed.list}; \textbf{Bottom}: SPIRE 250$\,\mu$m and PACS 100$\,\mu$m k-corrections in function of redshift colour-coded according to the SED best fit class.}
\label{sed.fit.plus}
\end{figure*}

Thanks to this multi-wavelength SED fitting we are able to also investigate the relation between monochromatic rest-frame luminosities at different wavelengths. As an example we report in Fig. \ref{sed.fit.plus} a comparison between SPIRE 250$\,\mu$m and PACS 100$\,\mu$m monochromatic rest-frame luminosities plotted against the IR bolometric luminosity. Historically, the monochromatic rest-frame luminosity at $60-100\,\mu$m has been considered a good indicator of the IR bolometric luminosity, due to a strong correlation between the two (e.g., \citealt{Patel2013} used the relation between MIPS 70$\,\mu$m and $L_{\mathrm{IR}}$). In Fig. \ref{sed.fit.plus} we show that we confirm this trend in our SED fitting results while, on the other hand, the SPIRE 250$\,\mu$m luminosity doesn't show a strong correlation with the IR bolometric luminosity and thus cannot be used as a reliable indicator of the total IR emission of the galaxy. As also confirmed by other HerMES works that have carefully studied the SED shape of the HerMES sources (e.g. \citealt{Symeonidis2013}) we find that the SEDs in the FIR regime of our local HerMES sample peak close to the PACS 100$\,\mu$m band and thus the monochromatic luminosity at this wavelength best traces the total IR bolometric luminosity integrated between 8 and 1000$\,\mu$m. It is also interesting to notice the very different behaviour of the k-corrections estimated at SPIRE 250$\,\mu$m and PACS 100$\,\mu$m (lowest panels of  \ref{sed.fit.plus}). The differences between these two are remarkable and this is reflected in the different behaviour of the resulting luminosities.

\begin{figure*}
\begin{center}
   {\myincludegraphics{width=0.45\textwidth}{\figdir{}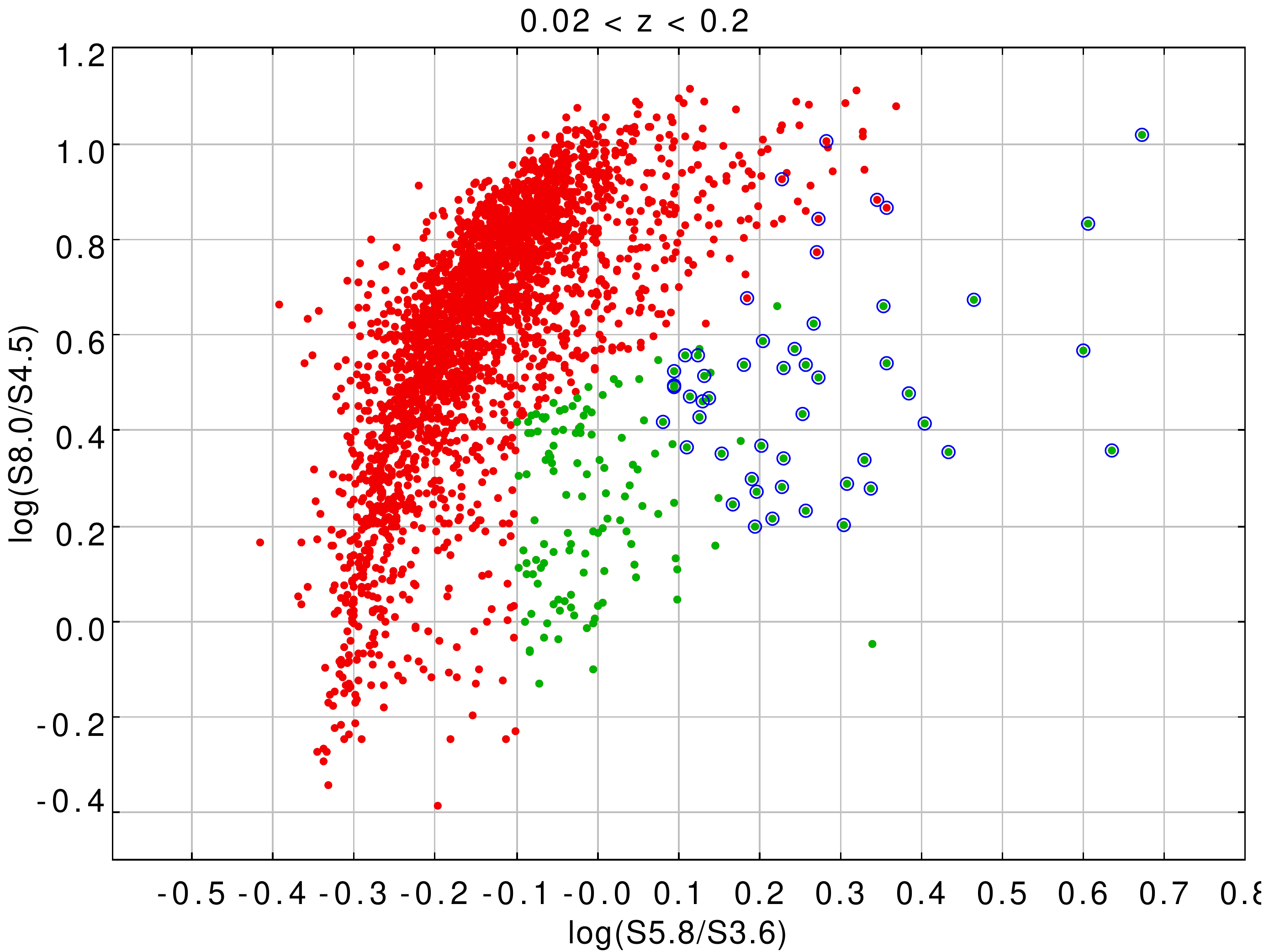}}
   {\myincludegraphics{width=0.45\textwidth}{\figdir{}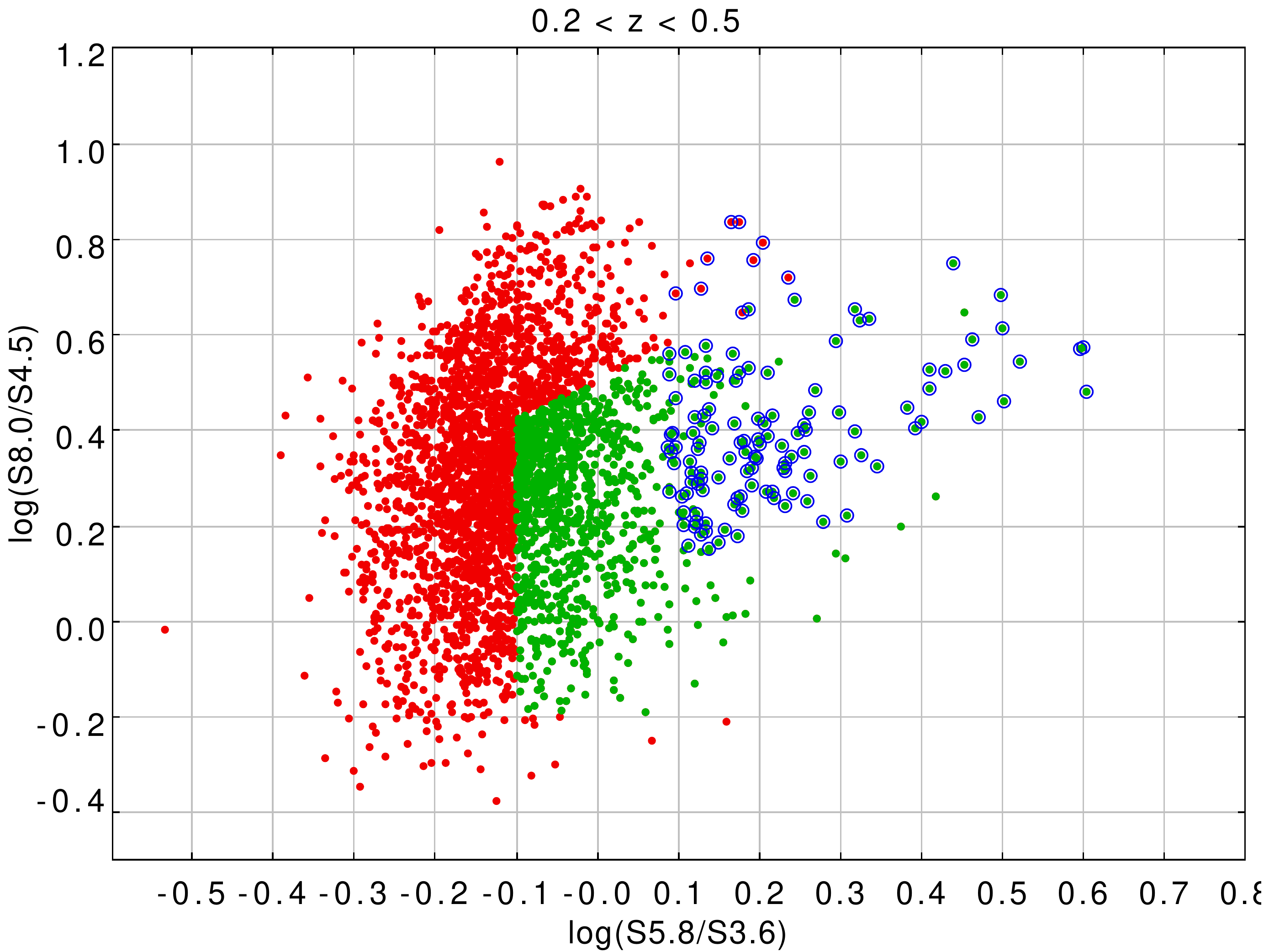}}
\end{center}
\caption[]{IRAC colour-colour plot as from \cite{Lacy2004} and \cite{Donley2012}. In the left and right panels respectively the complete samples of sources in the two redshift ranges $0.02<z\leq0.2$ and $0.2<z\leq0.5$ are reported. In both panels the over plotted solid green circles and blue open circles show the AGN-like objects selected using the \cite{Lacy2004} and \cite{Donley2012} criteria respectively, while the solid red circles represent the rest of the sample in each redshift bin.}
\label{AGN.colours}
\end{figure*}

While a detailed physical analysis of our sample is beyond the scope of this paper, we did exploit our SED fitting analysis and the IRAC colour-colour criteria by \cite{Lacy2004} and \cite{Donley2012} to search for any potential AGN contamination in our sample. On the whole, the vast majority of our sources shows galaxy- or starburst-like best fit SEDs with less than 10\,\% of the sample being best-fit by AGN-like SEDs (SED classes between 17 and 25 and between 28 and 31 as reported in Tab. \ref{sed.list}). These numbers do not change significantly even if we fit a single SED template to the whole range of available photometry (from optical to SPIRE bands). Fig. \ref{AGN.colours} confirms that our objects mostly lie within the starburst-dominated region of the IRAC colour-colour plot, with only a small fraction of the sources (mainly located at $z>0.25$) sitting in the area usually occupied by AGN-like objects. On the whole we find that about 20\% of our sources sit in the AGN region identified by \cite{Lacy2004}, with less than 6\% at $z\leq0.2$ and about 30\% at $0.2<z\leq0.5$. These fractions change significantly if we apply the selection reported in \cite{Donley2012} which is able to better discriminate pure bona-fide AGNs from samples that are contaminated by low- and high- redshift star forming galaxies as the one selected by Lacy's criterion. We find that only 3\% of our total sample is identified as AGN-dominated by Donley's criterion, less than 2\% at $z\leq0.2$ and 4\% at $0.2<z\leq0.5$.

\begin{table}
\centering
\begin{tabular}{cccc}
\hline
\textbf{Index} & \textbf{SED class} & \textbf{Spectral type} & \textbf{Reference}\\
\hline\hline
01 & Ell13 & Elliptical & Polletta+07 \\
02 & Ell5& Elliptical & Polletta+07 \\
03 & Ell2& Elliptical & Polletta+07 \\
04 & S0& Spiral & Polletta+07 \\
05 & Sa& Spiral & Polletta+07 \\
06 & Sb& Spiral & Polletta+07 \\
07 & Sc& Spiral & Polletta+07 \\
08 & Sd& Spiral & Polletta+07 \\
09 & Sdm& Spiral & Polletta+07 \\
10 & Spi4& Spiral & Polletta+07 \\
11 & N6090& Starburst & Polletta+07 \\
12 & M82& Starburst & Polletta+07 \\
13 & Arp220& Starburst & Polletta+07 \\ 
14 & I20551& Starburst & Polletta+07 \\
15 & I22491& Starburst & Polletta+07 \\
16 & N6240& Starburst & Polletta+07 \\
17 & Sey2& Obscured AGN & Polletta+07 \\
18 & Sey18& Obscured AGN & Polletta+07 \\
19 & I19254& Obscured AGN & Polletta+07 \\
20 & QSO2& Unobscured AGN & Polletta+07 \\
21 & Torus& Unobscured AGN & Polletta+07 \\
22 & Mrk231& Obscured AGN & Polletta+07 \\
23 & QSO1& Unobscured AGN & Polletta+07 \\
24 & BQSO1& Unobscured AGN & Polletta+07 \\
25 & TQSO1& Unobscured AGN & Polletta+07 \\
26 & Sb& Spiral & Gruppioni+10 \\
27 & Sdm& Spiral & Gruppioni+10 \\
28 & Sey2& Obscured AGN & Gruppioni+10 \\
29 & Sey18& Obscured AGN & Gruppioni+10 \\
30 & Mrk231& Obscured AGN & Gruppioni+10 \\
31 & qso\_high& Unobscured AGN & Gruppioni+10 \\
\hline
\end{tabular}
\caption[List of the SEDs used to perform the SED fitting analysis in the IR/submm regime]{List of the SEDs used to perform the SED fitting analysis in the IR/submm. The 'Spectral Type' columns shows the grouping procedure we implemented in order to collect together those SED classes with similar properties in terms of FIR colours.}
\label{sed.list}
\end{table}

\section{Statistical Methods} \label{stat.met.}
Accurately estimating the luminosity function (LF) is difficult in observational cosmology since the presence of observational selection effects like flux detection thresholds can make any given galaxy survey incomplete and thus introduce biases into the LF estimate.

Numerous statistical approaches have been developed to overcome this limit, but, even though they all have advantages, it is only by comparing different and complementary methods that we can be confident about the reliability of our results  
For this reason, to estimate the local luminosity functions (LLFs) in the SPIRE bands reported in this paper we exploit different LF estimators: the $1/V_{\rm max}$ approach of \cite{Schmidt1968} and the modified version $\phi_{\rm est}$ of \cite{PageCarrera2000}; the Bayesian parametric maximum likelihood method (ML) of \cite{Kelly2008} and \cite{Patel2013}; and the semi-parametric approach of \cite{Schafer2007}. All these methods are explained in the following sections.

\subsection{$1/V_{\rm max}$ Estimator}\label{Vmax}

\cite{Schmidt1968} introduced the intuitive and powerful $1/V_{\rm max}$ estimator for LF evaluation. The quantity $V_{\rm max}$ for each object represent the maximum volume of space which is available to such an object to be included in one sample accounting for the survey flux limits and the redshift bin in which the LF is estimated. $V_{\rm max}$ thus depends on the distribution of the objects in space and the way in which detectability depends on distance. 
Once the $V_{\rm max}$ (or  $V_{\rm max}(L_{i})$, since it depends on the luminosity of each object), is defined, the LF can be estimated as
\begin{equation}
\Phi(B_{j-1}< L \leqslant B_{j})=\sum_{B_{j-1}< L \leqslant B_{j}}\frac{1}{V_{\rm max}(L_{i})},
\end{equation}
in which its value is computed in bins of luminosity, within the boundary luminosities value of a defined bin $[B_{j-1} , B_{j}]$. It is usually expressed in the differential form as
\begin{equation} \label{diff.vmax}
\phi _{1/V_{\max } } (L,z) = \frac{1}{{\Delta L}}\sum\limits_{i = 1}^N {\frac{1}{{V_{\max ,i}}}},
\end{equation}
where $N$ is the number of objects within some volume-luminosity region.
Errors in the LF can be evaluated using Poisson statistics:
\begin{equation}\label{err.vmax}
\sigma_{\phi(L)}^2=\sum_{B_{j-1}< L \leqslant B_{j}}\frac{1}{(V_{\max}(L_{i}))^2}
\end{equation}
In our case there are three main selection factors that may constrain the $V_{\max}$ for each object in our sample: the limit in $r$ magnitude that guide the photometric redshift estimates in the SDSS survey, $r_{AB} < 22.2$; the MIPS 24$\,\mu$m flux limit that guides the SPIRE 250$\,\mu$m extraction, $S_{24} > 300$ $\mu$Jy; and finally the flux density limit in the SPIRE 250$\,\mu$m band, $S_{250} > 30$~mJy. Moreover, since we estimate the $1/V_{\max}$ in a number of redshift bins the $V_{\max}$ value is actually also limited by $z_{\min}$ and $z_{\max}$ for each $z$-bin. Taking into account all these considerations the $V_{\max}$ estimator used in the Eq. \ref{diff.vmax} is described by:
\begin{equation}\label{vmax.perbin}
V_{\max}=\frac{\Omega}{4\pi}\int_{z_{\min}}^{z_{\max}} \mathrm{d}z\frac{\mathrm{d}V}{\mathrm{d}z},
\end{equation}
where $z_{\min}$ and $z_{\max}$ are the redshift boundaries resulting from taking into account both the redshift bin range and the selection factors
\begin{eqnarray}\label{zmax.perbin.1}
z_{k,\min} &=& z_{\mathrm{bin}_k,\min} \\
z_{k,\max} &=& \mathrm{min}[z_{0,\max},.., z_{n,\max},z_{\mathrm{bin}_k,\max}]
\end{eqnarray}
for all $0,...,n$ selection factors and for each $k$ redshift bin. For instance, in the case of the SPIRE 250$\,\mu$m luminosity function estimate in $z$-bin $0.02<z<0.1$ the  conditions just shown become
\begin{eqnarray}\label{zmax.perbin.2}
z_{0.02<z<0.1,\min} &=& 0.02 \nonumber \\
z_{0.02<z<0.1,\max} &=& \mathrm{min}[z_{r_{AB},\max},z_{f24,\max},z_{f250,\max}, 0.1], \nonumber
\end{eqnarray}
where $z_{r_{AB},\max}$, $z_{f24,\max}$ and $z_{f250,\max}$ are the redshift at which a source in the sample reach the SDSS $r_{AB}$ magnitude limit (= 22), the 24$\,\mu$m flux limit (= 300 $\mu$Jy) and the SPIRE 250$\,\mu$m limit (= 30 mJy), respectively; 0.02 and 0.1 are the minimum and the maximum of the redshift bin. 

This method implies binning of the luminosity data, a non-parametric technique, and as such does not need to assume an analytic form. It does however contain the underlying assumption that galaxies have a uniform distribution in space. In principle this could be tested with the $V/V_{\max}$ distribution, but that still remains difficult when there are multiple selection factors limiting the sample. 

The simple $V_{\max}$ estimator have evolved, being improved and refined over the years to accommodate the many different types of survey that have steadily grown in size and complexity. One of these approaches is the one implemented in \cite{PageCarrera2000}, the so called $V_{\rm est}$ method, which we also used here to check whether with our $1/V_{\max}$ estimates we are ignoring any important incompleteness factor in our sample. \cite{PageCarrera2000} improved the method to take into account  systematic errors in the  $V_{\max}$ test introduced for objects close to the flux limit of a survey. This new method defines the value of the luminosity function $\phi(L)$ as $\phi_{\rm est}$, which assumes that $\phi$ does not change significantly over the luminosity and redshift intervals $\Delta L$ and $\Delta z$, respectively, and is defined as
\begin{equation} \label{phiest}
\phi_{\rm est}  = \frac{N}{{\int\limits_{L_{\min } }^{L_{\max } } \int\limits_{z_{\min } }^{z_{\max } (L)} {\frac{dV}{dz}dzdL} }},
\end{equation}
where $N$ is the number of objects within some volume-luminosity region.

Due to how the methods work in practice, for luminosity functions in most of the redshift intervals, the two will produce the same results, particularly for the highest luminosity bins of any given redshift bin. However, for the lowest luminosity objects in each redshift bin, which are close to the survey limit and occupy a portion of volume-luminosity space much smaller than the rectangular $\Delta L$ $\Delta z$ region, the two methods can produce the most discrepant results. Nevertheless in our case we do not find any substantial differences between the $1/V_{\max}$ and $1/V{\rm est}$ solutions, as shown in the following sections.

\subsection{Bayesian parametric maximum likelihood estimator}\label{bayes.MLE}

The Maximum Likelihood estimator has first been applied in studies of observational cosmology by \cite{STY1979}, the so called STY estimator. 
In maximum likelihood analysis, one is interested in finding the estimate that maximises the likelihood function of the data. For a given statistical model, parameterised by $\theta$, the likelihood function, $p(x|\theta)$, is the probability of observing the data, denoted by $x$, as a function of the parameters $\theta$. In Bayesian analysis, one attempts to estimate the probability distribution of the model parameters, $\theta$, given the observed data $x$. Bayes theorem states that the probability distribution of $\theta$ given $x$ is related to the likelihood function as
\begin{equation}
p(\theta|x) \propto p(x|\theta)p(\theta),
\end{equation}
where $p(x|\theta)$ is the likelihood function of the data, and the term $p(\theta)$ is the prior probability distribution of $\theta$; the result, $p(\theta,x)$, is called the posterior distribution. The prior distribution, $p(\theta)$, should convey information known prior to the analysis. In general, the prior distribution should be constructed to ensure that the posterior distribution integrates to 1, but does not have a significant effect on the posterior. In particular, the posterior distribution should not be very sensitive to the choice of prior distribution, unless the prior distribution is constructed with the purpose of placing constraints on the posterior distribution that are not conveyed by the data. The contribution of the prior to $p(\theta|x)$ should become negligible as the sample size becomes large.

From a practical standpoint, the primary difference between the maximum likelihood approach and the Bayesian approach is that the former is concerned with calculating a point estimate of $\theta$, while the latter is concerned with mapping out the probability distribution of $\theta$ in the parameter space. The maximum likelihood approach uses an estimate of the sampling distribution of $\theta$ to place constraints on the true value of $\theta$. In contrast, the Bayesian approach directly calculates the probability distribution of $\theta$, given the observed data, to place constraints on the true value of $\theta$.

In terms of LF evaluation, the LF estimate is related to the probability density of  $(L,z)$ 
\begin{equation} \label{eq.p}
p(L,z) = \frac{1}{N} \phi(L,z)\frac{\mathrm{d}V}{\mathrm{d}z},
\end{equation}
where $N$ is the total number of sources in the observable Universe and is given by the integral of $\phi$ over $L$ and $V(z)$.  The quantity $p(L,z)$d$L$d$z$ is the probability of finding a source in the range $L,L+$d$L$ and $z, z+$d$z$. Eq. \ref{eq.p} separates the LF into its shape, given by $p(L,z)$, and its normalisation, given by $N$.  Once we have an estimate of $p(L,z)$, we can easily convert this to an estimate of $\phi(L,z)$ using Eq. \ref{eq.p}. 

In general it is easier to work with the probability distribution of $L$ and $z$ instead of directly with the LF, because $p(L,z)$ is more directly related to the likelihood function. The function $\phi(L,z)$ can be described, as we have seen, by a parametric form with parameter $\theta$, so that we can derive the likelihood function for the observed data. The presence of flux limits and various other selection effects can make this difficult, since the observed data likelihood function is not simply given by Eq. \ref{eq.p}. In this case, the set of luminosities and redshifts observed by a survey gives a biased estimate of the true underlying distribution, since only those sources with $L$ above the flux limit at a given $z$ are detected. In order to derive the observed data likelihood function, it is necessary to take the survey's selection method into account. This is done by first deriving the joint likelihood function of both the observed and unobserved data, and then integrating out the unobserved data. The probability $p(L,z)$ (as reported in \citealt{Patel2013}) then becomes
\begin{equation}
p(L,z|\theta) = \frac{\phi(L,z|\theta)p(\rm selected|L,z)}{\lambda} \frac{\mathrm{d}V}{\mathrm{d}z},
\end{equation}
where $p(\rm selected|L,z)$ stands for the probability connected with the selection factors of the survey and $\lambda$ is the expected number of sources, determined by
\begin{equation}
\lambda =\iint \phi(L,z|\theta)p(\rm selected|L,z)\mathrm{d} \mathrm{log} L\frac{\mathrm{d}V}{\mathrm{d}z}\mathrm{d}z,
\label{lambda.patel}
\end{equation}
where the integrals are taken over all possible values of redshift and luminosity. 

This last equation gives the expected number of objects in a sample composed by sources of the same morphological type and collected in a single field survey. For our purposes we have to change the equation to the following:
\begin{equation}
\lambda = \sum_{\rm SED}\sum_{\rm fields} \iint\Phi(L,z|\theta)p(\rm selected|L,z)\mathrm{d} \mathrm{log} L\frac{\mathrm{d}V}{\mathrm{d}z}\mathrm{d}z ,
\label{lambda.sed.field}
\end{equation}
where we sum together the expected number of sources for each SED type, used for the SED fitting procedure, and survey areas that compose our HerMES Wide Fields sample.

Since the data points are independent, the likelihood function for all $N$ sources in the Universe would be
\begin{equation}
p(L,z|\theta) = \prod_{i=1}^N p(L_i,z_i| \theta).
\end{equation}
Indeed, we do not know the luminosities and redshifts for all $N$ sources, nor do we know the value of $N$, since our survey only covers a fraction of the sky and is subject to various selecting criteria. As a result, our survey only contains $n$ sources. For this reason the selection process must also be included in the probability model, and the total number of sources, $N$, is an additional parameter that needs to be estimated.
Then the likelihood becomes:
\begin{equation}
p(n|\theta) = p(N,\{L_i,z_i\}|\theta) = p(N| \theta)p(\{L_i,z_i\} | \theta),
\end{equation}
where $p(N|\theta)$ is the probability of observing $N$ objects and $p(\{L_i,z_i\} | \theta)$ is the likelihood of observing a set of $L_i$ and $z_i$, both given the model LF. Is it possible to assume that the number of sources detected follows a Poisson distribution \citep{Patel2013}, where the mean number of detectable sources is given by $\lambda$.
Then, the term $p(N,\{L_i,z_i\}|\theta)$ could be written as the product of individual source likelihood function, since each data point is independent:
\begin{eqnarray}
&p(N| \theta)p(\{L_i,z_i\}| \theta) = & \\ 
& = \frac{\lambda^N e^{-\lambda}}{N!}\displaystyle\prod_{i=1}^N \frac{\Phi(L,z|\{\theta\})p(\rm{selected}|L,z)}{\lambda}\frac{\mathrm{d}V}{\mathrm{d}z} \nonumber , &
\end{eqnarray}
Then we can use the likelihood function for the LF to perform Bayesian inference by combining it with a prior probability distribution, $p(\theta)$ to compute the posterior probability distribution, $p(\theta | {d_i})$, given by Bayes' theorem :
\begin{equation}
p(\theta|{d_i}) = \frac{p(\{d_i\}|\{\theta\})p(\{\theta\})}{\displaystyle\int p(\{d_i\}|\{\theta\})p(\{\theta\})\mathrm{d}\theta}
\end{equation}
The denominator of this equation represents the Bayesian evidence which is determined by integrating the likelihood over the prior parameter space. This last step is needed to normalise the posterior distribution.

Calculating the Bayesian evidence is computationally expensive, since it involves integration over $m$-dimensions for an $m$ parameter LF model. Therefore, Monte Carlo Markov Chain (MCMC) methods, used to examine the posterior probability, perform a random walk through the parameter space to obtain random samples from the posterior distribution. MCMC gives as a result the maximum of the likelihood, but an algorithm is needed to investigate in practice the region around the maximum. \cite{Kelly2008} suggested to use the Metropolis-Hastings algorithm (MHA; \citealt{Metropolis1953,Hastings1970}) in which a proposed distribution is used to guide the variation of the parameters. The algorithm uses a proposal distribution which depends on the current state to generate a new proposal sample. The algorithm needs to be tuned according to the results and the number of iterations, as well as the parameter step size change.
Once we obtain the posterior distribution, we have the best solution for each of the parameters describing the LF model that we have chosen at the beginning; we have the mean value and the standard deviation ($\sigma$) for each of the parameters that we can combine together to find the $\sigma$ of the parametric function chosen as the shape of our LF (see Sec. \ref{results} for further details on our calculation).

\subsection{A \textit{Semi-Parametric} Estimator} \label{schafer.sec}

\cite{Schafer2007} introduced the \textit{semi-parametric} method in order to estimate luminosity functions given redshift and luminosity measurements from an inhomogeneously selected sample of objects (e.g. a flux-limited sample).
In such a limited sample, like ours, only objects with flux within some range are observable. When this bound on fluxes is transformed into a bound in luminosity, the truncation limits take an irregular shape as a function of redshift; additionally, the k-correction can further complicates this boundary.

We refer the reader to the original paper, \cite{Schafer2007}, for a complete description of the method; here we report only the main characteristics of it. This method shows various advantages in comparison with the other techniques previously described: it does not assume a strict parametric form for the LF (differently to the parametric MLE); it does not assume independence between redshifts and luminosities; it does not require the data to be split into arbitrary bins (unlike for the non-parametric MLE); and it naturally incorporates a varying selection function. This is obtained by writing the luminosity function $\phi(z,L)$ as
\begin{equation}\label{lf.schafer}
\mathrm{log}\phi (z,L) = f(z) + g(L) + h(z,L,\theta),
\end{equation}
where $h(z,L,\theta)$ assumes a parametric form and is introduced to model the dependence between the redshift $z$, the luminosity $L$ and the real valued parameter $\theta$. The functions $f$ and $g$ are estimated in a completely free-form way. 

Nevertheless, it is important to notice that this method assumes a $complete$ data-set in the un-truncated region that requires some care when applying it to samples that may suffer some incompleteness. Discussion on how this issue may influence our results are reported in the later sections.

\subsection{Parametrising the Luminosity Function} \label{analy.form}

Using the classical maximum likelihood technique (STY), as well the one based on Bayesian statistics, implies the assumption of a parametric form able to describe the LF. This choice is not straightforward and over the years the selected LF models varied. In this work we decide to use the \textit{Log Gaussian Function} introduced by \cite{Saunders1990} to fit the \textit{IRAS} IR LF  and widely used for IR LF estimates (e.g. \citealt{Gruppioni2010}, \citealt{Gruppioni2013}, \citealt{Patel2013}). Usually this function is called the \textit{modified Schechter function} since its formalism is very similar to the one introduced by \cite{Schechter1976}. This parametric function is defined as
\begin{equation}\label{log.gaus}
\Phi(L)=\Phi^* \left(\frac{L}{L^{*}}\right)^{1-\alpha}\exp\left[-\frac{1}{2\sigma^{2}}\log^{2}\left(1+\frac{L}{L^{*}}\right)\right],
\end{equation}
where, $\Phi^*$ is a normalisation factor defining the overall density of galaxies, usually quoted in units  of  $h^3$Mpc$^{-3}$, and $L^*$ is the characteristic luminosity. The parameter $\alpha$ defines the faint-end slope of the LF and is typically negative, implying relatively large numbers of galaxies with faint luminosities.
We also checked whether another functional form was more suitable to describe our LFs, but we did not find any evidence of improvement or substantial differences by using e.g. a \textit{double power law} function (used by \citealt{Rush1993} or \citealt{Franceschini2001}). We therefore decide to report and discuss the estimates obtained by using only the \textit{Log Gaussian Function} in order to be able to compare our results with other more recent results that use the same parametrisation. This approach is well suited to describe the total galaxy population, but may be inadequate if we divide the population into sub-groups according, for example, to their optical properties (see Sec. \ref{discussion} for more details) as done by other authors while studying the behaviour of the local mass functions of galaxies (e.g. \citealt{Baldry2012}).

\section{Results} \label{results}

We estimate the LFs at SPIRE 250$\,\mu$m as well as at SPIRE 350 and 500$\,\mu$m by using the SPIRE 250$\,\mu$m selected sample and extrapolating the luminosities from the SED fitting results. The higher sensitivity of the SPIRE 250$\,\mu$m channel with respect to the 350 and 500$\,\mu$m channels largely ensures that we do not miss sources detected only at these longer wavelengths. Additionally we estimate the IR bolometric luminosity functions using the integrated luminosity between 8 and 1000$\,\mu$m and at 24, 70, 100 and 160$\,\mu$m; these last monochromatic estimates are also used to check our procedure against other published LFs. 

As a summary, in Tab. \ref{llf-values.tab} we report our $1/V_{\rm max}$ luminosity function values for each SPIRE band and the IR bolometric rest-frame luminosity per redshift bins. We exclude from the calculation the sources with $z<0.02$, as explained in Sec. \ref{sam.sec.}. The error associated with each value of $\Phi$ is estimated following Poissonian statistics, as shown in Eq. \ref{err.vmax}.\\ 
Since we use photometric redshifts in our sample, we quantify the redshift uncertainties that may affect our results by performing Monte Carlo simulations.
We created 10 mock catalogues based on our actual sample, allowing the photometric redshift of each source to vary by assigning a randomly selected value according to the Gaussian SDSS photometric error. For each source in the mock catalogues we performed the SED fitting and recomputed both the monochromatic and total IR rest-frame luminosities and the $V_{\rm max}$-based LFs, using the randomly varied redshifts.
The comparison between our real IR LF solution and the mean derived from the Monte Carlo simulations shows that the uncertainties derived from the use of the photometric redshifts do not significantly change the error bar estimated using the Poissonian approach and mainly after the lower luminosity bins at the lower redshifts ($z<0.1$). Even though the differences are really small, in Tab. \ref{llf-values.tab} we report the total errors, taking into account all these uncertainties. As an extra test we also check what happens if we estimate the LFs in each field using only spectroscopic redshifts and correct the solutions for the incompleteness effect due to this selection. The resulting LFs are effectively undistinguishable and thus confirms that the uncertainties introduced by the use of photometric redshifts are of the order of the Poissonian ones.

The errors that we quote in Tab. \ref{llf-values.tab} are the total errors, taking into account both Poissonian and redshift uncertainties associated with $\Phi$.

\begin{figure*}
\begin{center}
   {\myincludegraphics{width=0.25\textwidth}{\figdir{}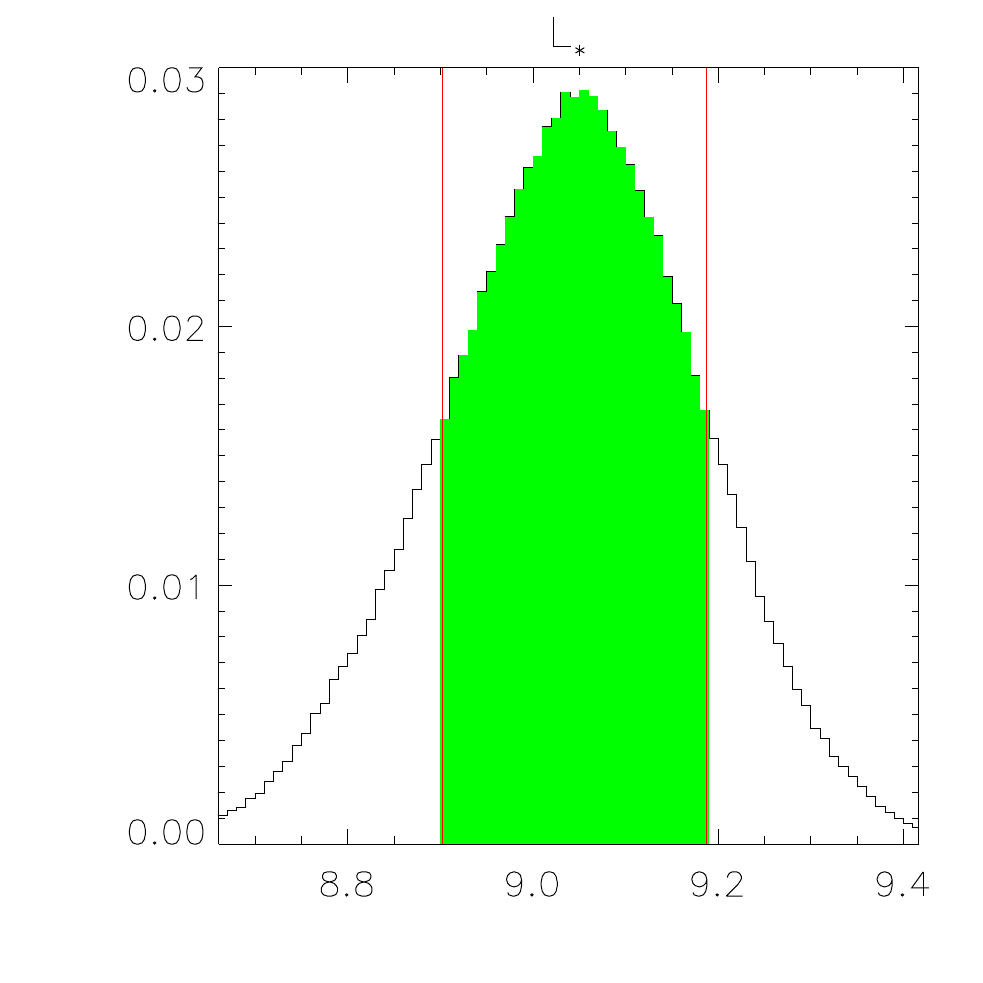}}
   {\myincludegraphics{width=0.25\textwidth}{\figdir{}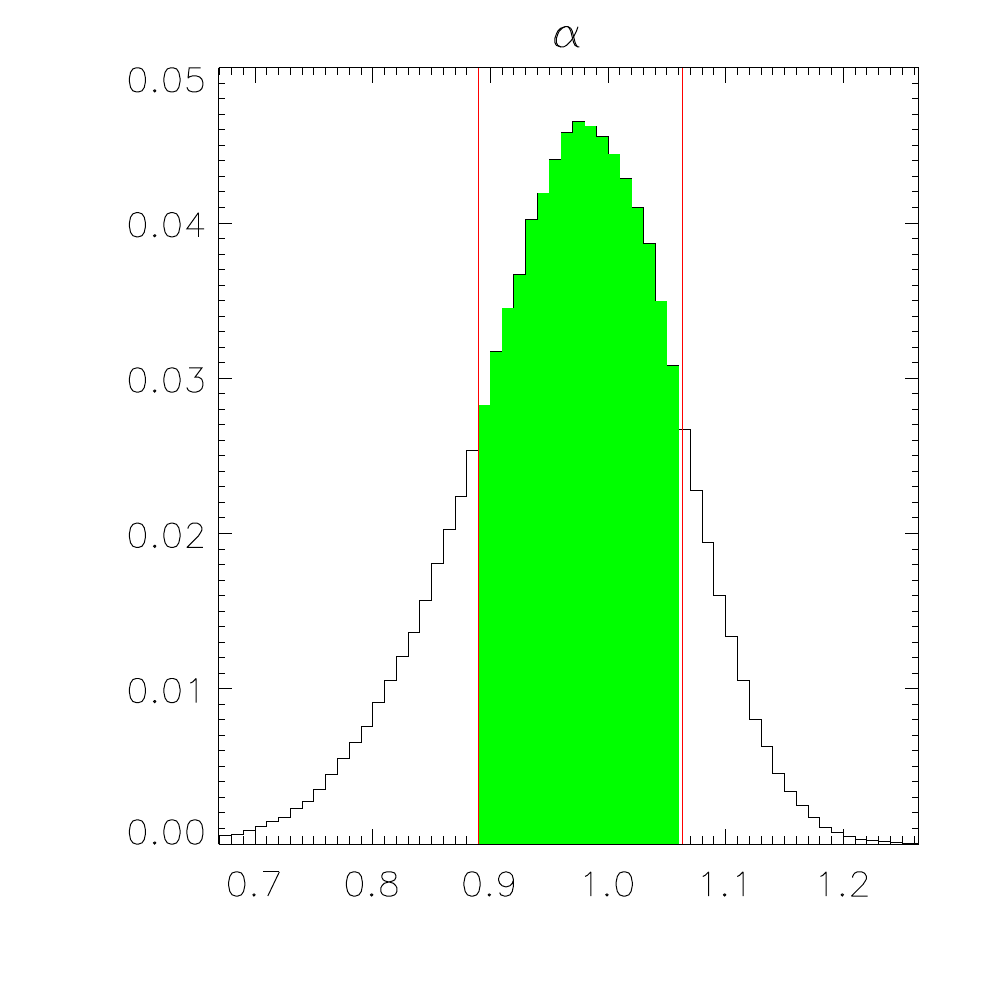}}
   {\myincludegraphics{width=0.25\textwidth}{\figdir{}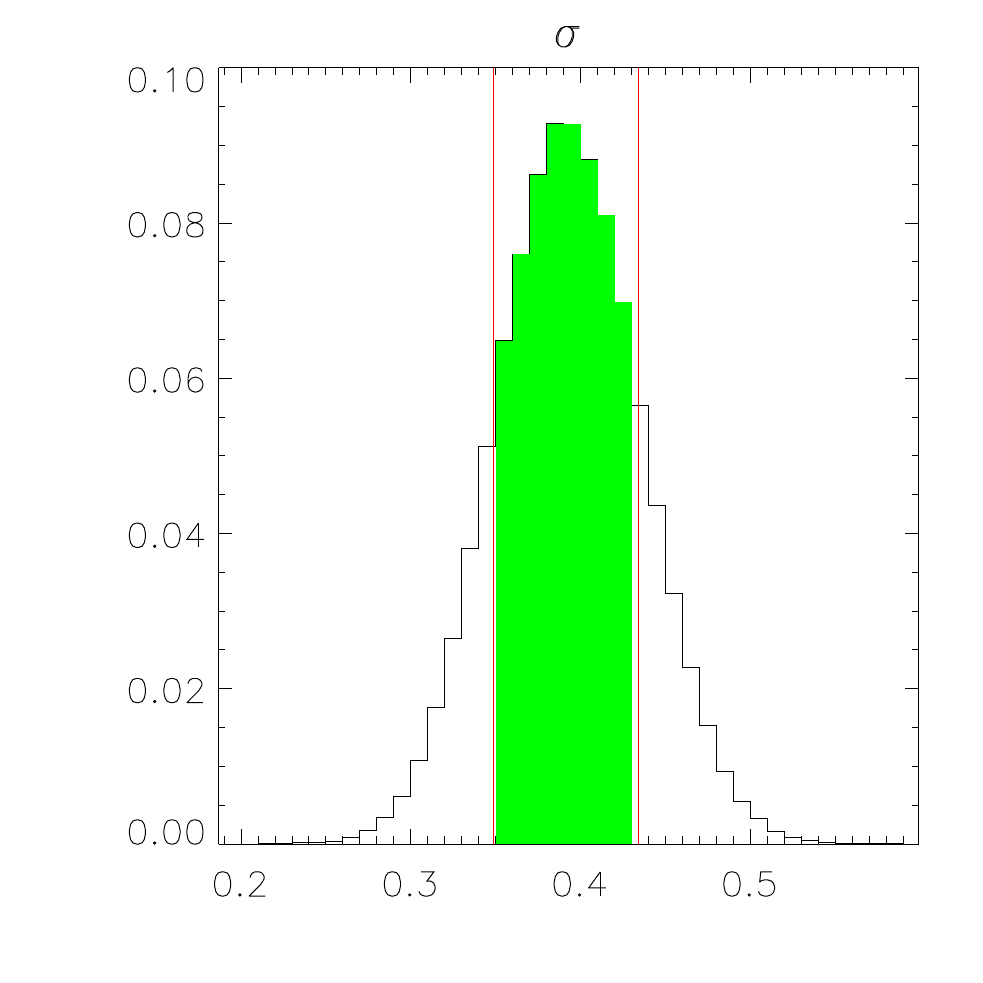}}
   {\myincludegraphics{width=0.25\textwidth}{\figdir{}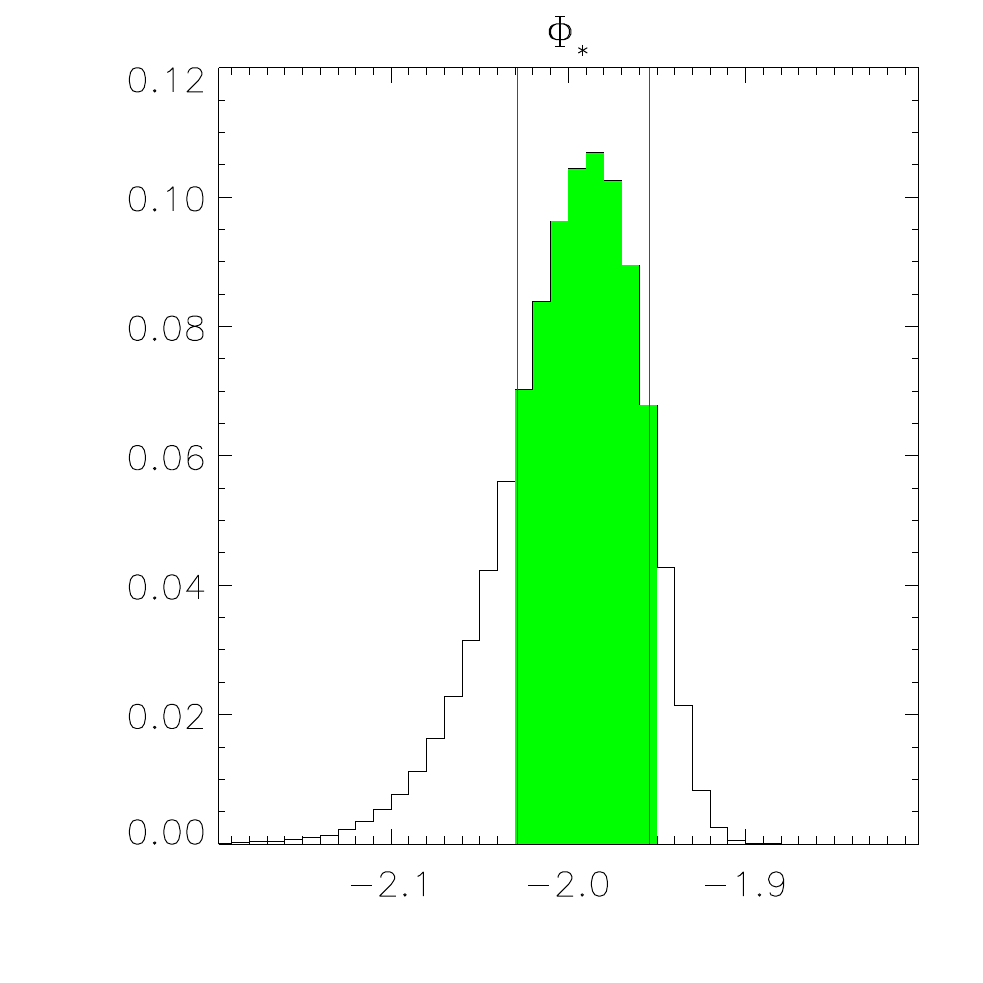}}
\end{center}
\caption[]{Probability histogram of the best fitting parameters ($L^*$, $\alpha$, $\sigma$ and $\Phi^*$) for the SPIRE 250$\,\mu$m local luminosity function within $0.02<z<0.1$, determined using the MCMC parametric Bayesian procedure performing $5\times10^6$ iterations. The highlighted area is the $\pm1\sigma$ confidence area for each parameter, as reported in Tab. \ref{mcmc.param}.}\label{mcmc.hist}
\end{figure*}

In Tab. \ref{mcmc.param} we report the values of the best parameter solutions of the parametric bayesian ML procedure (explained in Sec. \ref{bayes.MLE}) using the log-Gaussian functional form (Eq. \ref{log.gaus}). In Fig. \ref{mcmc.hist} we report the histograms representing the probability distribution of the best fit parameters produced by the MCMC procedures. To obtain these estimates we run an MCMC procedure with $5\times10^6$ iterations. This procedure is a highly time-consuming process, thus we focused our attention in the most local bin $0.02<z<0.1$ of our analysis where we want to obtain a precise estimate of the shape of the local LF observed by \textit{Herschel} at 250$\,\mu$m, which is our selection band. Such an estimate represents a fundamental benchmark to study the evolution of the luminosity function (e.g. Vaccari et al., in prep.) as discussed later in Sec. \ref{discussion}. 

\begin{table}
\centering
 \begin{tabular}{lc|c}
\multicolumn{2}{c|}{\textbf{Parameter}} & \textbf{$\langle \sigma \rangle$} \\
\hline
 log($L^*$) [L$_\odot$] &     $9.03^{+0.14}_{-0.13}$   &      0.14 \\
 $\alpha$                       &      $0.96 ^{+0.09}_{-0.07}$  &    0.08  \\             
  $\sigma$                     &       $0.39^{+0.04}_{-0.04}$  &     0.04 \\              
 log($\Phi^*$) [Mpc$^{-3}$dex$^{-1}$] &       $-1.99^{+0.04}_{-0.02}$ &       0.03 \\
\end{tabular}
\caption[]{Best fit parameter solution and uncertainties for the local SPIRE 250$\,\mu$m LF determined using the parametric Bayesian ML procedure. The redshift range for this solution is $0.02<z<0.1$.}
\label{mcmc.param}
\end{table}

A summary of the results is reported in the following figures. In Fig. \ref{local.whole} and \ref{local.perfield} we report the SPIRE 250$\,\mu$m rest-frame LF estimated by using the $1/V_{\rm max}$ and the parametric Bayesian ML, reporting both the solutions for the five fields togheter (see Tab. \ref{spire-llf-numbers.tab}) and for each field separately. The SPIRE LLFs in different fields do not show any field to field variations beyond what is expected from cosmic variance, i.e. about 15$\%$ as predicted by theoretical models (\citealt{Moster2011}). To report the confidence area of our Bayesian ML solution we estimate the standard deviation of the best fit model using the following equation: 

\begin{eqnarray}\label{log.gaus.err}
&&\sigma^2_{\Phi(x_{1}, x_{2},..x_{n})} =\displaystyle \sum_{j=1}^{n}\left( \frac{\partial \Phi}{\partial x_j} \sigma_{x_j}\right)^2 + \nonumber \\
&&+ 2 \sum_{j=1}^{n}\sum_{k=j+1}^{n} r_{x_jx_k}\left( \frac{\partial \Phi}{\partial x_j} \sigma_{x_j}\right) \left( \frac{\partial \Phi}{\partial x_k} \sigma_{x_k}\right).
\end{eqnarray}

This equation represents the general formula for the parametric standard deviation in the case of non-independent variables. The functional form of $\Phi$ is, as already stated, the log-Gaussian function described in Eq. \ref{log.gaus}, in which the parameters $L^*$, $\alpha$, $\sigma$ and $\Phi^*$ are in fact not independent from each other. Thus $\Phi(x_{1}, x_{2},..x_{n})$ reported in Eq. \ref{log.gaus.err} can be translated, into our specific case, as $\Phi(\rm{L}_*, \alpha, \sigma, \Phi^*)$, while $\sigma_{x_j}$ expresses the error associated to the $j$-th parameter in the sum (and the same with $\sigma_{x_k}$ for the $k$-th parameter).

\begin{figure*}
\begin{center}
   {\myincludegraphics{width=0.8\textwidth}{\figdir{}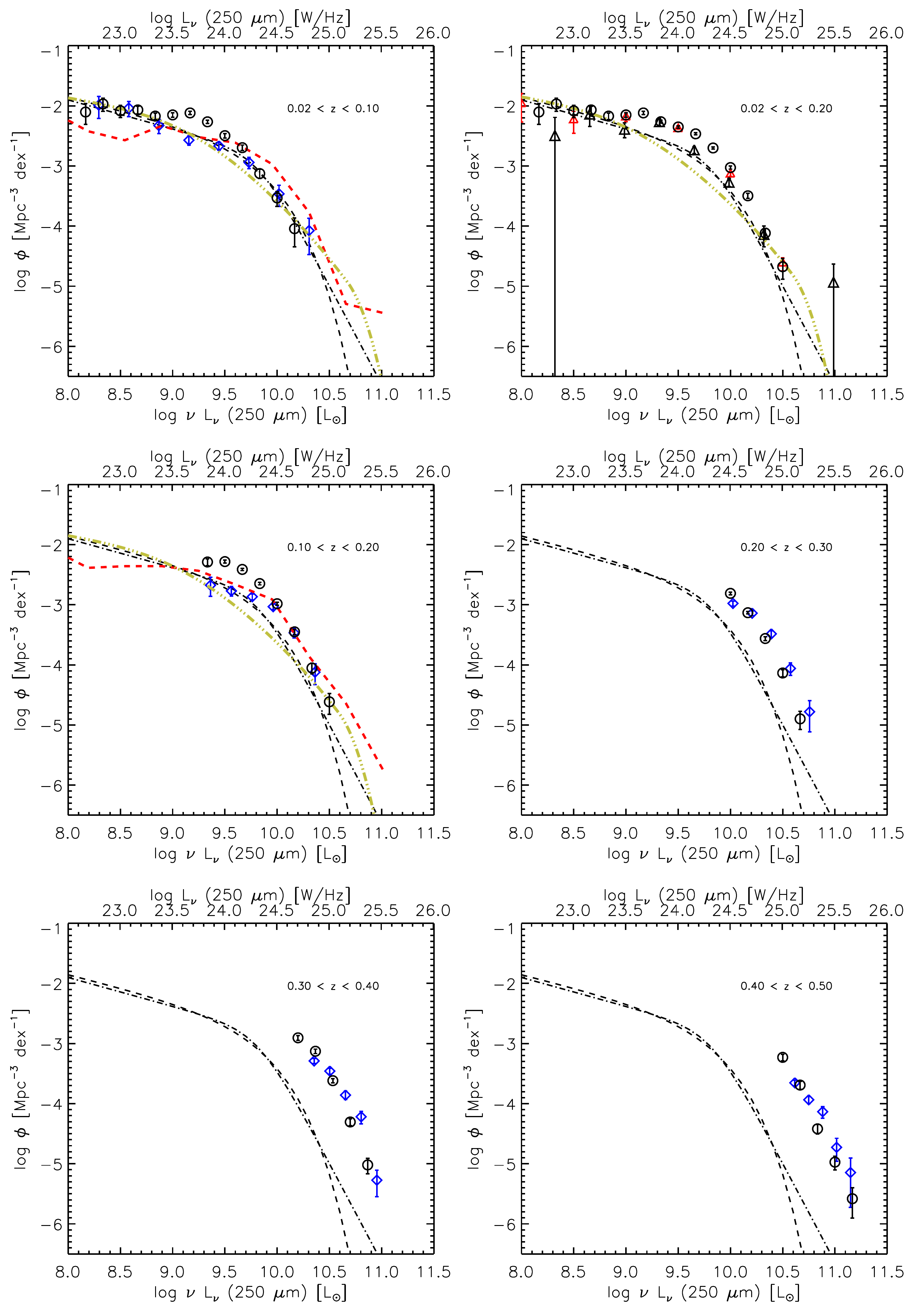}} 
\end{center}
\caption[]{SPIRE 250$\,\mu$m rest-frame local luminosity functions compared to the H-ATLAS estimate from \cite{Dye2010}. The black open circles are our $1/V_{\rm max}$ results; the blue open diamonds are the SDP H-ATLAS SPIRE 250$\,\mu$m rest-frame local luminosity function from \cite{Dye2010}; the red open triangles are the SDP HerMES SPIRE 250$\,\mu$m rest-frame local luminosity function of \cite{Vaccari2010}; the black open triangles are the the SDP HerMES SPIRE 250 rest-frame luminosity function of \cite{Eales2010b}; the red dashed line is the SPIRE 250$\,\mu$m luminosity function predicted by \cite{Fontanot2012}; the beige dashed-dot-dot-dot line is the \cite{Negrello2007} model; the black dot-dashed and dashed lines are local luminosity function prediction at 250$\,\mu$m from \cite{SerjeantHarrison2005}. \cite{Negrello2007} and \cite{SerjeantHarrison2005} estimates are reported at the same local ($z=0$) redshift in all panels.}
\label{local.dye}
\end{figure*}

\begin{figure*}
\begin{center}
   {\myincludegraphics{width=0.45\textwidth}{\figdir{}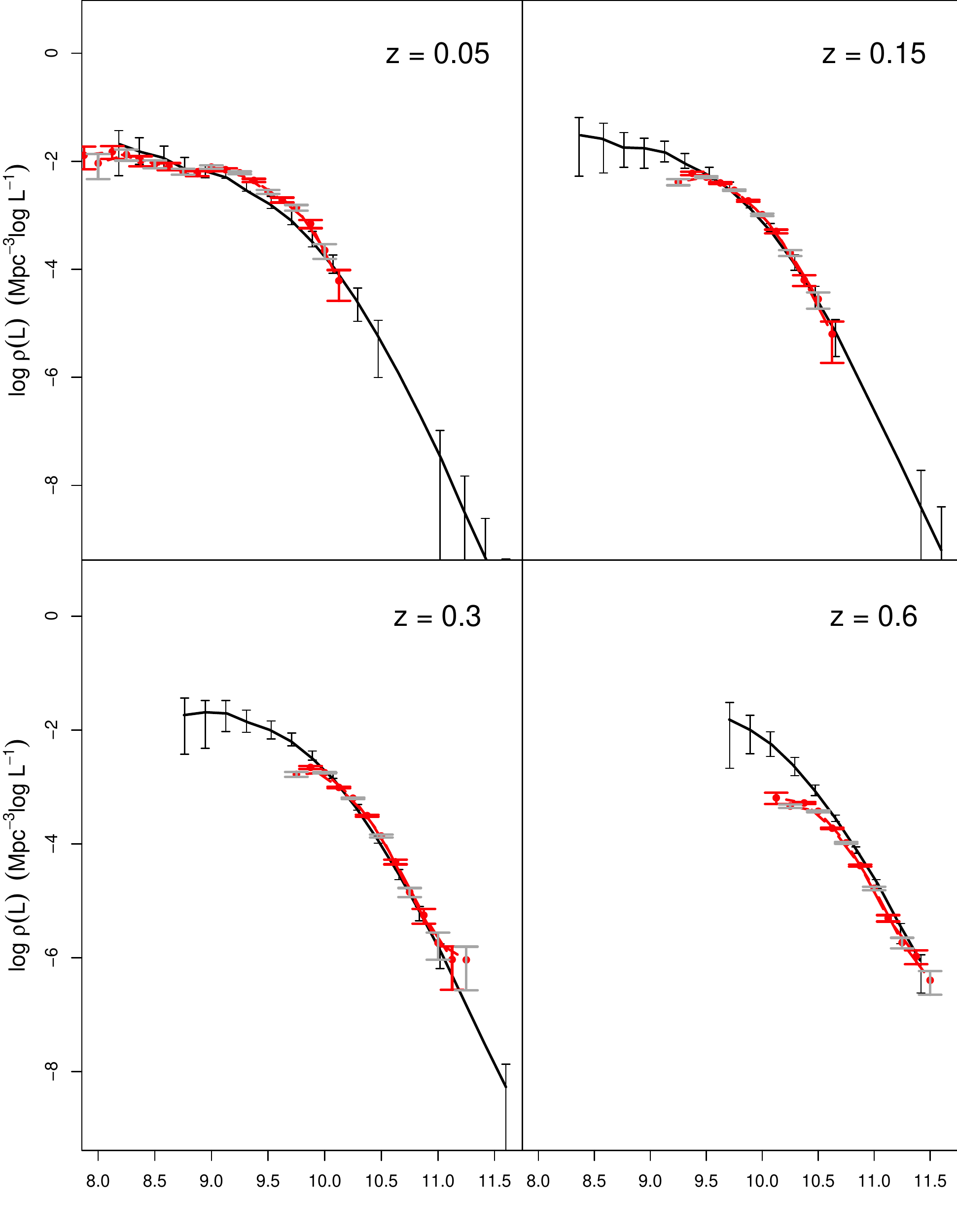}}
\end{center}
\caption[]{SPIRE 250$\,\mu$m rest-frame local luminosity function estimated using the semi-parametric method of \cite{Schafer2007} and the modified $1/V_{\rm max}$ approach of \cite{PageCarrera2000}. Our classic $1/V_{\rm max}$ estimate is shown in grey; in red is the estimate using the \cite{PageCarrera2000} method and in black the estimate using the \cite{Schafer2007} approach.}
\label{local.schafer}
\end{figure*}

\begin{figure*}
\begin{center}
   {\myincludegraphics{width=0.45\textwidth}{\figdir{}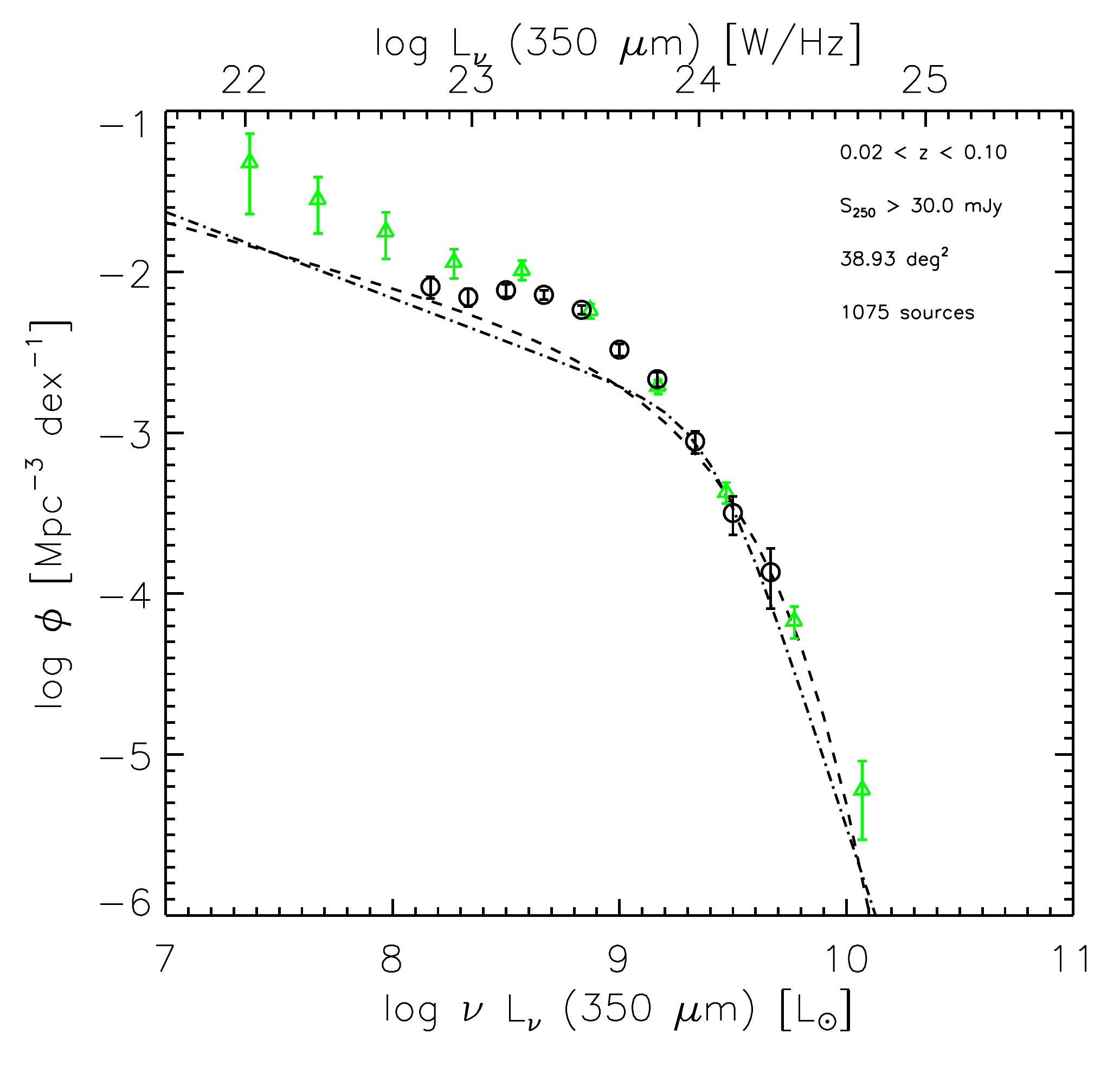}}
   {\myincludegraphics{width=0.45\textwidth}{\figdir{}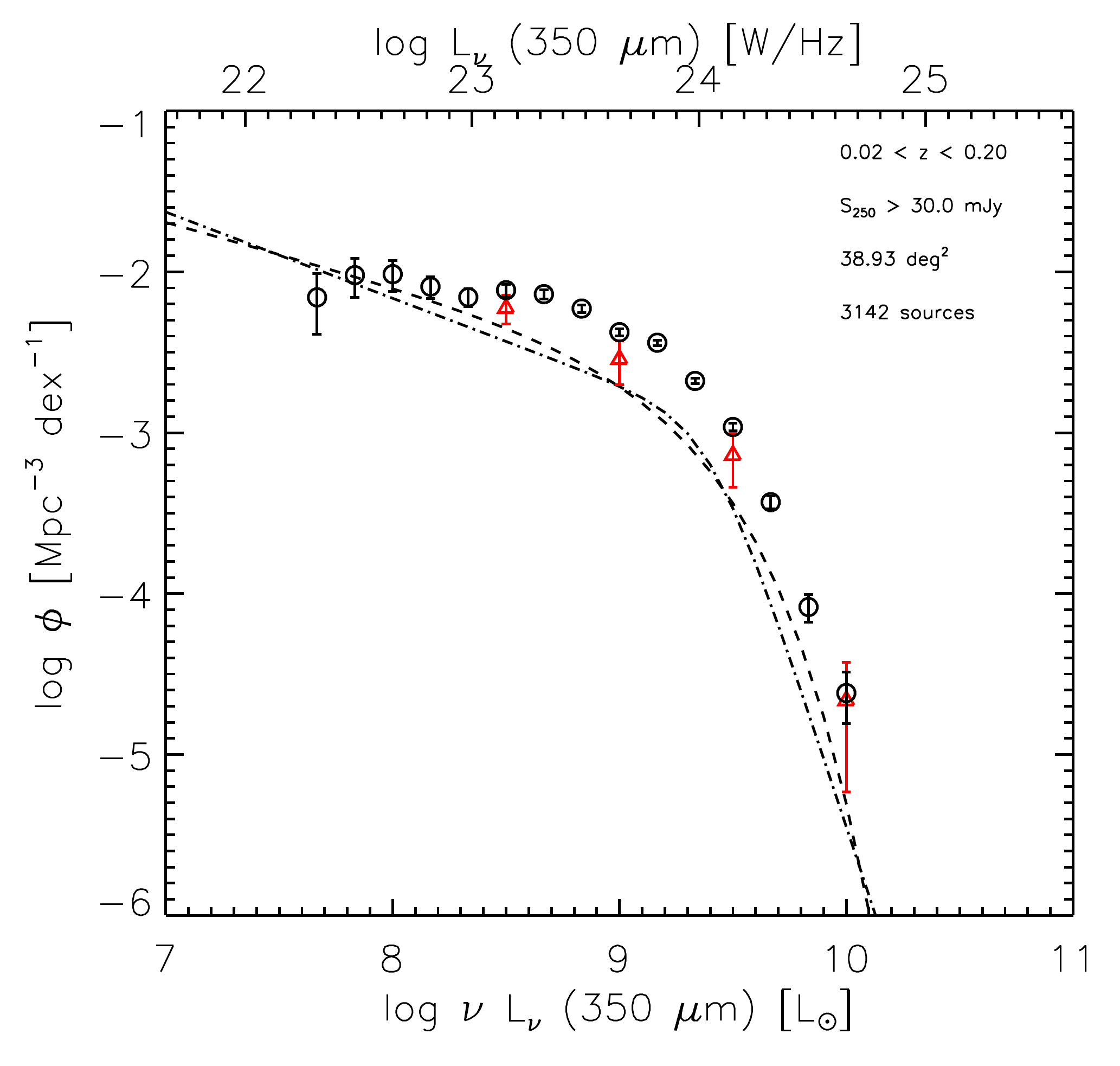}}
\end{center}
\caption[]{SPIRE 350$\,\mu$m rest-frame local luminosity function estimates. The black open circles are our $1/V_{\rm max}$; the red open triangles are the SDP HerMES SPIRE 350$\,\mu$m rest-frame local luminosity function from \cite{Vaccari2010}; the green open triangles are the Planck 857 GHz or 350$\,\mu$m local luminosity function estimate from \cite{Negrello2013}; the black dot-dashed and dashed lines are local luminosity function prediction at 350$\,\mu$m from \cite{SerjeantHarrison2005}.}
\label{local.350}
\end{figure*}

\begin{figure*}
\begin{center}
   {\myincludegraphics{width=0.45\textwidth}{\figdir{}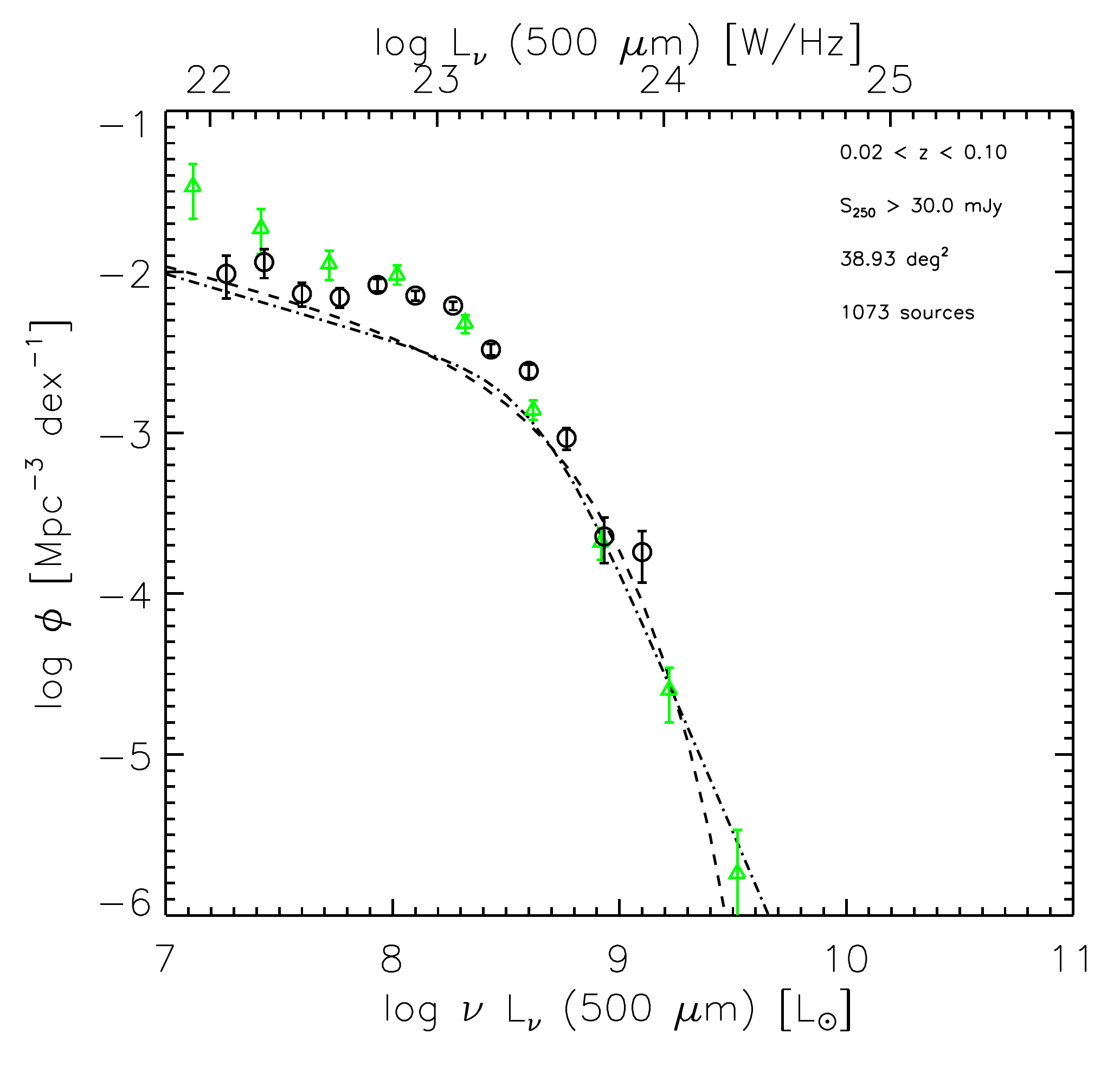}}
   {\myincludegraphics{width=0.45\textwidth}{\figdir{}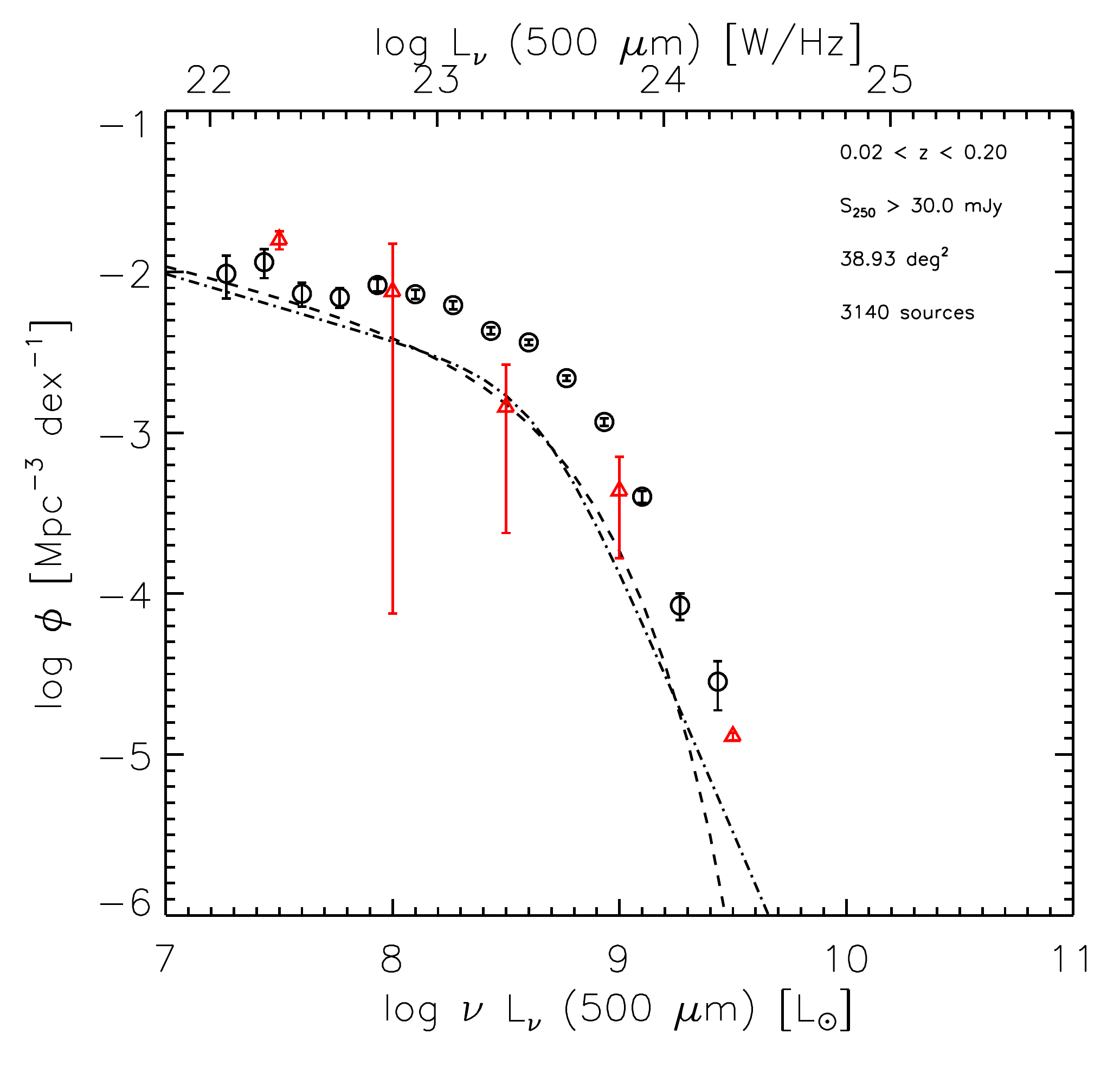}}
\end{center}
\caption[]{SPIRE 500$\,\mu$m rest-frame local luminosity function estimates. The black open circles are our $1/V_{\rm max}$; the red open triangles are the SDP HerMES SPIRE 500$\,\mu$m rest-frame local luminosity function from \cite{Vaccari2010}; the green open triangles are the Planck 545 GHz or 550$\,\mu$m local luminosity function estimate from \cite{Negrello2013} converted to our wavelength by using a spectral index of $\alpha = 2.7$; the black dot-dashed and dashed lines are local luminosity function prediction at 500$\,\mu$m from \cite{SerjeantHarrison2005}.}
\label{local.500}
\end{figure*}

\begin{figure*}
\begin{center}
   {\myincludegraphics{width=1.0\textwidth}{\figdir{}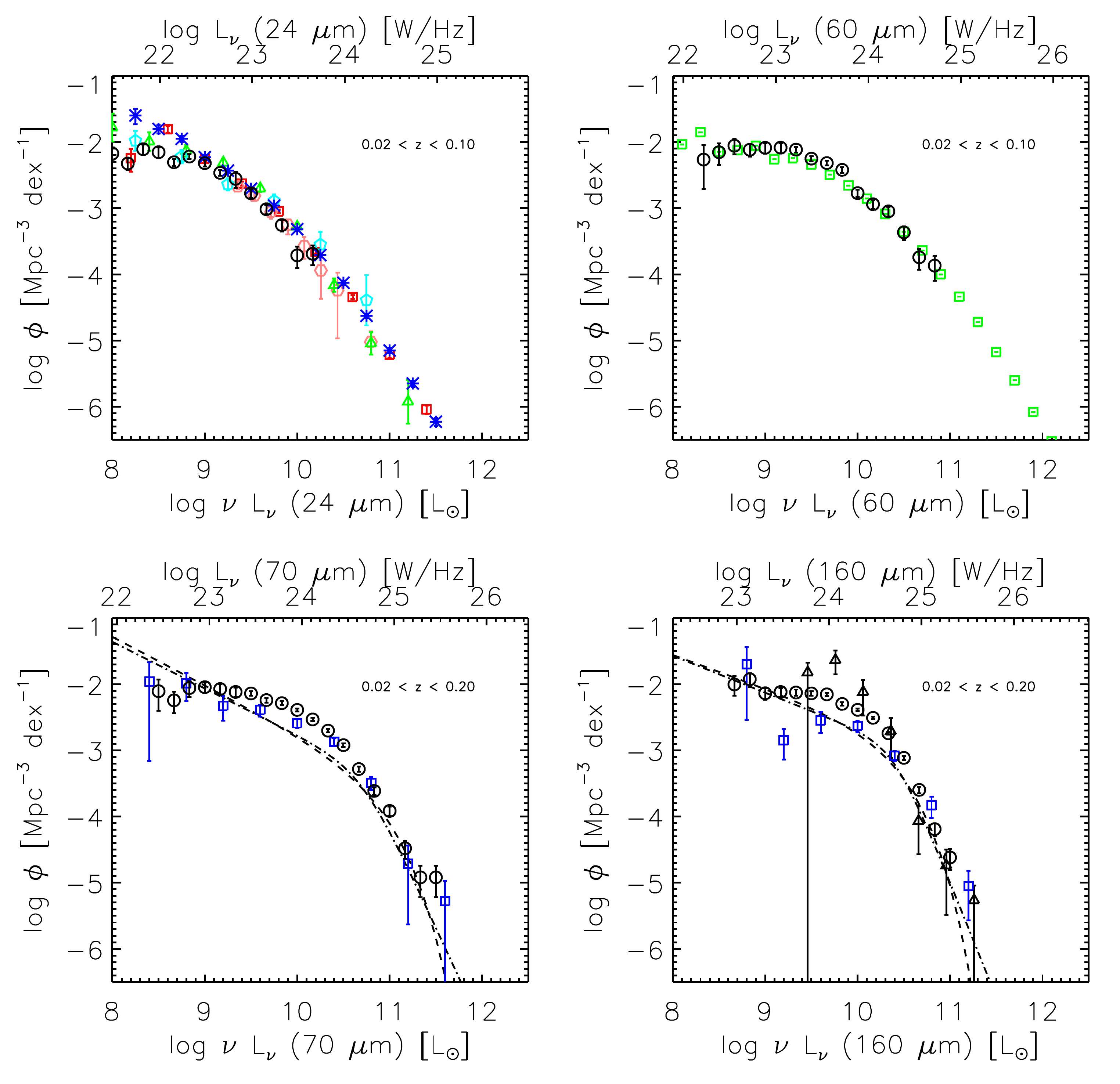}} 
\end{center}
\caption[]{MIPS 24/70/60$\,\mu$m and IRAS 60$\,\mu$m LFs as derived by our SPIRE 250$\,\mu$m sample. The black open circles are our $1/V_{\rm max}$ in all the panels. \textbf{Top left} The MIPS 24$\,\mu$m LF estimate. The open red squares are the IRAS 25$\,\mu$m LF from \cite{Shupe98}; the open green triangles are the MIPS 24$\,\mu$m LF from \cite{Marleau2007}; the open pink exagons are the MIPS 24$\,\mu$m LF of \cite{Babbedge2006}; the blue asterisks are the 25$\,\mu$m from IIFSCz by \cite{WangMRR2009} converted to MIPS 24$\,\mu$m; the open light blue pentagons are the MIPS 24$\,\mu$m LF from \cite{Rodighiero2010}. \textbf{Top right} The \textit{IRAS} 60$\,\mu$m LF estimate. The open green squares are the \textit{IRAS} 60$\,\mu$m LF from \cite{Saunders1990}. \textbf{Bottom left} The MIPS 70$\,\mu$m LF estimate. The open blues squares are the MIPS 70$\,\mu$m LF of \cite{Patel2013}; the dot-dashed and dashed line are the LF estimates from \cite{SerjeantHarrison2005}. \textbf{Bottom right} The MIPS 160$\,\mu$m LF estimate. The open blues squares are the MIPS 160$\,\mu$m LF of \cite{Patel2013}; the open black triangles are the ISO 170$\,\mu$m LF from \cite{Takeuchi2006} converted to MIPS 160$\,\mu$m; the dot-dashed and dashed line are the LF estimates from \cite{SerjeantHarrison2005}.}
\label{local.multilambda}
\end{figure*}

\begin{figure*}
\begin{center}
   {\myincludegraphics{width=0.8\textwidth}{\figdir{}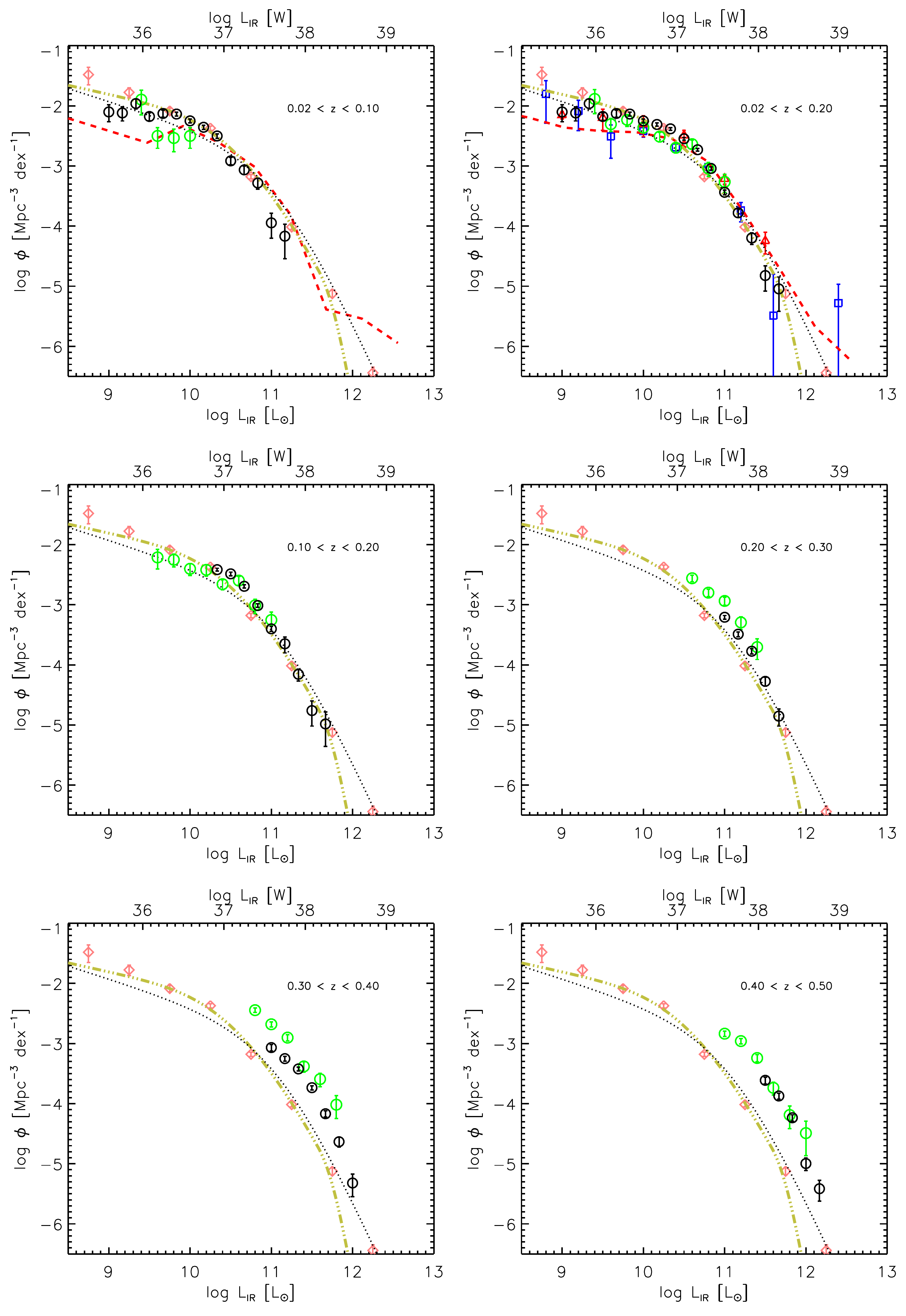}}
\end{center}
\caption[]{The IR bolometric rest-frame local luminosity functions. The black open circles are our $1/V_{max}$ results; the green open circles are the $1/V_{max}$ results using COSMOS data (area 1.7 deg$^2$ and flux limited $S250 > 10$~mJy, Vaccari et al., in prep.); the blue open squares are the SWIRE IR bolometric rest-frame luminosity function of \cite{Patel2013} using a MIPS 70 and 160$\,\mu$m selected sample in LH and XMM-LSS; the red open triangles are the IR bolometric rest-frame luminosity function estimate of \cite{Vaccari2010}; the red dashed line is the IR bolometric luminosity function predicted of \cite{Fontanot2012}; the pink open diamonds are the \textit{IRAS} IR bolometric rest-frame luminosity function of \cite{Sanders2003}; the beige dashed-dot-dot-dot line is \cite{Negrello2007} model; the black dotted line is \cite{Valiante2009} model. \cite{Sanders2003}, \cite{Negrello2007} and \cite{Valiante2009} estimates are reported at the same local ($z=0$) redshift in all panels.}
\label{local.LIR}
\end{figure*}

In Fig. \ref{local.schafer} we report the SPIRE 250$\,\mu$m rest-frame LF estimated by using the semi-parametric method described in Sec. \ref{schafer.sec} and the modified $1/V_{\rm max}$ estimates from \cite{PageCarrera2000} described in Sec. \ref{Vmax}. In Fig. \ref{local.dye} we compare our SPIRE 250$\,\mu$m $1/V_{\rm max}$ LF solution to the H-ATLAS results of \cite{Dye2010}. In Figs. \ref{local.350}, \ref{local.500} and \ref{local.LIR} we report the SPIRE 350/500$\,\mu$m and IR bolometric rest-frame LFs respectively. Finally, as a check on the robustness of our SPIRE 250$\,\mu$m selected sample we estimate the LFs also at other wavelengths, namely MIPS 24/70/160$\,\mu$m and PACS 70/100/160$\,\mu$m, and compare our results to others already published. In Fig. \ref{local.multilambda} we report the 24/70/90/160$\,\mu$m rest-frame LFs compared with local predictions at these wavelengths given by different authors.

\begin{figure*}
\begin{center}
   {\myincludegraphics{width=0.45\textwidth}{\figdir{}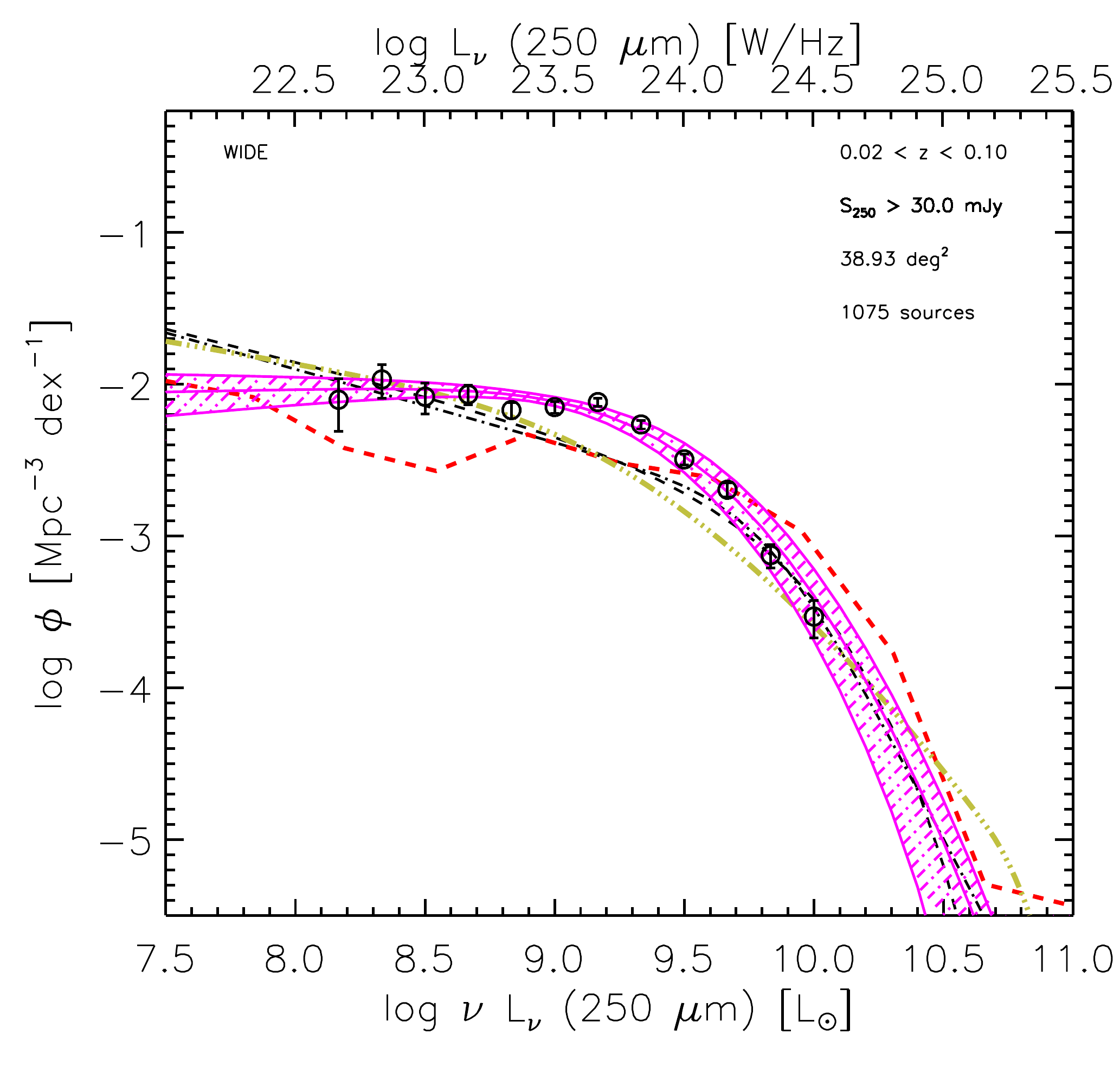}}
  {\myincludegraphics{width=0.45\textwidth}{\figdir{}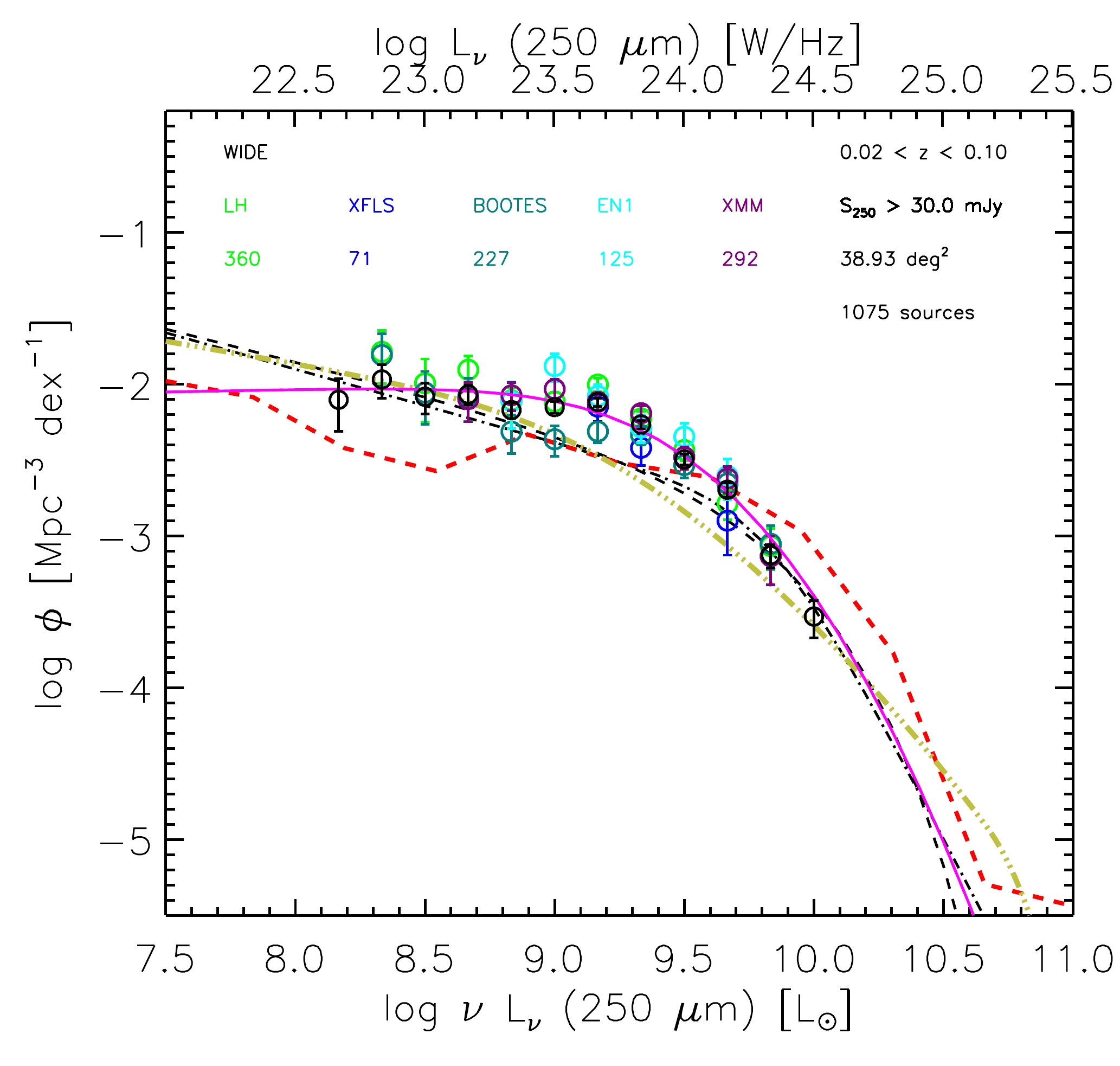}}
\end{center}
\caption[]{SPIRE 250$\,\mu$m rest-frame local luminosity function estimates. The black open circle are our $1/V_{max}$ estimates; the red dashed line is from the \cite{Fontanot2012} model; the beige dashed-dot-dot-dot line is from \cite{Negrello2007} model and the black dot-dashed and dashed lines are local luminosity function prediction at 250$\,\mu$m from \cite{SerjeantHarrison2005}. The magenta shaded region is the $\pm1\sigma$ best MCMC solution using the log-Gaussian functional form reported in the text. The magenta line in the right panel is  the mean from the MCMC solution plotted with the LFs estimates in each field (colour-coded as reported in the legend; the colour-coded number reported in the plot below each field's name is the number of sources in each field in the considered redshift bin).}
\label{local.whole}
\end{figure*}

\begin{figure*}
\begin{center}
   {\myincludegraphics{width=0.44\textwidth}{\figdir{}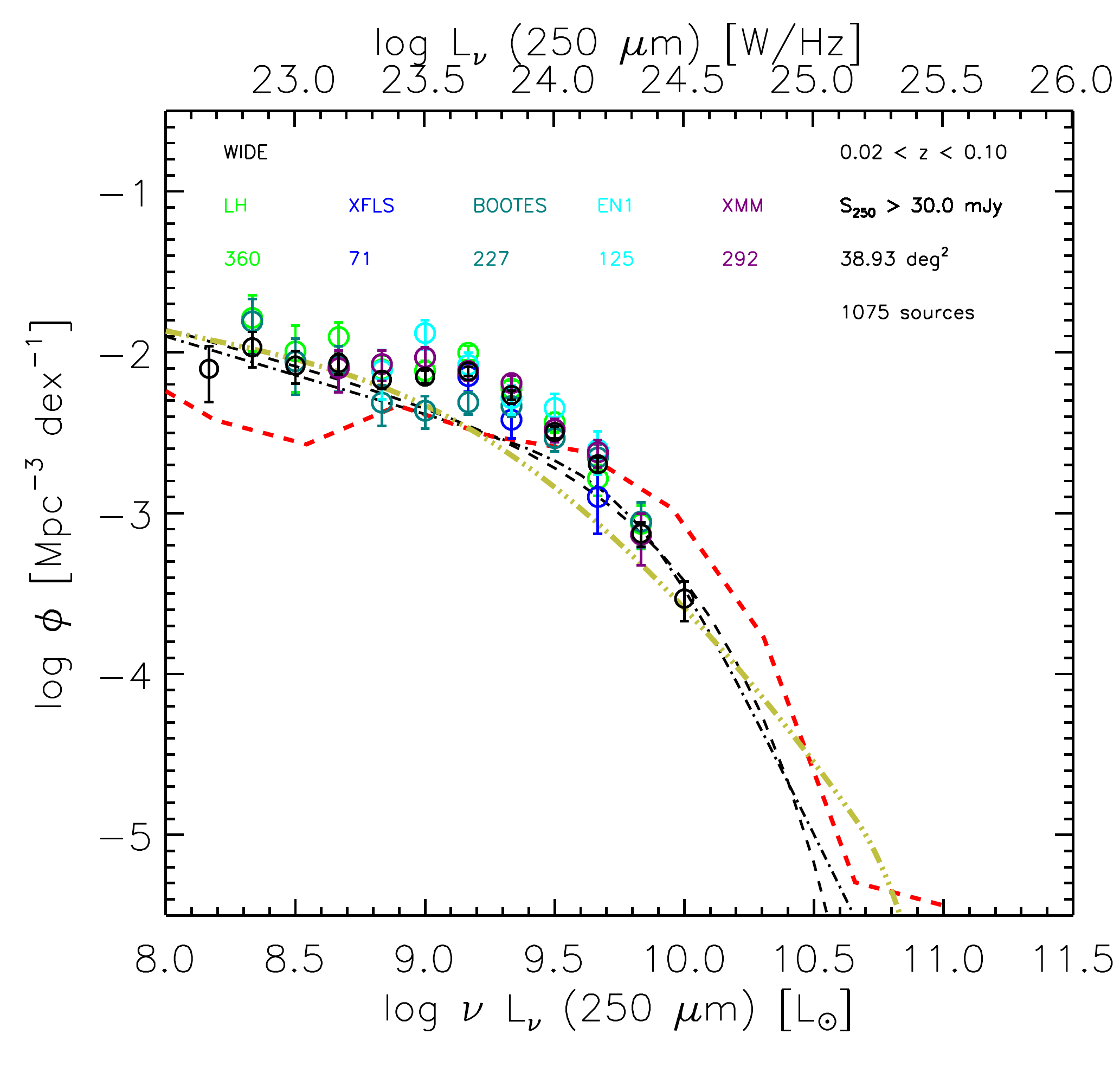}}
   {\myincludegraphics{width=0.44\textwidth}{\figdir{}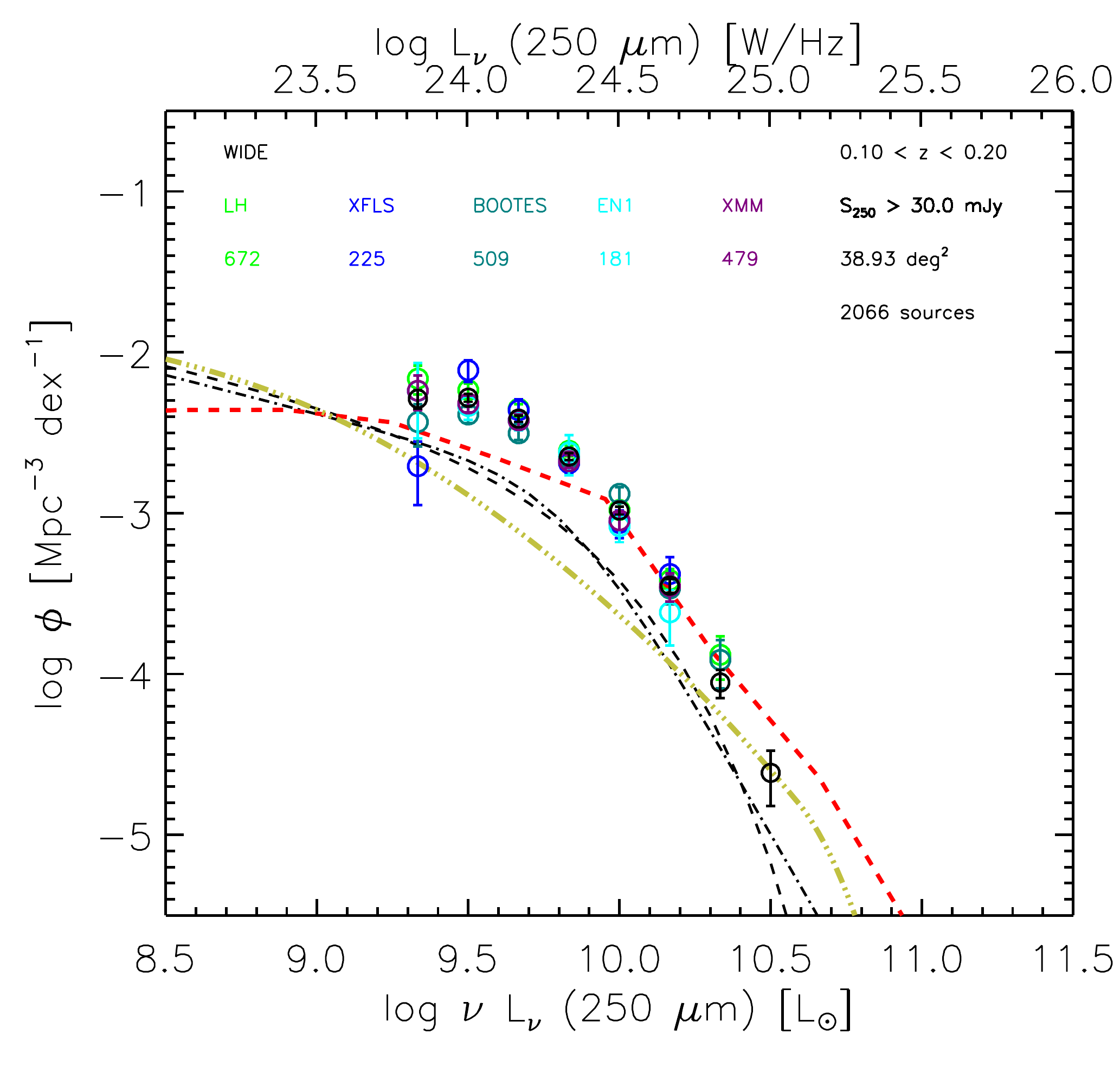}}
   {\myincludegraphics{width=0.44\textwidth}{\figdir{}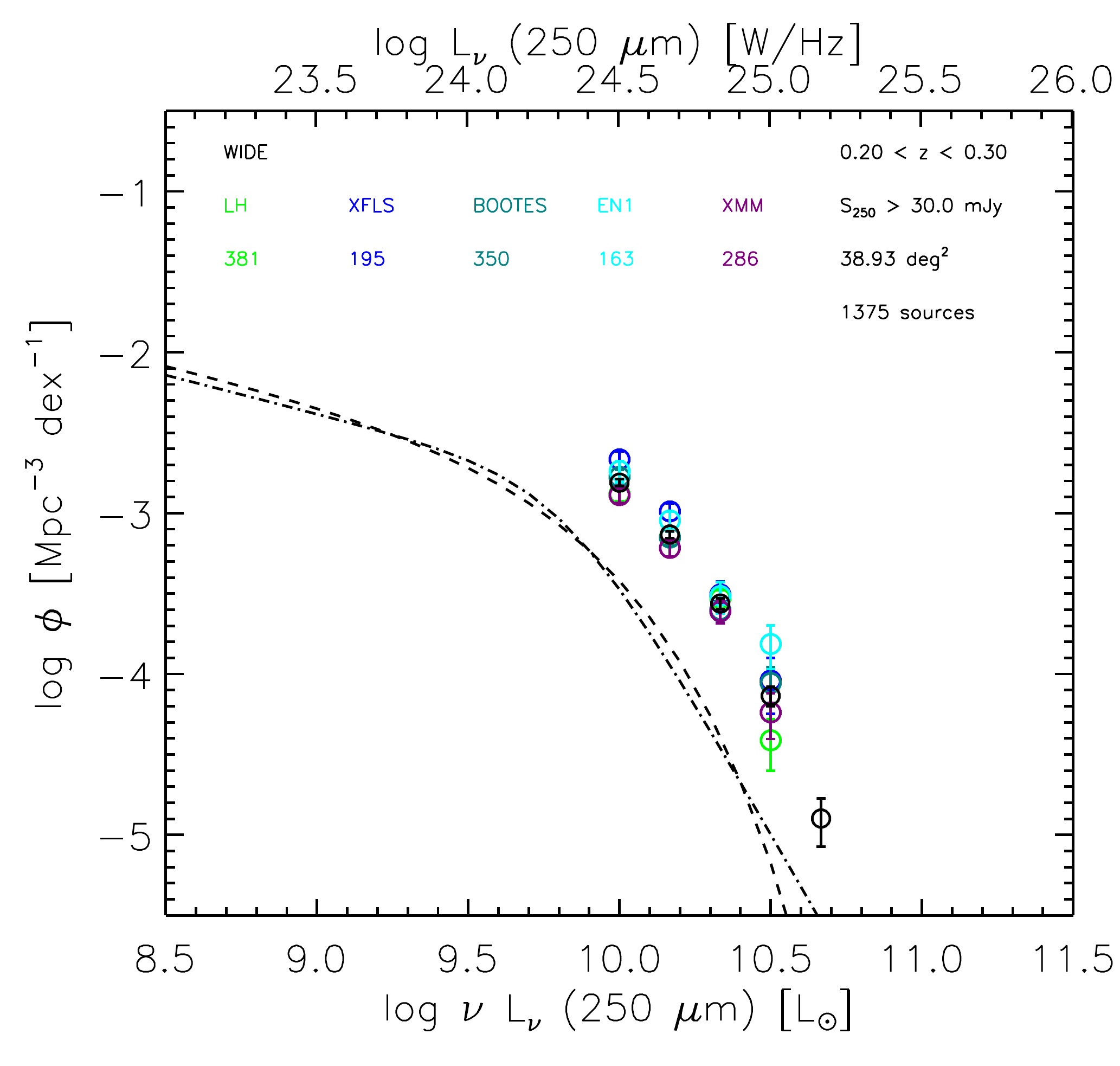}}
  {\myincludegraphics{width=0.44\textwidth}{\figdir{}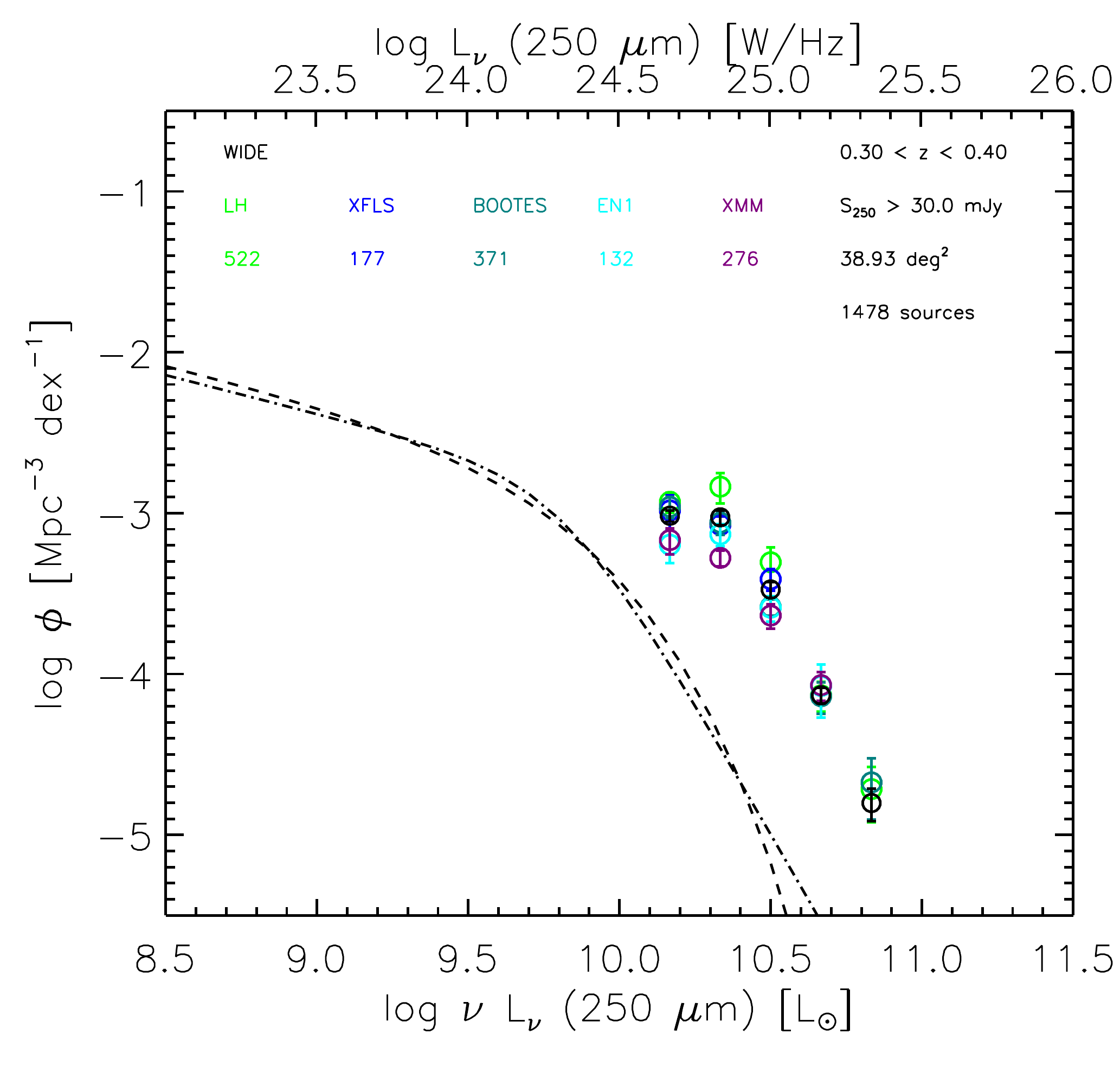}}
   {\myincludegraphics{width=0.44\textwidth}{\figdir{}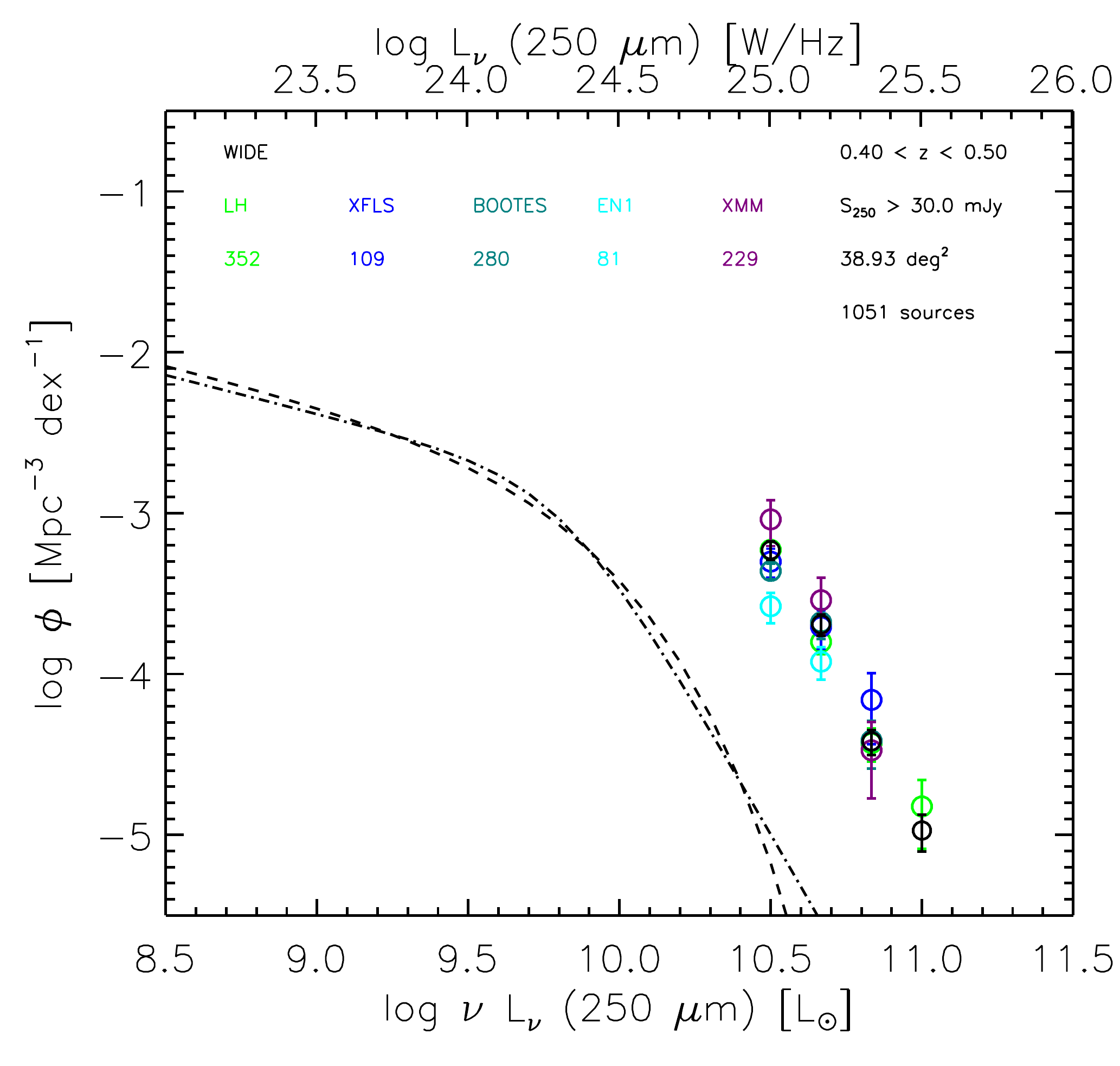}}
\end{center}
\caption[]{SPIRE 250$\,\mu$m rest-frame local luminosity function estimates from field to field. The Colour-coded open circles are our $1/V_{\rm max}$ results for each field (the black is the solution for all the five fields considered together); the red dashed line is the \cite{Fontanot2012} model; the beige dashed-dot-dot-dot line is the \cite{Negrello2007} model; the black dot-dashed and dashed lines are local luminosity function predictions at 250$\,\mu$m from \cite{SerjeantHarrison2005}. \cite{Negrello2007} and \cite{SerjeantHarrison2005} estimates are reported at the same local ($z=0$) redshift in all panels.}
\label{local.perfield}
\end{figure*}

\begin{figure*}
\begin{center}
   {\myincludegraphics{width=0.45\textwidth}{\figdir{}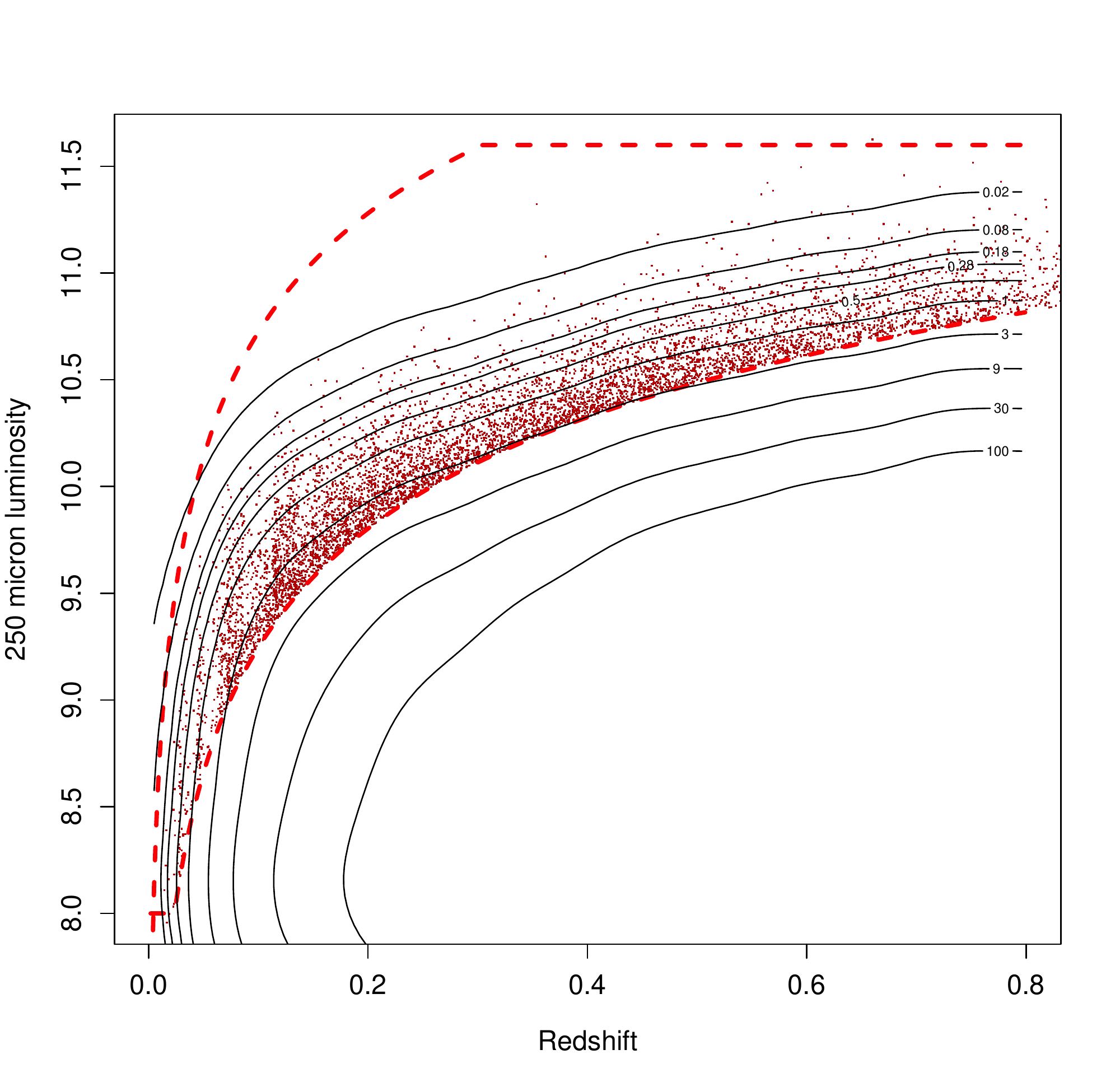}}
\end{center}
\caption[]{SPIRE 250$\,\mu$m luminosity distribution \textit{vs} redshift plane, as reconstructed using the \cite{Schafer2007} estimator. The red points are the data; the red dashed lines mark the flux limitations adopted in the application of the \textit{semi-parametric} LF estimator by \cite{Schafer2007} and the solid black lines are iso-density contours corresponding to the \textit{semi-parametric} reconstructions of the source volume density as a function of luminosity and redshift.}
\label{local.schafer.bivest}
\end{figure*}

\subsection{The IR local luminosity density and the IR local spectral energy distribution}

\begin{figure*}
\begin{center}
   {\myincludegraphics{width=0.45\textwidth}{\figdir{}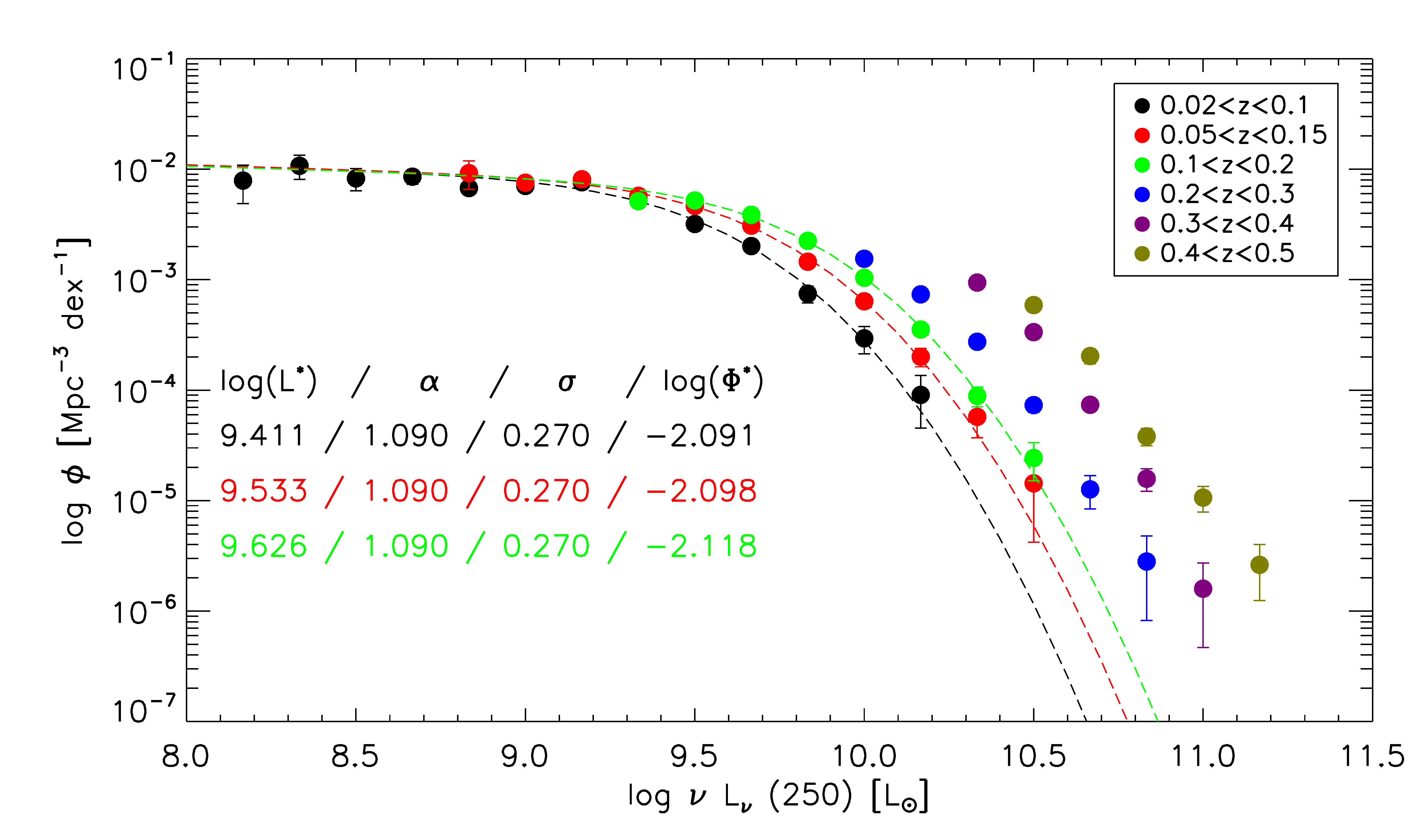}}
   {\myincludegraphics{width=0.45\textwidth}{\figdir{}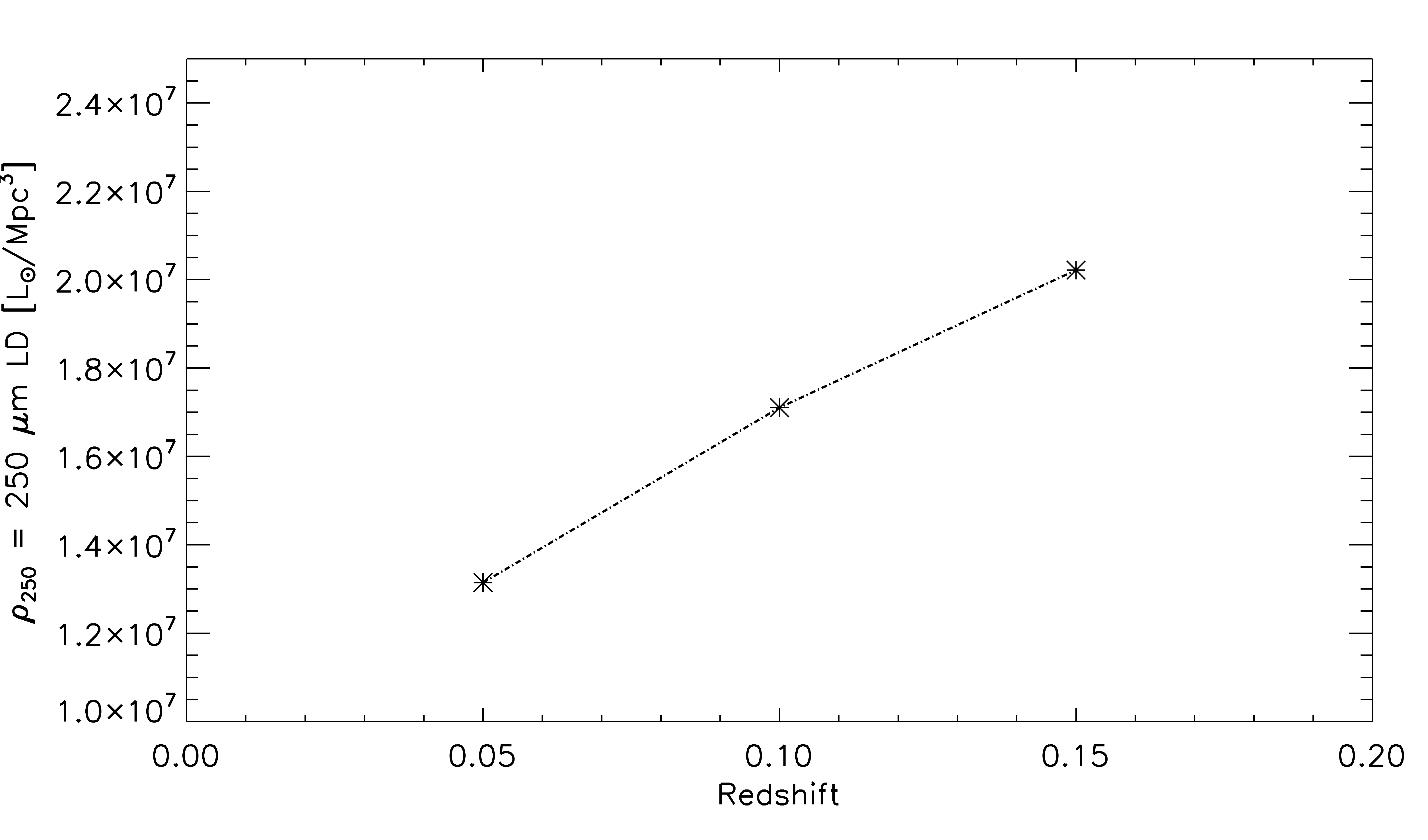}}
   {\myincludegraphics{width=0.45\textwidth}{\figdir{}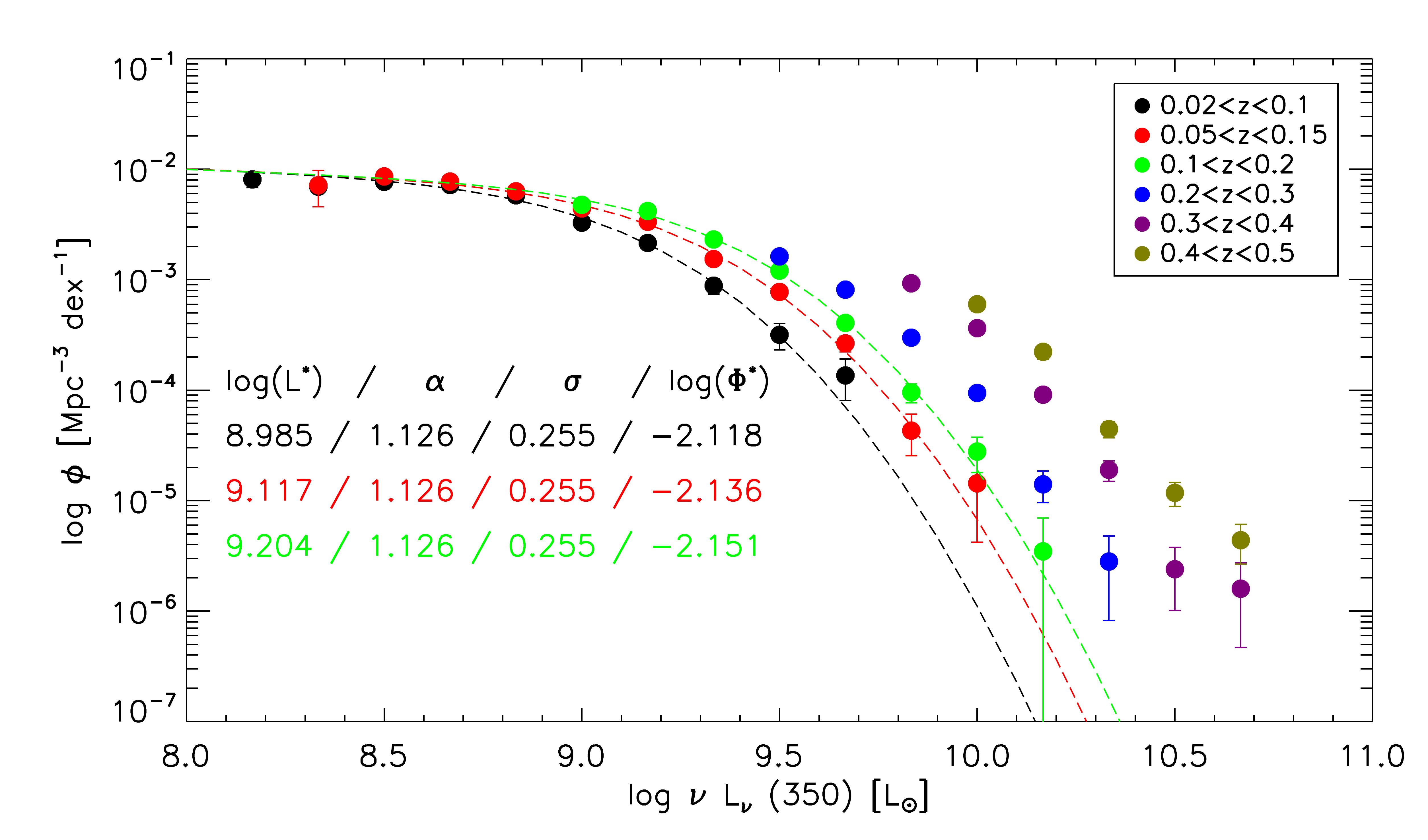}}
   {\myincludegraphics{width=0.45\textwidth}{\figdir{}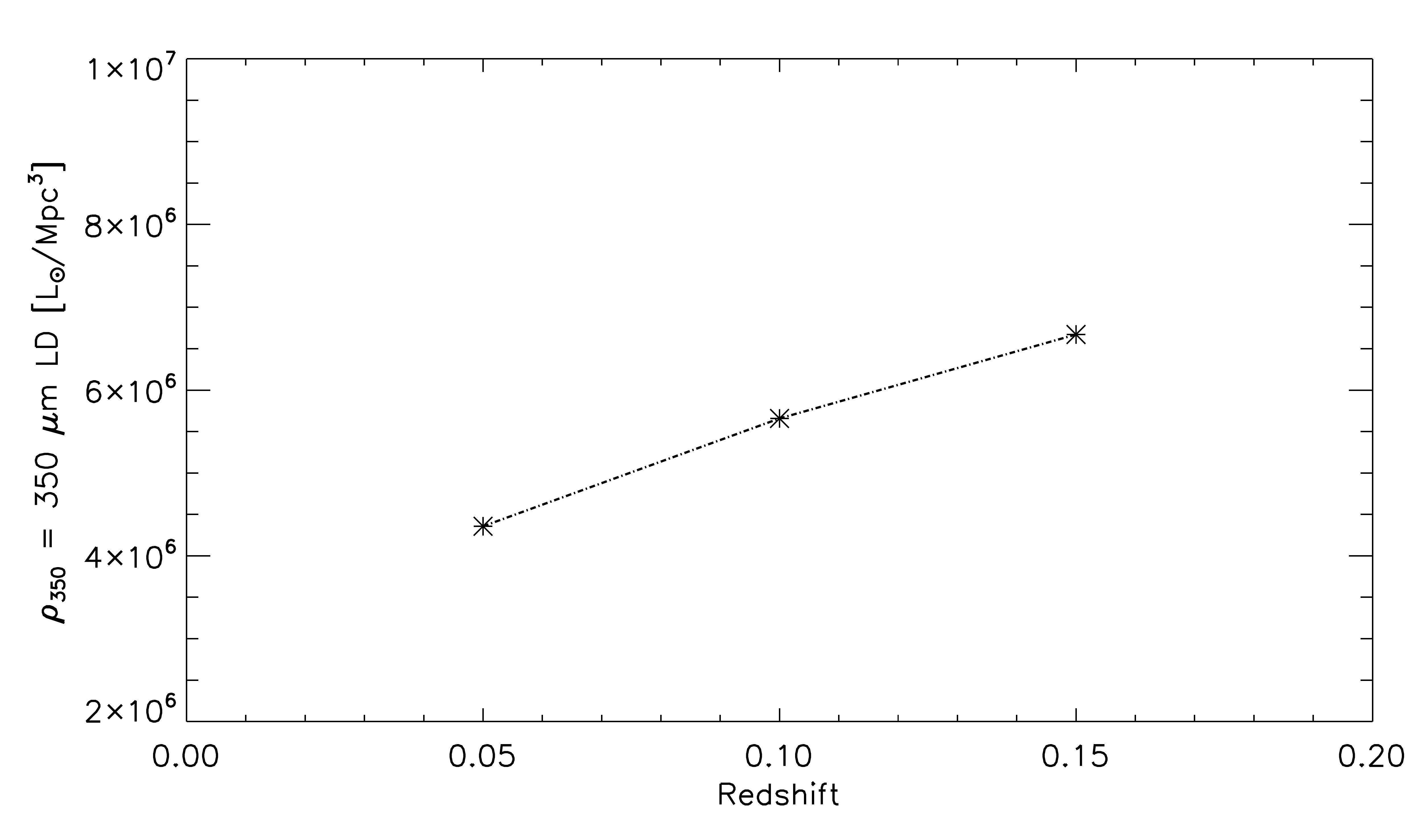}}
   {\myincludegraphics{width=0.45\textwidth}{\figdir{}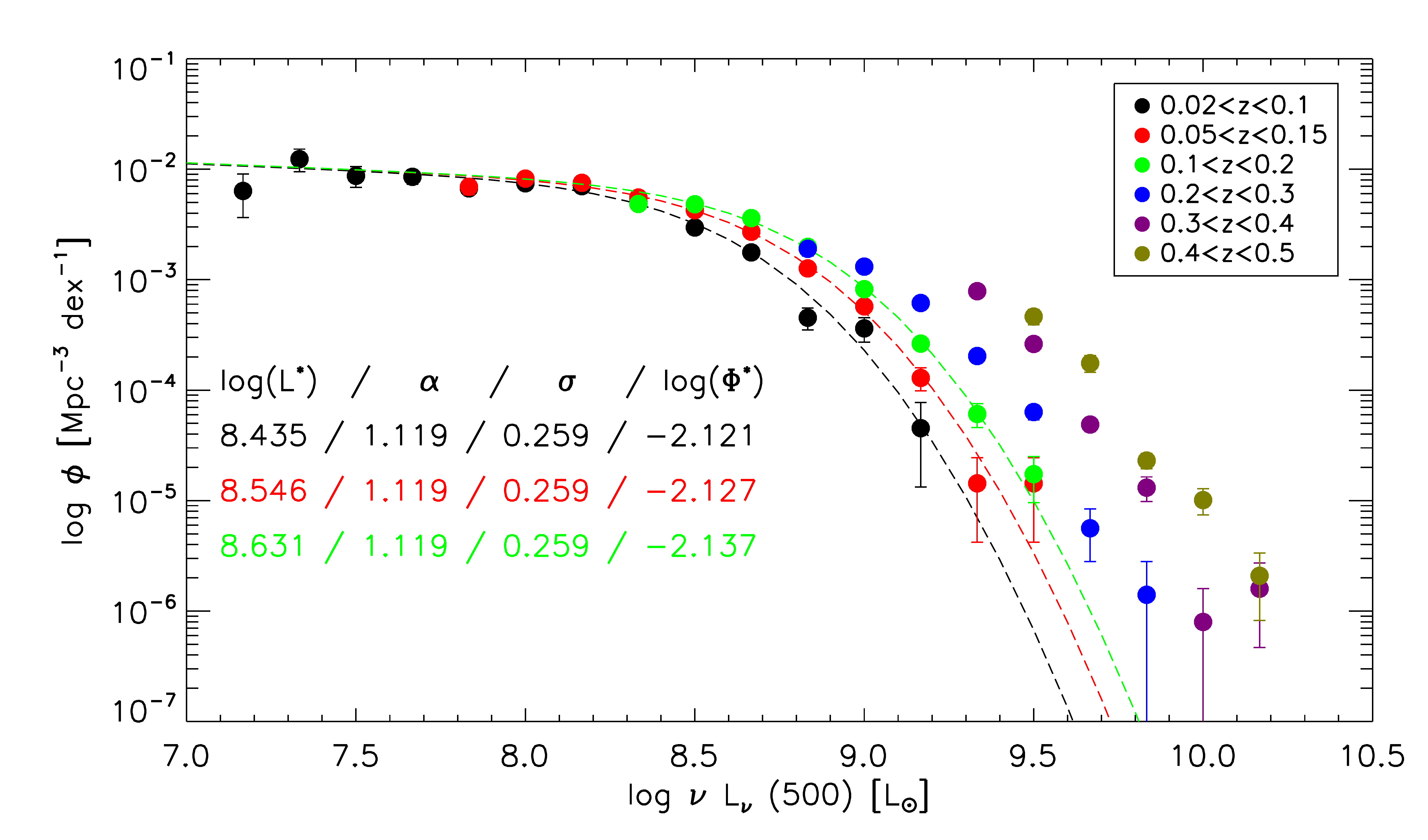}}
   {\myincludegraphics{width=0.45\textwidth}{\figdir{}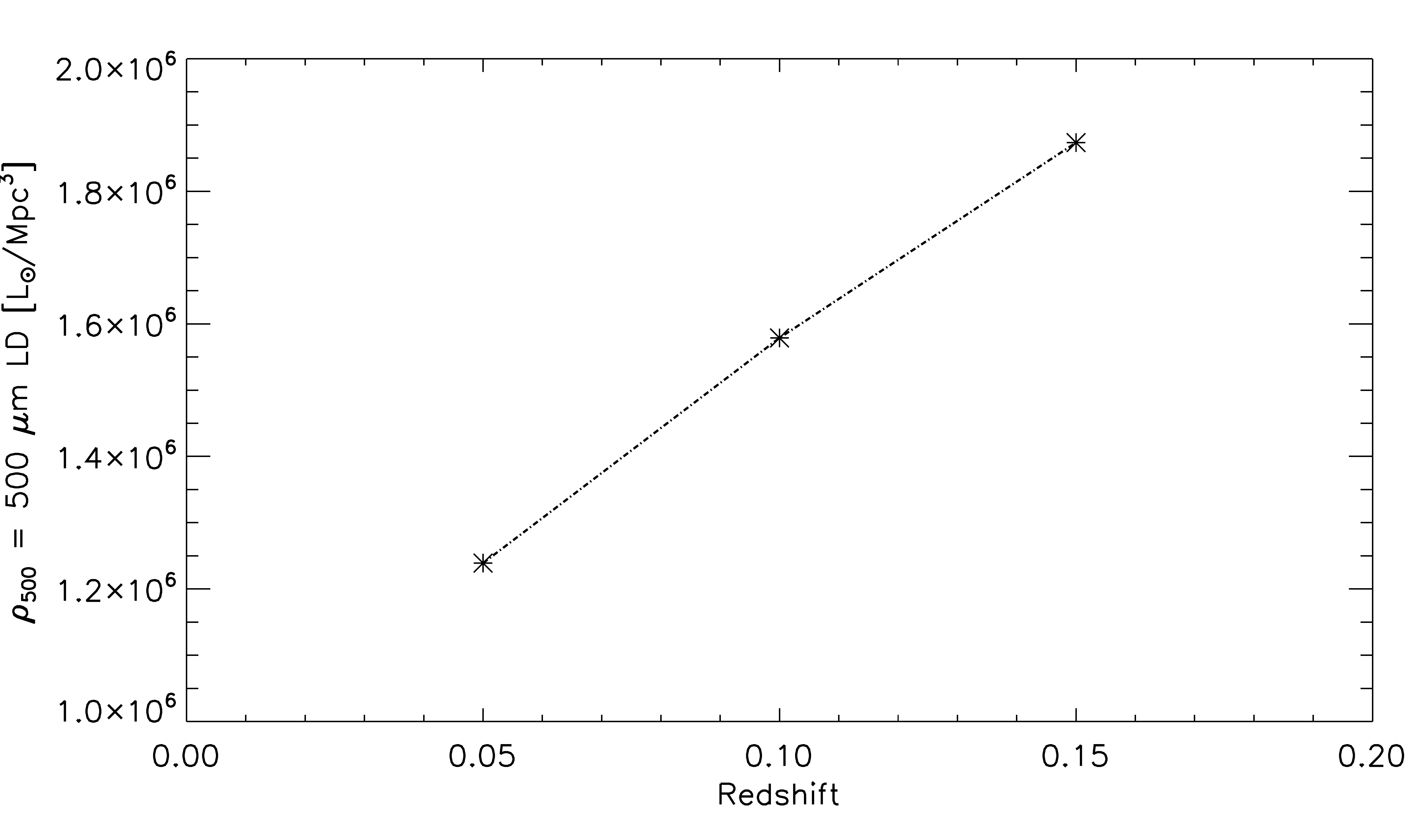}}
\end{center}
\caption[]{SPIRE 250/350/500$\,\mu$m rest-frame LFs evolution within $0.02<z<0.5$ along with the luminosity density estimates. \textbf{Left}: the SPIRE 250/350/500$\,\mu$m LFs evolution within $0.02<z<0.5$. The colour-coded full point are our $1/V_{\rm max}$ solution in each redshift bin while the solid curves represent the best fit solution to the first three redshift bins reported in the legend by using a modified Schechter function whose best-fit parameters are reported in the panel. \textbf{Right}: SPIRE 250/350/500$\,\mu$m luminosity density resulting from the fit of the LFs in the first three redshift bins reported on the left.}
\label{local.ld}
\end{figure*}

Once we obtains our LF solutions in each redshift bin and for each band, we can integrate them to find the luminosity density per redshift bin which is connected to the amount of energy emitted by the galaxies at each wavelengths and at each instant. To obtain this information we perform a $\chi^2$ fit to our $1/V_{\rm max}$ estimates, using the modified Schechter function described in Eq. \ref{log.gaus}. Since we are limited to a local sample, at $z>0.2$ we do not populate the low luminosity bins of our LFs and for this reason we cannot really constrain the integration at higher redshift. We thus report in Fig. \ref{local.ld}, \ref{SFR.fig} and in Tab. \ref{LLD} our luminosity density estimates for the SPIRE 250/350/500$\,\mu$m and the IR bolometric luminosity within $z<0.2$, reporting the results for three redshift bins whose mean redshifts are 0.05, 0.1 and 0.15.

\begin{figure*}
\begin{center}
   {\myincludegraphics{width=0.8\textwidth}{\figdir{}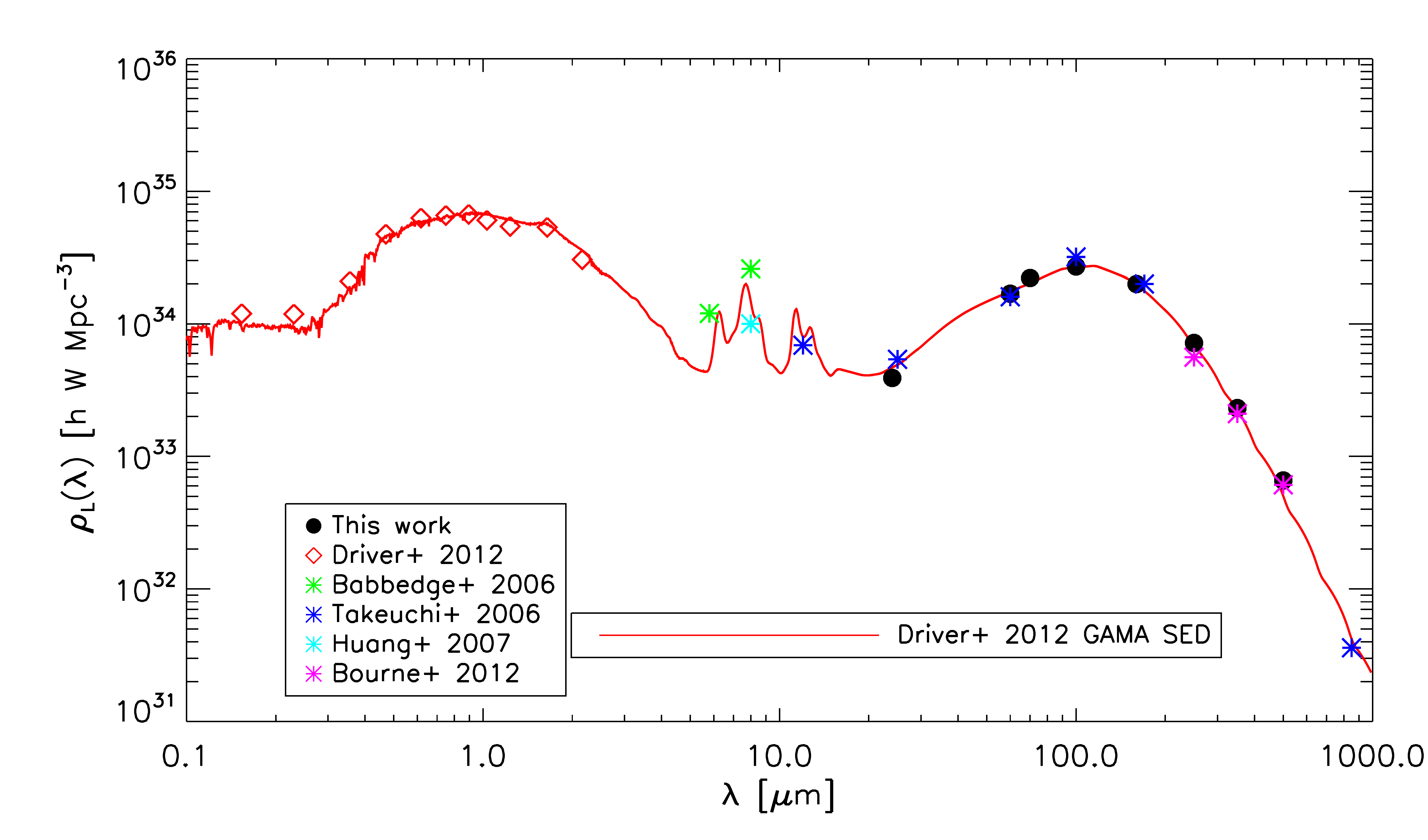}}
\end{center}
\caption[]{The multi-wavelengths energy output in the local Universe. The local luminosity density at different wavelengths was computed by integrating the relevant monochromatic local luminosity functions
over the $0.02<z<0.1$ bin. Plotted values of this work, solid black circles, are reported in Tab. \ref{local.energy}.}
\label{local.em}
\end{figure*}

In Fig. \ref{local.em} we report the conversion of our luminosity density estimates at SPIRE 250/350/500$\,\mu$m, as well at MIPS 24/70/160$\,\mu$m wavelengths to the energy output and we compare our result to those reported by \cite{Driver2012}. 
Our plotted estimates, together with others extrapolated at 90 and 170$\,\mu$m, are reported in Tab. \ref{local.energy}.

We find that, even though our sample is selected at 250$\,\mu$m, we can reproduce the energy density at all the other considered FIR bands in the very Local Universe. This confirms the shape of the energy density published by Driver at al. (2012) estimated using the GAMA I dataset combined with GALEX, SDSS and UKIRT. 

\begin{table}
\centering
\begin{tabular}{ccccc}
\multicolumn{5}{c}{\textbf{Local luminosity density}}\\
\hline
\textbf{$\langle z \rangle$} & log($\rho_{L}$,$\sigma$)$_{250}$ & log($\rho_{L}$,$\sigma$)$_{350}$ &  log($\rho_{L}$,$\sigma$)$_{500}$ &  log($\rho_{L}$,$\sigma$)$_{IR}$ \\
\hline\hline
0.05 & 7.11 , 0.02 & 6.64 , 0.02 & 6.09 , 0.02 & 7.92 , 0.02 \\
0.10 & 7.23 , 0.02 & 6.75 , 0.01 & 6.20 , 0.01 & 8.02 , 0.02 \\
0.15 & 7.31 , 0.02 & 6.82 , 0.02 & 6.27 , 0.02 & 8.07 , 0.02 \\
\hline
\end{tabular}
\caption[]{Local luminosity density estimates in the SPIRE 250/350/500$\,\mu$m bands and for the IR bolometric luminosity using the local SPIRE sample within $0.02<z<0.1$. The values are reported as log(LLD) and log(errors), expressed in L$_{\odot}$ Mpc$^{-3}$.}
\label{LLD}
\end{table}

\begin{table}
\centering
\begin{tabular}{cc}
\multicolumn{2}{c}{\textbf{Local energy output}}\\
\hline
$\lambda$ & $\rho_L(\lambda)\,\lambda$ \\
\hline
$\mu$m & $10^{33}$ h W Mpc$^{-3}$ \\
\hline
24 &  3.91 $\pm$ 0.69 \\
60 & 16.87 $\pm$ 3.47 \\
70 & 22.18 $\pm$ 4.77 \\
90 & 25.93 $\pm$ 5.59 \\
100 & 27.10 $\pm$ 5.79 \\
160 & 19.95 $\pm$ 4.27 \\
170 & 18.54 $\pm$ 4.00\\
250 & 6.98 $\pm$ 1.45 \\
350 & 2.32 $\pm$ 0.46 \\
500 & 0.58 $\pm$ 0.14 \\
\hline
\end{tabular}
\caption[]{Local energy output of the Universe at different wavelengths.}
\label{local.energy}
\end{table}

\subsection{The local star formation rate}

The estimate of the local luminosity function in the SPIRE bands is of fundamental importance for studying the evolution of the SPIRE LFs at higher redshift. In practice, local luminosity function estimates guide the priors on the parameters that define the LF shape that is adopted when fitting the LF also at higher redshifts (Vaccari et. al., in prep.). Additionally, thanks to the large volume sampled by shallow and wide area surveys, these estimates allow us to calculate the SFRD in the local Universe with small uncertainties. By integrating the luminosity function in different redshift bins, whenever the observed bands are related to the emission of the young stellar populations, like in this case, we can estimate the SFR at those redshifts. 
In this context, we can easily use the IR bolometric luminosity as a tracer of SFR and thus the IR bolometric luminosity density as tracer of the SFRD.

We thus fit our $1/V_{\rm max}$ local luminosity function estimates with a modified Schechter function described in Eq. \ref{log.gaus}, obtaining the estimate of the local luminosity density (LLD) reported in Tab. \ref{LLD}. The lower and upper limits that we used in the LFs integration to estimate the LLDs are $L=10^8 L_{\odot}$ and $L=10^{14} L_{\odot}$ respectively. These limits guarantee that we account for the bulk of the IR luminosity emitted by our sources.
 We then convert the estimate of the luminosity density into star formation rate density using the \cite{Kennicutt1998} relation (assuming a Salpeter IMF): $\psi(t) = \mathrm{SFR} = k(\lambda)\mathrm{L}(\lambda)$ where $\mathrm{k}( \mathrm{IR}) = 4.5 \times 10^{-44} [\mathrm{M}_\odot \mathrm{yr}^{-1}\mathrm{W} \mathrm{Hz}] $. 

We used our SED fitting analysis and the IRAC colour-colour criteria by \cite{Lacy2004} and \cite{Donley2012} to quantify the possible AGN contamination in our sample, as discussed in Sec. \ref{sedfit.sec}. We find that in our sample the fraction of objects showing AGN-like IRAC colours and AGN-like SEDs is very small and even if we discard from our results the total luminosity contribution of these sources, our LFs and thus SFR estimates do not significantly deviate from the results obtained using our total sample.
Even for these AGN-like sources (mainly located above $z\sim0.25$), the vast majority of the IR luminosity is still contributed by dust emission associated with ongoing star formation. This is also confirmed by \cite{Hatz09,Hatz10} and \cite{Bothwell2011} that show how AGN contribution to the FIR emission of the general extragalactic population is rather small. For these reasons we conclude that the AGN contribution does not significantly affect our LF and SFRD estimates.

\begin{figure*}
\begin{center}
   {\myincludegraphics{width=0.5\textwidth}{\figdir{}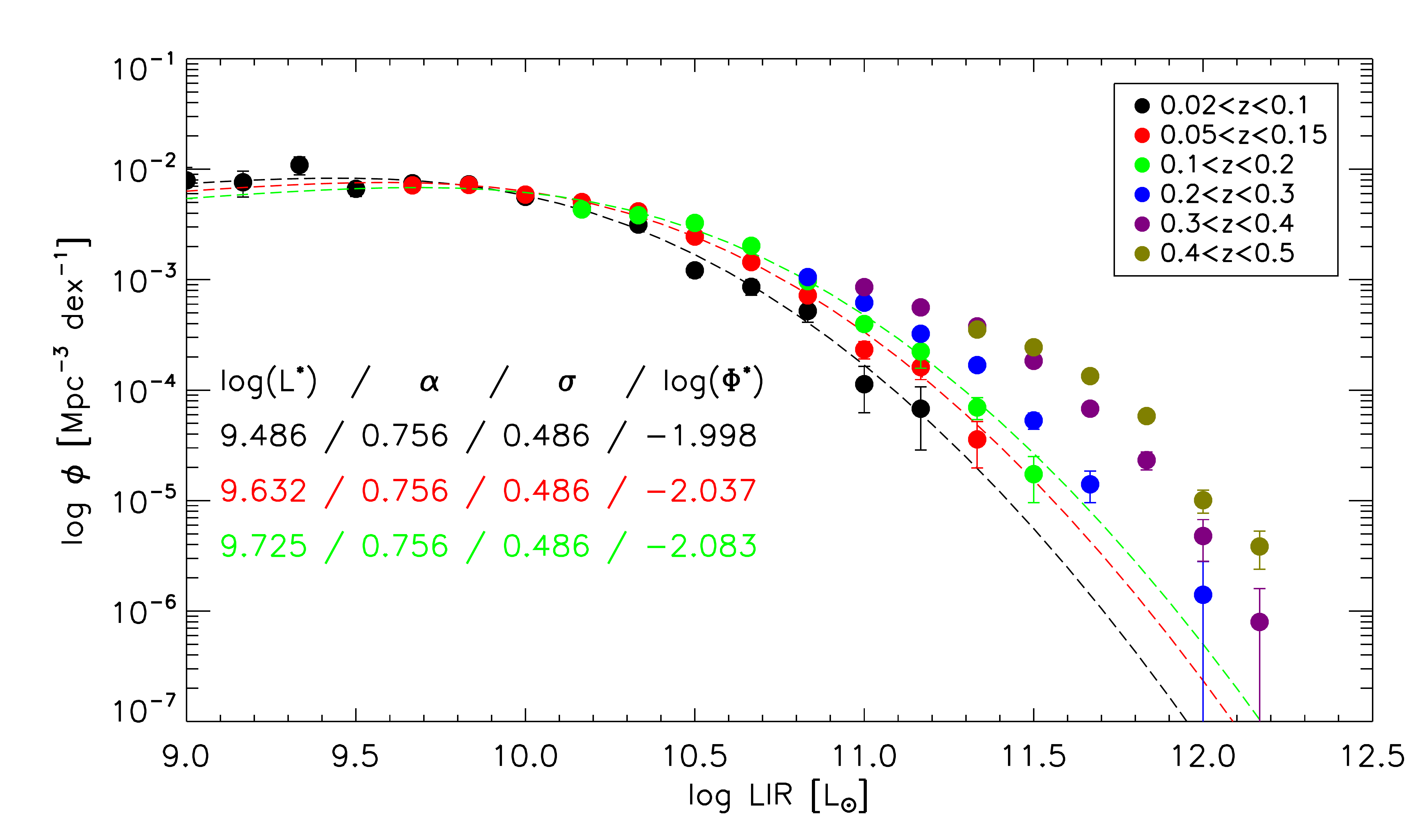}}
   {\myincludegraphics{width=0.4\textwidth}{\figdir{}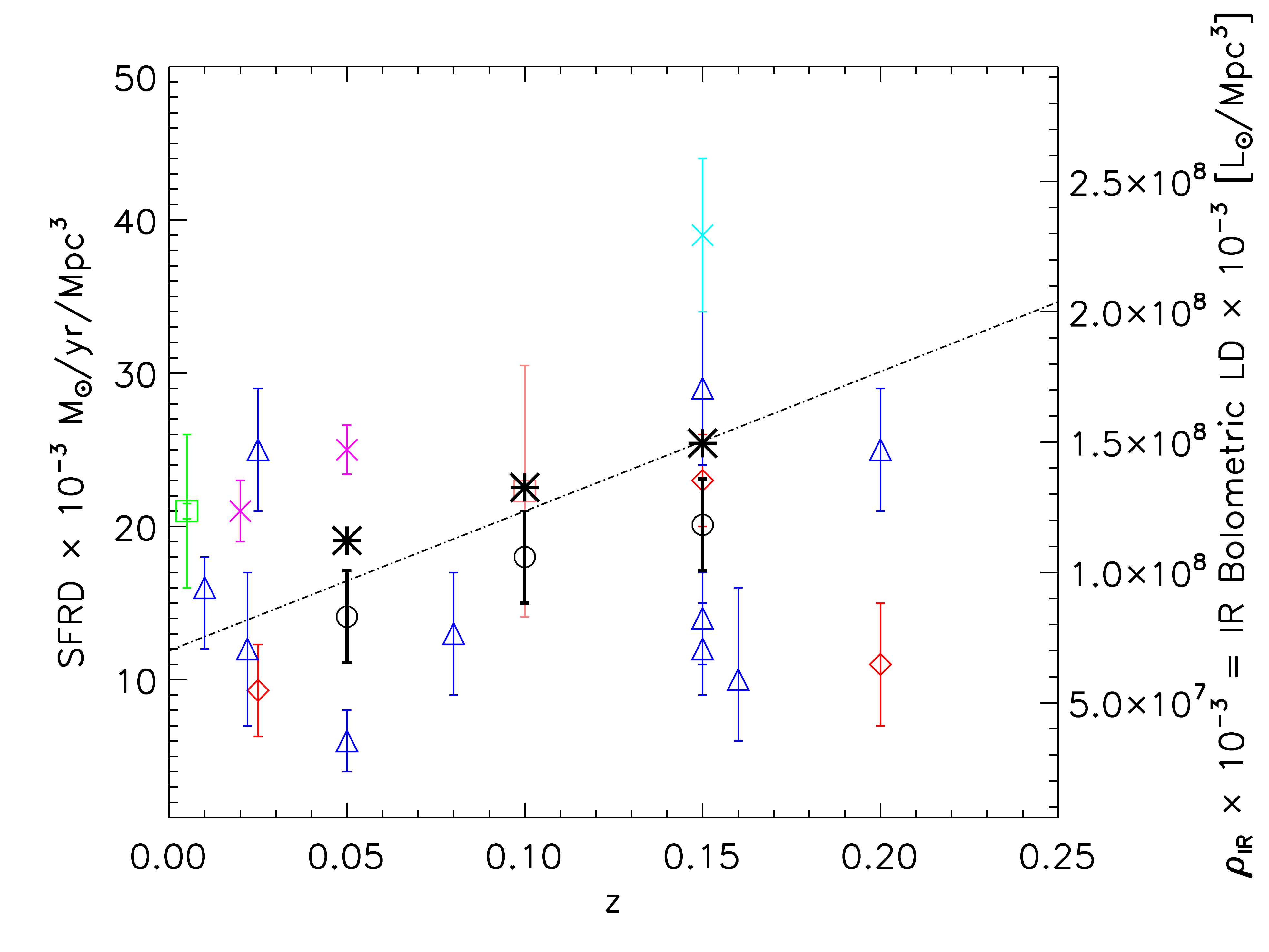}} 
\end{center}
\caption[]{The IR bolometric rest-frame luminosity function evolution within $0.02<z<0.5$ along with an illustration of the star formation rate density in the local Universe showing published results and from this work. \textbf{Left}: The infrared bolometric luminosity function within $0.02<z<0.5$, integrated in the first three redshift bins reported in the legend by using a modified Schechter functions; \textbf{Right}: The derived SFRD in the local Universe. Black open circles are our data as result of the integrations of the LFs on the left converted to SFRD by using the \cite{Kennicutt1998} relation (assuming a Salpeter IMF) and black asterisk are our results plus the contribution of the UV SFRD as estimated by Wider et al. (2005) at $\langle z \rangle = 0.05$ and \cite{Budavari05} at $\langle z \rangle = 0.1, 0.15$. This sum should represent the total SFRD in the Local Universe. The red open diamonds are OII estimates by \cite{Gallego2002}, \cite{Sullivan2000} and \cite{Hogg1998};  the blue open triangles are H$\alpha$ estimates by \cite{Gallego1995}, \cite{Tresse1998}, \cite{Sullivan2000}, \cite{PerezGonzalez2003}, \cite{Ly2007}, \cite{Hanish2006}, \cite{Brinchmann2004}, \cite{Dale2010} and \cite{Westra2010}; the green open square is Radio 1.4 GHz estimates by \cite{Serjeant2002} and \citealt{Condon1989}; the magenta crosses are FUV+IR estimates by \cite{Martin2005.59M} and \cite{Bothwell2011};  the cyan crosses are FUV estimates by \cite{Sullivan2000}; the pink open squares are IR estimate from \cite{Vaccari2010}; the black dashed line is from \cite{HB06}.}
\label{SFR.fig}
\end{figure*}

The SFRD estimate we obtain from the IR bolometric luminosity density (estimated at $0.02<z<0.1$, $0.05< z<0.15$ and $0.1<z<0.2$) are reported in Tab. \ref{SFR.comp}, together with other SFRD estimates obtained by various authors using different SFR tracers (all the results are converted to the same IMF and cosmology). These same data are also shown in Fig. \ref{SFR.fig}. The uncertainties reported in Tab. \ref{SFR.comp} are percentage errors.

\begin{table*}
\centering
\begin{small}
\begin{tabular}{lllc}
\hline
\textbf{Reference} & \textbf{SFR tracer}  & $<z>$ & SFRD \\
&&&($10^{-3}\;M_{\odot} \;\mathrm{yr}^{-1}$ Mpc$^{-1}$) \\
 \hline
 \hline
\cite{Gallego2002}& [OII] & $0.025$ & $9.3\pm3$ \\
\cite{Sullivan2000}& [OII] & 0.15 & $23\pm3$ \\
\cite{Hogg1998} & [OII] & 0.20 & $11 \pm 4$ \\
\cite{Gallego1995}& H$\alpha$ & $0.022$ & $12\pm5$ \\
\cite{Tresse1998}& H$\alpha$ &0.2 & $25\pm 4$\\
\cite{Sullivan2000}& H$\alpha$ & 0.15 & $14\pm3$ \\
\cite{PerezGonzalez2003} & H$\alpha$ &0.025 & $25\pm 4$\\
\cite{Ly2007}& H$\alpha$ &0.08 & $13\pm 4$\\
\cite{Hanish2006} & H$\alpha$ &0.01 & $16^{+2}_{-4}$\\
\cite{Brinchmann2004} & H$\alpha$ & 0.15& $29 \pm 5$ \\
\cite{Dale2010}& H$\alpha$ & 0.16& $10^{+6}_{-4} $ \\
\cite{Westra2010}& H$\alpha$ & 0.05& $6 \pm 2 $ \\
\cite{Westra2010}& H$\alpha$ & 0.15& $12 \pm 3 $ \\
\cite{Serjeant2002}& 1.4 GHz & $0.005$ & $21\pm5$ \\
\cite{Condon1989}& 1.4 GHz & $0.005$ & $21\pm0.5$ \\
\cite{Sullivan2000}& FUV & 0.150 & $39\pm5$ \\
\cite{Martin2005.59M}& FUV+IR & 0.02 & $21\pm 2$\\
\cite{Bothwell2011} & FUV+IR & 0.05  &$25 \pm 1.6$ \\
\cite{Vaccari2010} & IR & 0.1 & $22.3 \pm 8.2$ \\
This work & IR & 0.05 & $14.11 \pm 2.4$ \\
This work & IR & 0.10 & $18.00 \pm 2.9$ \\
This work & IR & 0.15 & $20.10 \pm 2.2$ \\
This work & FUV+IR & 0.05 & $19.07 \pm 2.4$ \\
This work & FUV+IR & 0.10 & $22.53 \pm 2.9$ \\
This work & FUV+IR & 0.15 & $25.42 \pm 2.2$ \\
\hline
\end{tabular}
\end{small}
\caption[Star formation rate density in the local Universe: literature results and from this work]{Star formation rate density in the local Universe: literature results and from this work. This table is an updated version of the one reported in \cite{Bothwell2011}. The FUV unobscured SFRD added to our IR results and quoted in this Table are from: \cite{Wyder05} at z=0.05 and \cite{Budavari05} at z=0.1 and z=0.15.}
\label{SFR.comp}
\end{table*}

\begin{figure*}
\begin{center}
   {\myincludegraphics{width=0.45\textwidth}{\figdir{}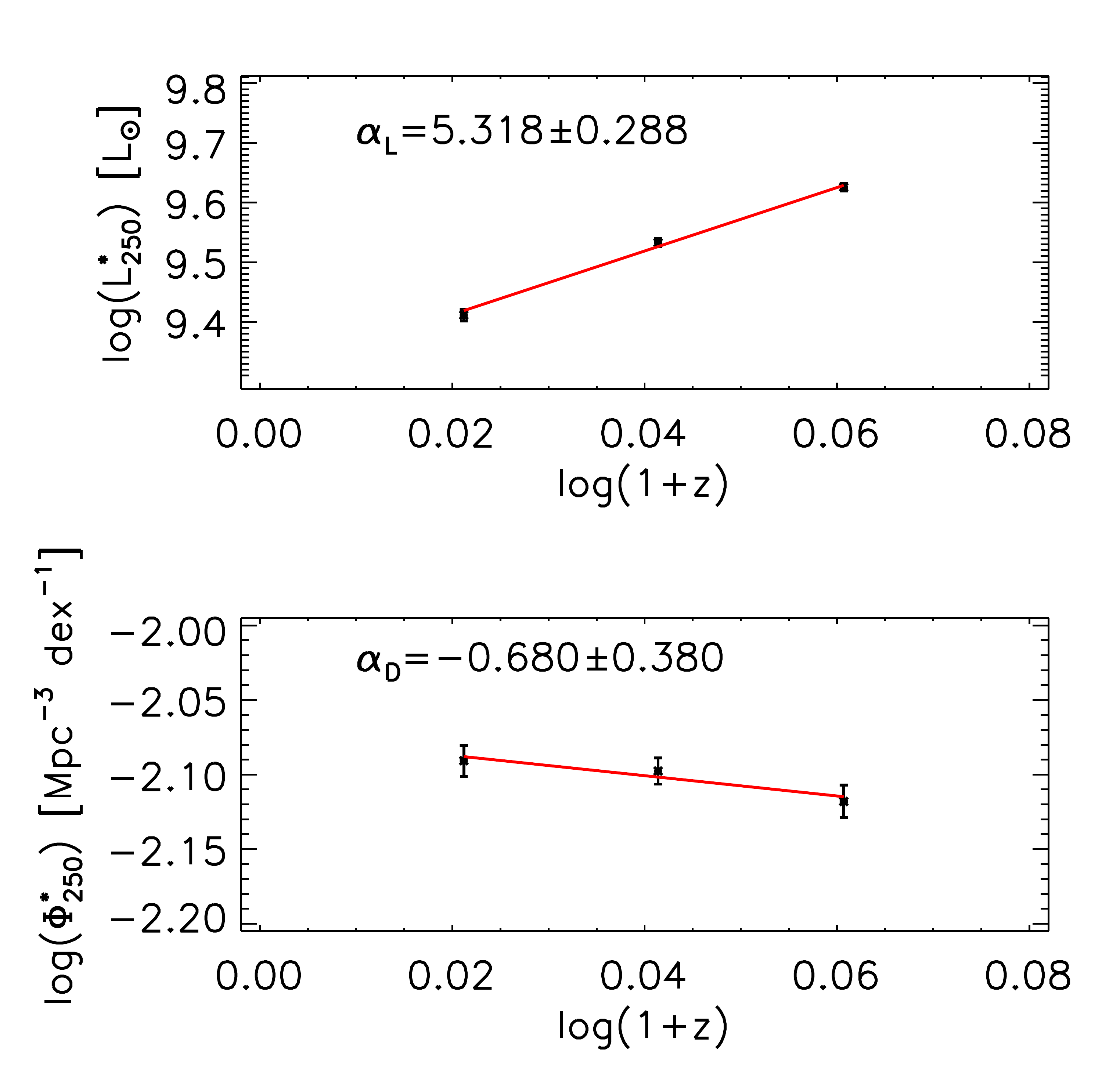}}
   {\myincludegraphics{width=0.45\textwidth}{\figdir{}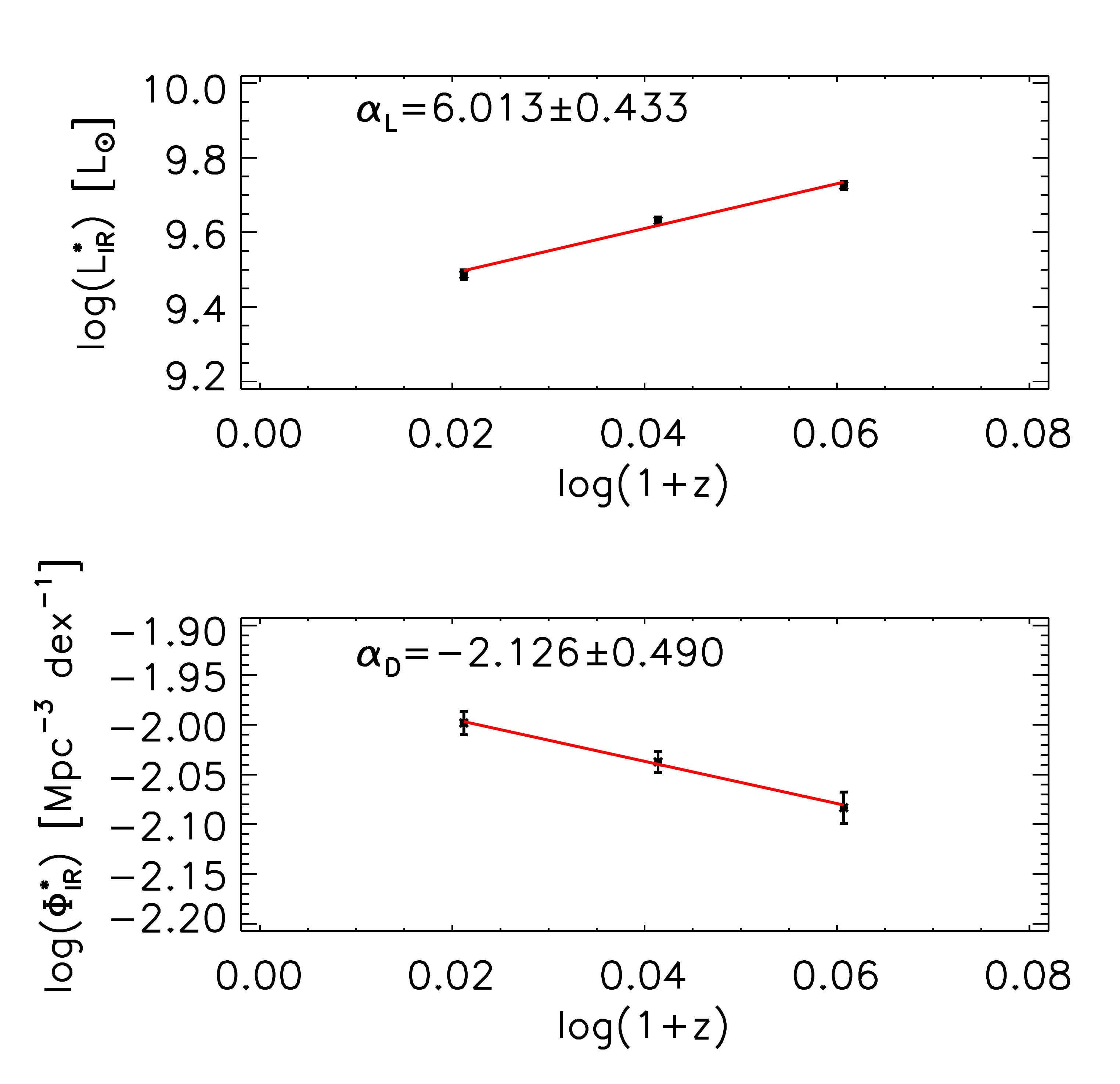}}
\end{center}
\caption[]{Evolution of L$^{*}$ and $\Phi^{*}$ as function of $z$ (L$_* \propto (1+z)^{\alpha_L}$ and $\Phi^* \propto (1+z)^{\alpha_D}$) estimated for the LF at 250 $\mu$m (left panels) and for the IR bolometric rest-frame local luminosity function (right panels) within $0.02 < z < 0.15$.}
\label{evo}
\end{figure*}



\section{Discussion}\label{discussion}
Using some of the widest-area surveys performed by \textit{Spitzer} and \textit{Herschel}, in this paper we have studied in details the local luminosity functions of SPIRE sources. Our LLFs at 250/350/500$\,\mu$m (SPIRE) strongly constrain the local luminosity density of the Universe throughout the FIR/submm wavelength range. 

Our estimates mostly confirm and improve upon the HerMES SDP results published in \cite{Vaccari2010}, thanks to our increased statistics; this is particularly visible in the 500$\,\mu$m LF solution, which shows strongly reduced uncertainties. \cite{Dye2010} used \textit{Herschel} SDP data to compute the H-ATLAS (\citealt{Eales2010a}) SPIRE local luminosity function. This analysis was carried out very early during the Herschel mission and relied on shallower SPIRE observations, much fewer ancillary data and smaller coverage area than our own. We thus judge that the H-ATLAS analysis is likely to have suffered from detection and cross-identification incompleteness and that the discrepancy we find between our results and their is therefore likely to be due to the H-ATLAS analysis. Our results are in fact broadly in agreement with their estimates in the highest luminosity bins where the uncertainties on the H-ATLAS SPIRE flux estimates were possibly smaller and their sample more complete, but at the lowest luminosities and redshifts they found LF values lower by up to 50\%, as shown in Fig. \ref{local.dye}. 

The semi-parametric method for the luminosity function estimate of Schafer (2007, see Fig. \ref{local.schafer}) is in perfect agreement with other classical estimators at low redshifts, $z<0.2-0.3$. At higher redshifts the agreement becomes poorer, being acceptable at high luminosities but degrading at lower luminosity values, where the semi-parametric estimate is always in excess of the $1/V_{\rm max}$ values. The precise origin of this problem is not fully understood, but it is clear that it happens in regions of the data-space that are poorly sampled by the observations or where the data are scattered e.g. by the effects of the K-correction.

From the IR bolometric luminosity function we can estimate the SFRD of the Local Universe in various redshift bins. In Fig. \ref{SFR.fig} we report our SFRD solutions and we compare them to others already published in the same redshift range. We see a large scatter in the Local SFRD estimates using different SFR diagnostics. In particular the H$\alpha$ measurements present the largest scatter between different published results. Our new data are entirely consistent with \cite{Vaccari2010} and show good agreement also with OII-based estimates (except perhaps at $z=0.2$ by \citealt{Hogg1998}). Instead, our SFRD based on the FIR/SMM bolometric flux is systematically lower than the radio 1.4$\,$GHz estimates and those combining IR and UV data by \cite{Martin2005.59M} and \cite{Bothwell2011}. In principle the Radio flux should be unaffected by dust extinction and thus a more faithful representation of the total SFR than either the IR or IR+UV values. Nevertheless the Radio flux can be more affected by the AGN activity than the IR/submm ones. If we include the UV-uncorrected portion of the SFRD mapped by short-wavelength UV spectral data to our FIR estimate, we find that our total SFRD UV+IR is comparable, within the errors, to the radio estimates, thus confirming that the UV+IR SFRD estimate is a good proxy for the total SFRD in the local Universe and the contamination from AGN in the radio derivation is negligible.

The analysis reported in this paper represents a fundamental local benchmark to study the evolution of the LF and, consequently, of the derived SFR with cosmic time. Studying the evolution of the luminosity function requires very deep data that are then limited to very small areas of the sky and thus it is difficult to constrain the local shape of the LF where a large statistical sample of local galaxies (like ours) is required. This can be seen in Figs. \ref{local.LIR} 
where we compare our local analysis with the one done using the deep COSMOS data (area 1.7 deg$^2$ and flux limited $S_{250} > 10$~mJy) 
that will be reported in Vaccari et al. (in prep.). Only the large area surveyed by our sample enables us to really study the local shape of the LF, while the deep sample allows us to populate only a few luminosity bins. On the other hand, deep data become more and more important with increasing redshift where, our sample soon starts being limited to the higher luminosity bins.

Our luminosity function estimates show significant and rapid luminosity evolution already at low redshifts. In Fig. \ref{evo} we report our results about the redshift evolution of the parameters expressing the spatial density dependence of the LFs ($\Phi^*$) and the luminosity dependence (L$^*$) estimated for the IR bolometric at the 250$\,\mu$m luminosity functions. We found positive evolution in luminosity and negative evolution in density with L$_{IR}^* \propto (1+z)^{6.0\pm0.4}$, $\Phi_{IR}^* \propto (1+z)^{-2.1\pm0.4}$ for the IR bolometric LF and L$_{250}^* \propto (1+z)^{5.3\pm0.2}$, $\Phi_{250}^* \propto (1+z)^{-0.6\pm0.4}$ for the 250$\,\mu$m LF. The high values of the evolution rates that we find (both positive and negative) for the luminosity and density parameters are however consistent with previous results based on previous and more limited datasets from Spitzer (\citealt{Patel2013}) and from IRAS (\citealt{Hacking1987}; \citealt{Lonsdale1990}).Similar, although slightly lower, trends for positive luminosity and negative density evolution are found by \cite{Gruppioni2013}. \cite{Gruppioni2013} used a sample deeper and over a much smaller area than ours. Their sample includes sources as faint as ours but they are very few in the local Universe since they suffer from a small sample variance due to the little areas targeted. For this reason we are able to get a more accurate estimate of the LFs down to similar luminosities in the local Universe.

Interesting for our analysis is the comparison with \cite{Negrello2013} reported in Figs. \ref{local.350} and \ref{local.500} which show a steep LF in the lowest luminosity bins while our estimate remains flat down to $L_{350}\sim10^8[L_\odot]$  and $L_{500}\sim10^7[L_\odot]$ respectively. In general, our low-z luminosity functions is computed at $z>0.02$, while the Planck sources used by \cite{Negrello2013} are located at a mean redshift value of $z\sim0.01$. This means that our analysis is based on a deeper sample somehow complementary to the Planck's one. Our sample therefore does not suffer from contamination from either the Local Super Cluster or the Virgo Cluster (like Planck and thus potentially the \citealt{Negrello2013} estimates) while representing the LF of typical galaxies in the not-so-nearby Universe (unlike Planck). Moreover, it can be argued that our measurement averages over any local inhomogeneity by sampling a larger cosmic volume than Planck ($\sim10$ times larger at $z\sim0.2$ over 39 deg$^2$ than Planck at $z\sim0.01$ over 30,000 deg$^2$). Indeed over a much smaller area, but with a much deeper sample, the flatness of the slope is also confirmed by \cite{Gruppioni2013} when measuring the $0<z<0.3$ IR LF. In any case at values of $L_{350}$ brighter than $\sim10^8 [L_\odot]$ and $L_{500}$ brighter than $\sim10^7 [L_\odot]$, where we are $\sim100\%$ complete and where the Planck sample is less affected by the presence of local structures and inhomogeneities, we find that our results are in overall agreement with \cite{Negrello2013} at both 350 and 500 $\mu$m. Similar considerations can be made when we compare our IR bolometric LF with the previous estimate obtained by \cite{Sanders2003}, which appears to be slightly steeper than ours in the lowest-luminosity bins (see Fig.\ref{local.LIR}). The mean and median redshifts of the entire IRAS sample used by \cite{Sanders2003} are in fact $z=0.0126$ and $z=0.0082$ respectively and their LF estimate can therefore be affected by the Virgo cluster in the same way as Planck's estimates discussed above.

\begin{figure*}
\begin{center}
  {\myincludegraphics{width=0.45\textwidth}{\figdir{}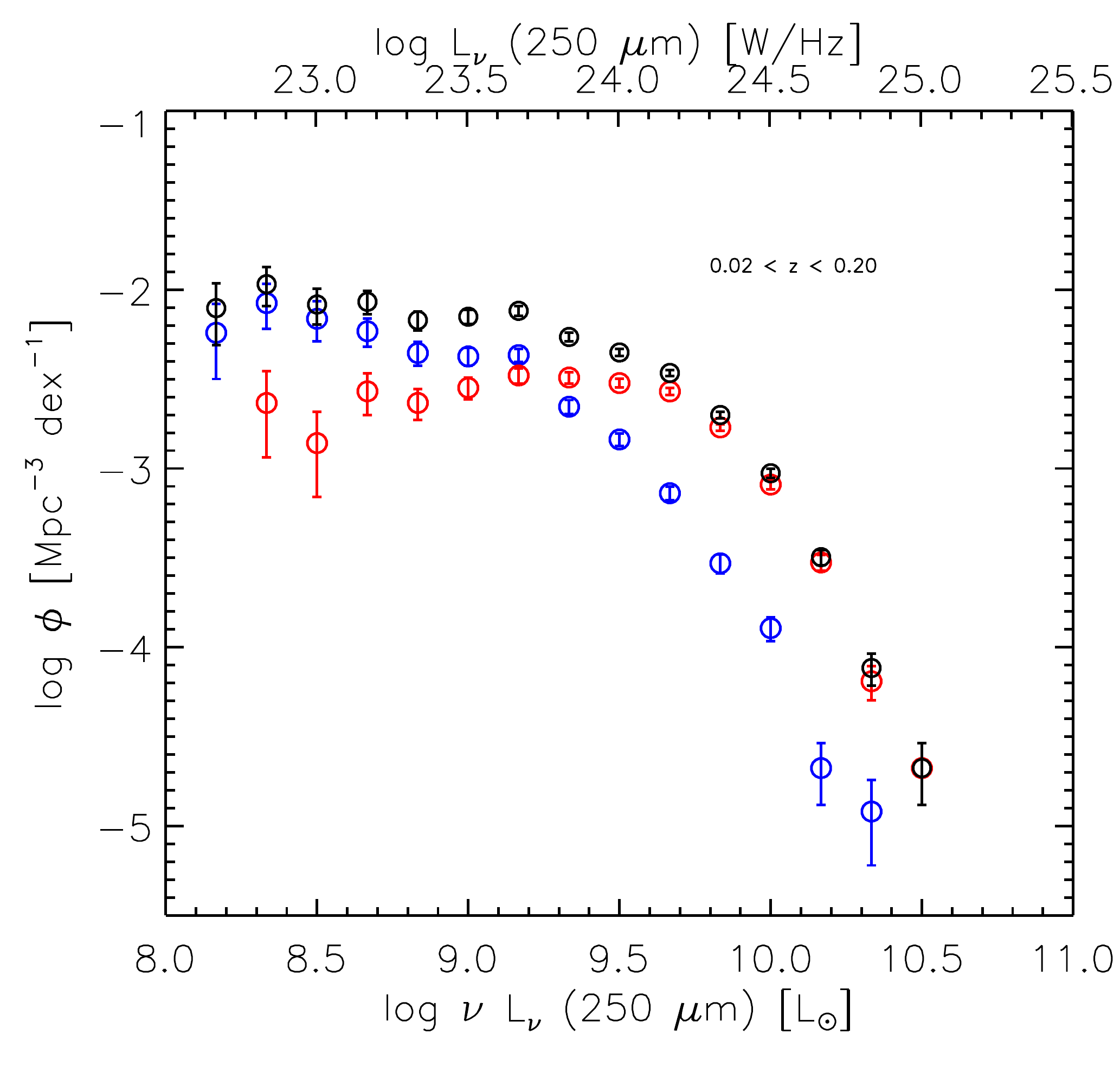}}
   {\myincludegraphics{width=0.45\textwidth}{\figdir{}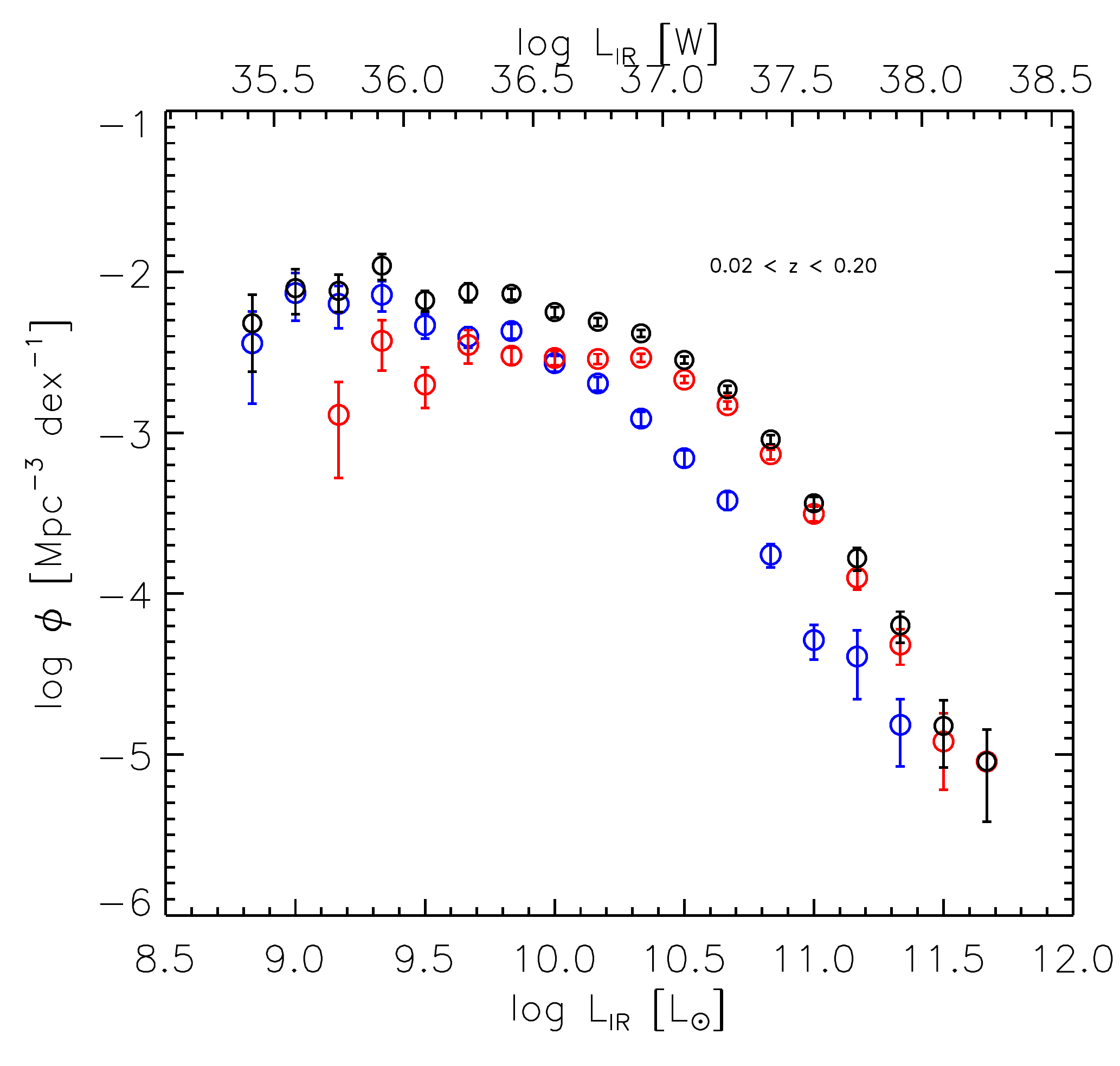}}
\end{center}
\caption[]{250$\,\mu$m and IR bolometric LFs for the blue and red populations (reported with blue and red open circles, respectively) compared with the total of the two populations (black open circles) in the redshift range $0.02 < z < 0.2$.}
\label{bluered}
\end{figure*}

Our ability to map the local LF with good precision has revealed a wiggle in the shapes of the functions, with a local maximum at $\mathrm{log} L_{250}\sim 9.5$ and $\mathrm{log} L_{IR}\sim 10.5$, respectively. This feature, which appears relatively stable with wavelength, is reminiscent of similar behaviour found in the local mass functions of galaxies (\citealt{Moustakas2013}, \citealt{Baldry2012}, \citealt{Ilbert2013}), and interpreted as due to the summed contributions of red and blue galaxies, having Schechter functions with different slopes and cutoff masses. Given the known relationship between stellar mass and IR luminosity, it may not come as a surprise that a similar feature appears in our IR luminosity functions. To test this possibility, we have divided our sample into red and blue sub-populations, following the recipe of \cite{Baldry2012}, and separately calculated the LFs for the two classes. The results, reported in Fig. \ref{bluered}, confirm that red galaxies have an IR LF peaking at $\mathrm{log} (L_{250})\sim 9.5$ and $\mathrm{log} (L_{IR})\sim 10.5$ and decreasing at higher and lower $L$, while blue galaxies have steep Schechter slopes and lower characteristic luminosities. These are purely observational results and further analysis would be required to better constrain this feature, but this goes beyond the scope of this paper. At any rate our findings seem to indicate that massive early-type spirals dominate the high-IR-luminosity end of the LF, while bluer lower-mass late-type spirals and irregulars dominate its low-luminosity end.

We also performed a preliminary comparison with semi-analytical models of galaxy formation available in literature, focusing our attention on the redshift range between $z=0.02$ and $z=0.2$. From these preliminary comparisons we notice that the \cite{Fontanot2012} predictions (using the MORGANA code by \citealt{Monaco2007}) seem to broadly reproduce the shape of the LF within the uncertainties, but they underestimate the LF at lower luminosities when compared to our IR bolometric LF estimates. Other predictions done by e.g., \cite{Negrello2007}, \cite{SerjeantHarrison2005} and \cite{Valiante2009} at different wavelengths also show good agreement with our results at higher luminosities, but most of them seem to underestimate the LF when compared to what we obtain at lower luminosities. A more careful and systematic analysis of existing and improved models is required to properly address this issue (e.g., \citealt{Gruppioni2015}, Franceschini et al., submitted).

\section{Conclusions}\label{conclusions}

The determination of the galaxy luminosity function is often hampered by the difficulties of covering a wide area down to faint fluxes on the one hand, and determining counterparts and redshifts for detected sources in a complete and reliable manner on the other hand. In this work we have thus assembled and exploited the widest area \textit{Spitzer} and \textit{Herschel} extragalactic surveys to select IR galaxy samples in a complete and reliable manner, and the best UV/Optical/NIR ancillary data to identify them. Thanks to \textit{Spitzer} and \textit{Herschel} observations we are now able to reliably sample the IR bolometric luminosity of local sources and thus provide important insights into dust obscured star formation activity across cosmic time. Even with the best data sets, however, accurately constructing the luminosity function remains a tricky pursuit, since the presence of observational selection effects due to e.g. detection thresholds in apparent magnitude, colour, surface brightness or some combination thereof can make any given galaxy survey incomplete and thus introduce biases in the luminosity function estimates. Only a comparison of results coming from different luminosity function estimators applied to the same samples can ensure we can assess the impact of these biases in a robust manner.

Armed with the \catnameshort, we were able to describe the $0.02 < z < 0.5$ local luminosity function of sources selected in wide fields by \textit{Herschel} SPIRE imaging. We fully exploited the multi-wavelength information collected within the \catnameshort\ to perform a SED fitting analysis of SPIRE sources and thus estimate the monochromatic rest-frame luminosities at 250, 350 and 500$\,\mu$m as well as the IR luminosity between 8 and 1000$\,\mu$m. We then implemented a number of different statistical estimators to evaluate the local luminosity functions of flux-limited samples in these bands: the classical $1/V_{\rm max}$ estimator of \cite{Schmidt1968} and the modified $1/V_{\rm est}$ version of \cite{PageCarrera2000}; a parametric maximum likelihood technique (ML) based on a Bayesian approach as described in \cite{Kelly2008}; and finally a semi-parametric approach introduced by \cite{Schafer2007}.

Our high quality determinations of the IR luminosity functions have revealed for the first time some previously unidentified features in their shape, that we interpret as due to the contributions of red (possibly early-type) and 
blue (possibly late-type) galaxy populations, with their different Schechter forms. By means of this analysis we find that the luminosity functions show significant and rapid luminosity evolution already at low redshifts, $0.02 < z < 0.2$. Converting our IR LD estimate into an SFRD we can determine the SFRD of the local Universe up to redshift 0.2, where the integration of the LF solution is more reliable given that our data set fails to populate the low luminosity bins of the LF at higher $z$. Summing over our IR SFRD estimate of the unobscured contribution based on the UV dust-uncorrected emission from local galaxies, we estimate that SFRD $\simeq $ SFRD$_0+0.08 z$,  where SFRD$_0\simeq (1.9\pm 0.03)\times 10^{-2} [\mathrm{M}_\odot \mathrm{Mpc}^{-3}]$ is our total SFRD estimate at $z\simeq0.02$. This analysis represents a local benchmark for studying the evolution of the infrared luminosity function and star formation rate function with cosmic time.

\section*{Acknowledgements}
LM acknowledges support from the Science and Technology Facilities Council under grant ST/J001597/1.
LM, MV and AF acknowledges support from ASI ''\textit{Herschel} Science'' Contracts I/005/07/1 and I/005/11/0.
Mattia Negrello produced additional predictions based on his models.
JW acknowledges the Dark Cosmology Centre funded by the Danish National Research Foundation.
MV acknowledges support from the Square Kilometre Array South Africa project,
the South African National Research Foundation and Department of Science and
Technology (DST/CON 0134/2014), the European Commission Research Executive
Agency (FP7-SPACE-2013-1 GA 607254) and the Italian Ministry for Foreign Affairs
and International Cooperation (PGR GA ZA14GR02).
NS is the recipient of an ARC Future Fellowship. AF acknowledges support from the ERC via an Advanced Grant under grant agreement no. 321323-NEOGAL.
This work makes use of STILTS \url{http://www.starlink.ac.uk/stilts/} and TOPCAT \citep{taylor05}.
SPIRE has been developed by a consortium of institutes led by Cardiff University (UK) and including Univ. Lethbridge (Canada); NAOC (China); CEA, LAM (France); IFSI, Univ. Padua (Italy); IAC (Spain); Stockholm Observatory (Sweden); Imperial College London, RAL, UCL-MSSL, UKATC, Univ. Sussex (UK); and Caltech, JPL, NHSC, Univ. Colorado (USA). This development has been supported by national funding agencies: CSA (Canada); NAOC (China); CEA, CNES, CNRS (France); ASI (Italy); MCINN (Spain); Stockholm Observatory (Sweden); STFC (UK); and NASA (USA).
The authors would like to thank the anonymous referee for helpful comments.

\clearpage
\newpage
\begin{table}
\centering
\begin{small}
\begin{tabular}{|c|c|c|c|c|}
\hline
$\log L$ & $\log\,(\Phi,\sigma)_{250}$ & $\log\,(\Phi,\sigma)_{350}$ & $\log\,(\Phi,\sigma)_{500}$ & $\log\,(\Phi,\sigma)_{IR}$ \\
\hline
\multicolumn{5}{|c|}{\textbf{$0.02<z<0.1$ Luminosity Functions}}\\
\hline
7.16  & - & - & -2.19 , -2.07 & - \\
7.33  & - & - & -1.90 , -2.54 & - \\
7.49  & - & - & -2.06 , -2.72 & - \\
7.66  & - & - & -2.06 , -2.90 & - \\
7.83  & - & - & -2.17 , -3.11 & - \\
8.00  & - & - & -2.12 , -3.21 & - \\
8.16  & -2.10 , -2.06 & -2.09 , -2.89 & -2.15 , -3.34 & - \\
8.33  & -1.96 , -2.57 & -2.15 , -3.05 & -2.30 , -3.47 & - \\
8.49  & -2.08 , -2.72 & -2.11 , -3.17 & -2.52 , -3.58 & - \\
8.66  & -2.06 , -2.89 & -2.14 , -3.31 & -2.75 , -3.69 & - \\
8.83  & -2.17 , -3.08 & -2.23 , -3.43 & -3.34 , -3.99 & - \\
9.00  & -2.15 , -3.20 & -2.48 , -3.56 & -3.44 , -4.04 & -2.10 , -2.60 \\
9.16  & -2.11 , -3.31 & -2.66 , -3.65 & -4.34 , -4.49 & -2.11 , -2.69 \\
9.33  & -2.26 , -3.45 & -3.05 , -3.84 & - & -1.96 , -2.69 \\
9.49  & -2.49 , -3.57 & -3.49 , -4.07 & - & -2.17 , -3.00 \\
9.66  & -2.69 , -3.67 & -3.86 , -4.25 & - & -2.12 , -2.99 \\
9.83  & -3.12 , -3.88 & - & - & -2.13 , -3.24 \\
10.00 &-3.53 , -4.08 & - & - & -2.25 , -3.35 \\
10.16  &  -4.04 , -4.34 & - & - & -2.35 , -3.47 \\
10.33  & - & - & - & -2.50 , -3.56 \\
10.49  & - & - & - & -2.91 , -3.77 \\
10.66  & - & - & - & -3.06 , -3.85 \\
10.83  & - & - & - & -3.28 , -3.96 \\
11.00  & - & - & - & -3.94 , -4.29 \\
11.16  & - & - & - & -4.16 , -4.40 \\
\hline
\multicolumn{5}{|c|}{\textbf{$0.1<z<0.2$ Luminosity Functions}}\\
\hline
8.33  & - & - & -2.31 , -3.23 & - \\
8.49  & - & - & -2.31 , -3.52 & - \\
8.66  & - & - & -2.44 , -3.76 & - \\
8.83  & - & - & -2.70 , -4.05 & - \\
9.00  & - & - & -3.08 , -4.27 & - \\
9.16  & - & - & -3.57 , -4.51 & - \\
9.33  & -2.28 , -3.14 & -2.63 , -4.01 & -4.21 , -4.82 & - \\
9.49  & -2.28 , -3.49 & -2.91 , -4.18 & -4.76 , -5.10 & - \\
9.66  & -2.41 , -3.76 & -3.39 , -4.42 & - & - \\
9.83  & -2.64 , -3.95 & -4.02 , -4.73 & - & - \\
10.00 & -2.98 , -4.22 & -4.55 , -5.00 & - & - \\
10.16 & -3.45 , -4.45 & -5.45 , -5.44 & - & -2.36 , -3.37 \\
10.33 & -4.05 , -4.74 & - & - & -2.41 , -3.66 \\
10.49 & -4.61 , -5.03 & - & - & -2.48 , -3.64 \\
10.66  & - & - & - & -2.69 , -3.92 \\
10.83  & - & - & - & -3.01 , -4.18 \\
11.00 & - & - & - & -3.40 , -4.41 \\
11.16 & - & - & - & -3.64 , -4.16 \\
11.33 & - & - & - & -4.15 , -4.80 \\
11.49 & - & - & - & -4.76 , -5.10 \\
\hline
\end{tabular}
\end{small}
\caption[]{SPIRE 250, 350, 500$\,\mu$m and IR bolometric rest-frame $1/V_{\rm max}$  luminosity function estimates in the redshift ranges between 0.02 and 0.5, using the HerMES Wide Fields sample. $L$ indicates $\nu\,L_\nu$ for the monochromatic LFs and $L_{\rm IR}$ indicates the integrated luminosity between 8 and 1000$\,\mu$m for the IR bolometric rest-frame LF. These $L$ is expressed in units of $\rm L_\odot$ and LLF estimates and their errors are in $\mathrm{[Mpc^{-3}\,dex^{-1}]}$. The quantity $\sigma$ is the total error (Poissonian error + redshift uncertainties, estimated as explained in the text) associated with $\Phi$ in each band and luminosity/redshift bin.}
\label{llf-values.tab}
\end{table}

\begin{table}
\centering
\begin{small}
\begin{tabular}{|c|c|c|c|c|}
\hline
$\log L$ & $\log\,(\Phi,\sigma)_{250}$ & $\log\,(\Phi,\sigma)_{350}$ & $\log\,(\Phi,\sigma)_{500}$ & $\log\,(\Phi,\sigma)_{IR}$ \\
\hline
\multicolumn{5}{|c|}{\textbf{$0.2<z<0.3$ Luminosity Functions}}\\
\hline
8.83  & - & - & -2.72 , -3.57 & - \\
9.00  & - & - & -2.88 , -4.14 & - \\
9.16  & - & - & -3.21 , -4.13 & - \\
9.33  & - & - & -3.69 , -4.75 & - \\
9.49  & - & -2.78 , -4.02 & -4.19 , -4.93 & - \\
9.66  & - & -3.09 , -4.11 & -5.25 , -5.54 & - \\
9.83  & - & -3.52 , -4.66 & -5.85 , -5.82 & -  \\
10.00 & -2.81 , -4.05 & -4.02 , -4.88 & - & -  \\
10.16 & -3.13 , -4.21 & -4.85 , -5.33 & - & -  \\
10.33 & -3.56 , -4.65 & -5.55 , -5.69 & - & -  \\
10.49 & -4.13 , -4.92 & - & - &  - \\
10.66 & -4.89 , -5.37 & - & - & -2.81 , -3.62  \\
10.83  & -5.55 , -5.69 & - & - & -2.97 , -4.06 \\
11.00 & - & - & - & -3.208 , -4.19 \\
11.16 & - & - & - & -3.48 , -4.52 \\
11.33 & - & - & - & -3.77 , -4.77 \\
11.49 & - & - & - &  -4.27 , -4.85 \\
11.66 & - & - & - & -4.85 , -5.25 \\
\hline
\multicolumn{5}{|c|}{\textbf{$0.3<z<0.4$ Luminosity Functions}}\\
\hline
9.33  & - & - & -3.10 , -3.92 & - \\
9.49  & - & - & -3.58 , -4.44 & - \\
9.66  & - & - & -4.31 , -4.75 & - \\
9.83  & - & -3.03 , -3.92 & -4.88 , -5.33 & -  \\
10.00 & - & -3.43 , -4.40 & -6.09 , -6.05 & -  \\
10.16 & - & -4.04 , -4.74 & -5.79 , -5.94 & -  \\
10.33 & -3.02 , -3.87 & -4.72 , -5.29 & - & - \\
10.49 & -3.47 , -4.40 & -5.62 , -5.84 & - & - \\
10.66 & -4.13 , -4.74 & -5.79 , -5.94 & - & - \\
10.83 & -4.80 , -5.31 & - & - & -  \\
11.00 & -5.79 , -5.92 & - & - & -3.06 , -3.96 \\
11.16 & - & - & - & -3.25 , -4.05 \\
11.33 & - & - & - & -3.42 , -4.49 \\
11.49 & - & - & - &  -3.73 , -4.81\\
11.66 & - & - & - & -4.16 , -4.10 \\
11.83 & - & - & - & -4.63 , -5.36 \\
\hline
\multicolumn{5}{|c|}{\textbf{$0.4<z<0.5$ Luminosity Functions}}\\
\hline
9.66  & - & - & -3.75 , -4.43 & - \\
9.83  & - & -3.03 , -3.92 & -4.63 , -4.71 & -  \\
10.00 & - & -3.43 , -4.40 & -4.99 , -5.38 & -  \\
10.16 & - & -4.04 , -4.74 & -5.68 , -5.89 & -  \\
10.33 & - & -4.35 , -4.70 & - &  - \\
10.49 & -3.23 , -4.11 & -4.92 , -5.37 & - & -  \\
10.66 & -3.69 , -4.41 &-5.35 , -5.76 & - &  - \\
10.83 & -4.41 , -4.70 & - & - &  - \\
11.00 & -4.97 , -5.38 & - & - &  - \\
11.16 & -5.58 , -5.85 & - & - & -  \\
11.33 & - & - & - & -3.45 , -4.42 \\
11.49 & - & - & - & -3.61 , -4.42 \\
11.66 & - & - & - & -3.87 , -4.65 \\
11.83 & - & - & - & -4.23 , -5.10 \\
12.00 & - & - & - & -4.99 , -5.60 \\
\hline
\end{tabular}
\end{small}
\end{table}

\clearpage
\newpage

%

%
\bsp
\label{lastpage}
\end{document}